\documentclass[onecolumn,draftclsnofoot,11pt,dvipsnames]{IEEEtran}
\usepackage{cite}
\usepackage{cancel}
\usepackage{soul}
\usepackage{acro}
\usepackage{academicons}

 \usepackage{lineno}
\usepackage{url}
\usepackage{hyperref} 
\hypersetup{
    colorlinks=true, % Enable colored links
    linkcolor=blue, % Color of internal links
    citecolor=blue, % Color of citation links
    urlcolor=blue % Color of external hyperlinks
}
\usepackage{verbatim}

\usepackage{graphicx}
\graphicspath{../Figure/}
\DeclareGraphicsExtensions{.pdf,.jpeg,.png,.jpg, .svg}
\usepackage[font=footnotesize]{caption}
\usepackage{subcaption}

\ifundef{\XeTeXcolorgray}{\usepackage{xcolor}}{}
\usepackage{tikz}
\usepackage{makecell}
\usepackage{pgfplots}
\pgfplotsset{compat=1.14} 
\usepgfplotslibrary{groupplots}
\usetikzlibrary{calc}
\usepackage{psfrag}
\usepackage{amsmath}
\usepackage{amsfonts}
\usepackage{mathtools}
\usepackage{amssymb}
\usepackage{mathrsfs}
\usepackage{steinmetz} % To declare phase operator
\usepackage{bm}
\usepackage{bbm}
\usepackage{stackengine}
\usepackage{array}
\usepackage{theoremref}
\usepackage{textcomp}
\usepackage{stfloats}
\usepackage{algorithmic}
\usepackage[algo2e]{algorithm2e}
\usepackage{algorithm}

\newlength\myindent
\newlength\myindentt
\usepackage{lipsum}
\usepackage{mathtools}
\DeclareAcronym{AIR}{short=AIR,long= achievable information rate,}

\DeclareAcronym{AWGN}{short=AWGN,long= additive white Gaussian noise,}

\DeclareAcronym{ASE}{short=ASE,long= amplified spontaneous emission,}

\DeclareAcronym{BW-CMA}{short=BW-CMA,long= block-wise CMA,}

\DeclareAcronym{BW-DDLMS}{short=BW-DDLMS,long= DDLMS,}

\DeclareAcronym{CMA}{short=CMA,long= constant modulus algorithm,}
\DeclareAcronym{CW}{short =CW, long= continuous wave,}
\DeclareAcronym{DD}{short=DD,long= decision-directed,}

\DeclareAcronym{DD-Kabsch}{short=DD-{K}absch,long= decision-directed {K}absch,}

\DeclareAcronym{DD-Czegledi}{short=DD-{C}zegledi,long=  decision-directed {C}zegledi,}

\DeclareAcronym{DDLMS}{short=DDLMS,long= decision-directed least mean squares,}

\DeclareAcronym{DP}{short=DP,long= dual-polarization,}

\DeclareAcronym{DP-PDL}{short=DP-PDL,long= dual-polarization PDL,}

\DeclareAcronym{DSP}{short=DSP,long= digital signal processing,}

\DeclareAcronym{DOF}{short=DOF,long= degrees of freedom,}
\DeclareAcronym{EO}{short=EO-comb,long=  electro-optic frequency comb,}
\DeclareAcronym{EDFA}{short=EDFA,long=  erbium-doped fiber-amplifier,}
\DeclareAcronym{GD}{short=GD,long= gradient descent,}

\DeclareAcronym{GN}{short=GN,long= Gaussian Noise,}

\DeclareAcronym{GMI}{short=GMI,long= generalized mutual information,}
\DeclareAcronym{iid}{short = i.i.d.,long = identically and independently distributed,}
\DeclareAcronym{LMS}{short=LMS,long= least mean square,}

\DeclareAcronym{LS}{short=LS,long= least-square error,}

\DeclareAcronym{LS-DDLMS}{short= LS-DDLMS,long= LS-DDLMS,}

\DeclareAcronym{LS-SW-ULS}{short= LS-SW-ULS,long = LS-SW-ULS,}
\DeclareAcronym{LHS}{short=LHS, long=left-hand side,}
\DeclareAcronym{MCMA}{short= MCMA,long= modified {CMA},}

\DeclareAcronym{MI}{short=MI,long= mutual information,}

\DeclareAcronym{MIMO}{short=MIMO,long=  multiple-input multiple-output,}

\DeclareAcronym{ML}{short= ML,long= maximum likelihood}

\DeclareAcronym{MMA}{short=MMA,long= multi-modulus algorithm,}

\DeclareAcronym{MMSE}{short=MMSE,long= minimum mean square error,}
\DeclareAcronym{NLI}{short=NLI,long= nonlinear impairments,}

\DeclareAcronym{OFC}{short=OFC,long= optical frequency comb,}

\DeclareAcronym{pdf}{short=pdf,long= probability density function,}

\DeclareAcronym{PDL}{short=PDL,long= polarization-dependent loss,}

\DeclareAcronym{PDM}{short= PDM,long= polarization-division multiplexed,}

\DeclareAcronym{PM}{short= PM,long= polarization multiplexed,}

\DeclareAcronym{PM-16-QAM}{short=PM-$16$-QAM,long= polarization-multiplexed $16$ quadrature amplitude modulation,}

\DeclareAcronym{PMD}{short=PMD,long= polarization-mode dispersion,}

\DeclareAcronym{PN}{short=PN,long= phase noise,}

\DeclareAcronym{PTC}{short=PTC,long= polarization-time code,}

\DeclareAcronym{QAM}{short=QAM,long= quadrature amplitude modulation,}

\DeclareAcronym{QPSK}{short=QPSK,long= quadrature phase-shift keying,}

\DeclareAcronym{RDE}{short=RDE,long= radially directed equalizer,}

\DeclareAcronym{RF}{short=RF,long=radio-frequency,}

\DeclareAcronym{RHS}{short=RHS,long=right-hand side,}
\DeclareAcronym{SNR}{short=SNR,long=signal-to-noise-ratio,}

\DeclareAcronym{SOP}{short=SOP,long= state of polarization,}

\DeclareAcronym{SVD}{short= SVD,long= singular value decomposition}

\DeclareAcronym{SER}{short=SER,long= symbol error rate,}

\DeclareAcronym{SW-Kabsch}{short= SW-Kabsch,long= sliding window Kabsch,}

\DeclareAcronym{SW-LS}{short= SW-LS,long= sliding window least squares,}

\DeclareAcronym{SW-ULS}{short= SW-ULS,long= sliding window unitary least square error,}
\DeclareAcronym{WSS}{short=WSS,long=  wavelength selective switches,}
\DeclareAcronym{WDM}{short=WDM,long=  wavelength-division multiplexing,}
    \newcounter{counter}
    \newcounter{counterDefs}
    \newcounter{theoremsCounter}
    \newtheorem{theorem}[theoremsCounter]{Theorem}
    
    \newtheorem{lemma}[counter]{Lemma}
    \newtheorem{definition}[counterDefs]{Definition}

\definecolor{myMagneta}{rgb}{1, 0, 1}
\newcommand{\mbf}[1]{\mathbf{#1}}
\newcommand{\bs}[1]{\boldsymbol{#1}}
\renewcommand{\rm}[1]{\mathrm{#1}}
\newcommand{\mr}[1]{\mathrm{#1}}

\newcommand{\thetab}{\bs{\theta}}

\newcommand{\thetacb}[1]{\thetab_{#1}^{\mr{c}}}
\newcommand{\thetarb}[1]{\thetab_{#1}^{\mr{r}}}

\newcommand{\Deltab}{\bs{\Delta}}

\newcommand{\Deltacb}[1]{\Deltab_{#1}^{\mr{c}}}
\newcommand{\Deltarb}[1]{\Deltab_{#1}^{\mr{r}}}

\newcommand{\omegac}{\omega^{\mr{c}}}
\newcommand{\omegar}{\omega^{\mr{r}}}

\newcommand{\Ec}{E_{\mr{c}}}
\newcommand{\Er}{E_{\mr{r}}}

\DeclareMathOperator{\wrap}{wrap_{\pi}} % Trace
\DeclareMathAlphabet\mathbfcal{OMS}{cmsy}{b}{n}

\newcommand\norm[1]{\left\lVert#1\right\rVert}
\newcommand{\abs}[1]{\left\vert#1\right\vert}
\renewcommand{\vec}[1]{\underline{#1}} % Italic Underlined
\newcommand{\rvec}[1]{\underline{\mbf{#1}}} % Bold Vector (Random)
\newcommand{\tvec}[1]{\vec{\tilde{#1}}} % Tilde Vector

\newcommand{\CN}{\mathcal{CN}} % Complex Normal
\newcommand{\trG}{\mathcal{G}_\mathrm{tr}}

\newcommand{\CSGtr}{\mathcal{CSG}_\mathrm{tr}}

\newcommand{\SGtr}{\mathcal{SG}_\mathrm{tr}}
\newcommand{\Uniform}{\mathcal{U}} % Uniform Normal

\newcommand{\note}[2]{\overset{\text{(#1)}}{#2}}
\newcommand\oversetpi[1]{\mathstrut\mkern-1mu#1\mkern-18mu\raise1.25ex%
 \hbox{$\scriptscriptstyle 2\pi$}\mkern3mu}
\newcommand{\modplus}{\mathrel{\stackon[0pt]{$+$}{$\scriptscriptstyle \pi$}}}
\newcommand{\modsub}{\mathrel{\stackon[0pt]{$-$}{$\scriptscriptstyle \pi$}}}

\newcommand{\qedmark}{\ensuremath{\blacktriangle}}

\newcommand{\qed}{\hfill\qedmark}

\newcommand{\I}{\mr{I}} % Identity Matrix
\newcommand{\E}{\mathbb{E}}
\newcommand{\Eb}[1]{\mathbb{E}\left[#1\right]}

\newcommand{\given}{\,\vert\,}
\newcommand{\bgiven}{\,\bigl\vert\,}

\newcommand{\fparen}[1]{\left(#1\right)}
\newcommand{\paren}[1]{\big(#1\big)}

\renewcommand{\brack}[1]{\left[#1\right]}

\newcommand{\fbrace}[1]{\left\{#1\right\}}
\renewcommand{\brace}[1]{\big\{#1\big\}}
\newcommand{\moved}[1]{#1}
\newcommand{\added}[1]{#1}
\newcommand{\deleted}[1]{}

\begin{document} %All 
%\title{On the Capacity of Correlated MIMO Phase-Noise Channels with Fiber-Optic Applications}
\title{On the Capacity of Correlated Phase-Noise
Channels: An Electro-Optic Frequency Comb Example}
% =================== author names and IEEE memberships ============

 \author{Mohammad~Farsi,~\IEEEmembership{Student Member,~IEEE,}
% Viswanathan~Ramachandran,~\IEEEmembership{Member,~IEEE,}
 Hamdi~Joudeh,~\IEEEmembership{Member,~IEEE,}
 Gabriele~Liga,~\IEEEmembership{Member,~IEEE,}
 Alex~Alvarado,~\IEEEmembership{Senior Member,~IEEE,}
 Magnus~Karlsson,~\IEEEmembership{Fellow,~IEEE} and Erik~Agrell,~\IEEEmembership{Fellow,~IEEE} 
\thanks{M. Farsi and E. Agrell are with the Department
 of Electrical Engineering, Chalmers University of Technology, SE-41296 Gothenburg, Sweden (e-mail: farsim@chalmers.se; agrell@chalmers.se)}
 \thanks{M. Karlsson is with the Department of Microtechnology
 and Nanoscience, Chalmers University of Technology, SE-41296 Gothenburg,
Sweden (e-mail: magnus.karlsson@chalmers.se).}
\thanks{H. Joudeh, G. Liga, and Alex Alvarado are with the Department of Electrical Engineering, Eindhoven University of Technology (TU/e), Eindhoven,
Netherlands (e-mail: h.joudeh@tue.nl,
g.liga@tue.nl, a.alvarado@tue.nl).}
 \thanks{This work was supported by the Knut and Alice Wallenberg Foundation under grant 2018.0090. }}
\maketitle
%============================ Abstract ============================
%\input{src/abstract}

\begin{abstract}
The capacity of a discrete-time channel with correlated phase noises is investigated. In particular, the electro-optic frequency comb system is considered, where the phase noise of each subchannel is a combination of two independent Wiener phase-noise sources. Capacity upper and lower bounds are derived for this channel and are compared with lower bounds obtained by numerically evaluating the achievable information rates using quadrature amplitude modulation constellations. Capacity upper and lower bounds are provided for the high signal-to-noise ratio (SNR) regime. The multiplexing gain (pre-log) is shown to be $M-1$, where $M$ represents the number of subchannels. A constant gap between the asymptotic upper and lower bounds is observed, which depends on the number of subchannels $M$. For the specific case of $M=2$, capacity is characterized up to a term that vanishes as the SNR grows large.
\end{abstract}
% Note that keywords are not normally used for peer-reviewed papers.
%===================== INTRODUCTION =====================
\section{Introduction}

\IEEEPARstart{P}{hase} noise is a major issue in certain communication systems. It manifests as unwanted fluctuations in the signal phase and can severely degrade the quality and reliability of data transmission. One of the challenges in achieving higher throughputs involves the utilization of high-order constellations, which can make the entire system highly susceptible to the effects of phase noise. 

To assess the effect of phase noise on the throughput of communication systems, an essential approach is to analyze the Shannon capacity. However, determining the exact capacity of a phase-noise channel, even for simple channel models, remains an open challenge. While capacity bounds and their high-\ac{SNR} approximations have been documented in the literature, a closed-form solution for the capacity of phase-noise channels is currently unavailable. %Applying the duality approach and "escape-to-infinity" property of the channel input, 
The capacity of the general class of stationary phase-noise channels, including the widely used Wiener model \cite{colavolpe:2012}, was characterized in the high-\ac{SNR} regime by Lapidoth \cite{lapidoth:2002}.  
Later, Katz and Shamai \cite{katz:2004} derived upper and lower bounds on the capacity of the memoryless phase-noise channel and established that the capacity-achieving input distribution is in fact discrete. 

The capacity of the single-channel Wiener phase-noise model has been extensively studied in the context of wireless and optical fiber communications (see \cite{barletta:2012,barletta:2015,khanzadi:2015,ghozlan_pn_modles:2017} and references therein). The capacity results in \cite{lapidoth:2002} were extended to the \ac{MIMO} phase-noise channel in \cite{durisi_MIMO_Cap:2013,durisi:2014,khanzadi_simo:2015}. The existing research covers outcomes at opposite ends of the phase-noise model spectrum. On one end, \cite{durisi_MIMO_Cap:2013} characterizes the capacity of the \ac{MIMO} phase-noise channel with a common phase noise using the duality approach and the ``escape-to-infinity" property. On the other end, \cite{yang:2017} derives upper bound and pre-log expressions on the capacity of \ac{MIMO} phase-noise channel with separate phase noises. However, there remains an unaddressed area between \cite{durisi_MIMO_Cap:2013} and \cite{yang:2017}, where correlated phase noises impact \added{sub}channels and multiple phase-noise sources are present. To the best of our knowledge, the capacity of such channels has never been studied in the literature.

Recently, using \acp{EO} as a light source in optical communication systems has led to a new variant of \label{r2:c1_b} phase-noise channels \cite{ishizawa_pn_comb:2013}, in which the phase noise is correlated across different \added{sub}channels.
An \ac{EO} is a collection of equally spaced and precisely controlled optical frequencies resembling the teeth of a comb when displayed on a graph \cite{ataie_ultrahigh:2015}. 
This unique property of \acp{EO} enables them to encode and transmit vast amounts of information simultaneously on different wavelengths. Unlike traditional approaches that rely on individual laser modules, frequency combs provide equidistant frequency tones, eliminating the need for precise wavelength control and inter-\added{sub}channel guard bands.  Moreover, sharing a single light source results in a strong phase correlation between the comb lines (comb lines are phase-locked to each other) which can be utilized to either increase the phase-noise tolerance \cite{mazur_high_spectral:2018} or decrease the complexity of the digital signal processing \cite{lundberg_comb:2018}. 

Recent experiments have shown that in \acp{EO}, the phase noises are more than $99.99\%$ correlated between the channels \cite{lundberg_phase_corr:2018}, confirming the theoretical predictions \cite{ishizawa_pn_comb:2013} indicating the presence of two distinct phase noise terms that impact the comb. The first term arises from the \ac{CW} laser, which emits a constant and uninterrupted beam of coherent light, and affects all comb lines (carrier frequencies) uniformly. The second term originates from the \ac{RF} oscillator and increases linearly with the number of comb lines \cite{ishizawa_pn_comb:2013}. The phase-noise model of \acp{EO} falls in the unaddressed area between \cite{durisi_MIMO_Cap:2013} and \cite{yang:2017} where more than one phase noise source is present and the phase noises are correlated between the \added{sub}channels. {Intuitively, when all \added{sub}channels share the same source of phase noise as in \cite{durisi_MIMO_Cap:2013}, they can collectively enhance the \ac{AIR} through joint processing. Conversely, when each \added{sub}channel experiences independent phase noise as explored in \cite{yang:2017}, joint processing offers no advantage. Hence, investigating capacity in scenarios where multiple \added{sub}channels encounter correlated phase noise from multiple sources is important as it can provide insights into optimizing joint processing techniques and their effectiveness.}

\subsection{Contributions}
In this work, we investigate the capacity of a \added{parallel} channel affected by correlated phase noises originating from the transmitter and the receiver \acp{EO} and \ac{AWGN} from the amplifiers. In particular, the phase noise of each \added{sub}channel (comb line) is a combination of two independent Wiener phase-noise sources. Our contributions are as follows:
\begin{itemize}
    \item We derive capacity upper and lower bound for \added{a parallel} channel affected by multivariate correlated Wiener phase noises originating from the transmitter and the receiver \acp{EO}. To derive the upper bound, we use the duality approach \cite{lapidoth_duality:2003} considering a specific distribution for the output of the channel. For the lower bound, we determine a family of input distributions that results in a tight lower bound in the high-\ac{SNR} regime.
    
    \item We provide high-\ac{SNR} capacity upper and lower bounds that are obtained by modifying the derived upper and lower bounds. These bounds are derived through modifications to the originally derived upper and lower bounds. In particular, we show that the pre-log is one less than the number of \added{sub}channels. We also show that there is a constant gap between the high-\ac{SNR} upper and lower bounds, where the gap is a function of the number of \added{sub}channels. 
    \item We compare our bounds with lower bounds obtained by evaluating the information rates achievable with \ac{QAM} constellations numerically. 
    \item For the $2\times2$ \ac{EO} channel, the constant gap {between the high-\ac{SNR} lower and upper bounds} vanishes asymptotically as \ac{SNR} grows large.  This gives the capacity characterization in the high-\ac{SNR} regime. 

\end{itemize}

The remainder of the paper is organized as follows.  The notation and system model are presented in  Section~\ref{sec:system_model}. The main results are provided in Section~\ref{sec:main_results}. Numerical results and concluding remarks are given in Section~\ref{sec:conclusion}. The proofs to the theorems and lemmas within are presented in the Appendices~\ref{app:sec:preliminaries}--\ref{app:sec:proof_lemmas}.   

%============================= SYSTEM MODEL=============
\section{System Model}\label{sec:system_model}
\subsection{Notation}\label{subsec:notation}
Throughout the paper, we use the following notational conventions. All the vectors in the paper are $M$-dimensional and denoted by underlined letters, e.g., $\vec{x} = (x_0,\dots,x_{M-1})$ and $\vec{x}_k = (x_{k,0},\dots,x_{k,M-1})$. The $M$-dimensional vector of ones is denoted by $\mathbf{1}_{M} = (1,1,\dots,1)$. Matrices are denoted by uppercase Roman letters, and the $M$-dimensional identity matrix is denoted by $\mathrm{I}_M$. Bold-face letters $\mathbf{x}$ are used for random quantities and their corresponding nonbold counterparts $x$ for their realizations.  An $N$-tuple or a column vector of $(x_{m+1},\dots,x_{m+N})$ is denoted by $\{x_{i}\}^{m+N}_{i=m+1}$ or $\{x_{i}\}^{m+N}_{m+1}$ whenever it is clear from the context; similarly, a $K$-tuple of vectors $(\vec{x}_{k+1},\dots,\vec{x}_{k+K})$ is denoted by $\vec{x}^{k+K}_{k+1}$ . Random processes are considered as ordered sequences and indicated inside braces, i.e., $\{\mbf{x}_k\} = \{\mbf{x}_k\}_{1}^{\infty}$ is a random process. 

For any $a>0$ and $d>0$, the upper incomplete gamma function is denoted by \(\Gamma (a,d)=\textcolor{white}{,}\allowbreak \int_{d}^{\infty} \exp(-u)u^{a-1}du\), and $\Gamma(a)= \Gamma(a,0)$ denotes the gamma function. The $\log$ function refers to the base-2 logarithm. The argument (phase) of a complex value is denoted by $\phase{x}\in [-\pi,\pi)$. The $\wrap(\theta)$ function wraps $\theta$ into $[-\pi,\pi)$ and is defined as $\wrap(\theta)= \!\mod\!_{2\pi}(\theta+\pi)-\pi$. Moreover, $\theta\modplus\phi$ and  $\theta\modsub\phi$ denote $\wrap(\theta+\phi)$ and $\wrap(\theta-\phi)$, respectively. 
We denote Hadamard's (component-wise) product with $\circ$ and conjugate transpose operation with $(\cdot)^\dag$. \added{We define $E_1(x) = -\mathrm{Ei}(-x)$, where the exponential integral function is given by  $
\mathrm{Ei}(x) = - \int_{-x}^{\infty} e^{-t}/{t} \,\mathrm{d}t, \quad x > 0$
as stated in \cite[Section 8.21]  {gradshteyn2014table}.\label{eq:exponential_integral}}
We use $|\cdot|$ to denote absolute value, and $\norm{\cdot}$ to denote Euclidean norm. Whenever a scalar function is applied to a vector, e.g., $|\vec{x}|$, $\phase{\vec{x}}$, $\max(\vec{x})$, etc., it stands for applying the function to each element of the vector. Whenever inequalities are applied to a vector, e.g., $\vec{x}>c$, it stands for applying the inequalities to each element of the vector, i.e., $\vec{x}>c \iff x_i>c,~\forall i$.

\Acp{pdf} are denoted by $f_{\mbf{x}}(x)$ and conditional \acp{pdf} by $f_{\mbf{y|x}}(y|x)$, where the arguments or subscripts may sometimes be omitted if clear from the context. Expectation over random variables is denoted by $\E[\cdot]$. Sets and distributions are indicated by uppercase calligraphic letters, e.g., $\mathcal{X}$. The uniform distribution on the range $[a,b)$ is denoted by $\Uniform[a,b)$. The Gaussian and wrapped-$\pi$ Gaussian distributions with mean $\mu$ and variance $\sigma^2$ are denoted by $\mathcal{N}(\mu, \sigma^2)$ and $\mathcal{WN}(\mu, \sigma^2)$, respectively. We denote the standard zero-mean complex circularly symmetric Gaussian distribution for a scalar by $\CN(0,1)$, and for an $M$-dimensional vector by $\CN(\vec{0},\mathrm{I}_M)$. The von Mises (also known as Tikhonov) distribution with mean $\mu$ and scaling factor $\kappa$ is denoted by $\mathcal{VM}(\mu,\kappa)$. {The differential entropy rate of a stochastic process $\{\mbf{x}_k\}$ is defined as $h(\{\mbf{x}_k\}) = \lim_{k\to \infty} \frac{1}{k}h(\mbf{x}_1,\dots,\mbf{x}_k)$.} 

The truncated $M$-dimensional gamma distribution $\trG(\mu,\vec{\alpha},\gamma)$ denotes the distribution of a real vector $\rvec{r}$ with independent elements $\mbf{r}_m$ and the \ac{pdf}
\begin{equation}
f_{\rvec{r}}(\vec{r}) =
\prod_{m=0}^{M-1} \frac{e^{-r_m/\mu}}{\Gamma(\alpha_m,\gamma)}\mu^{-\alpha_m}r_{m}^{\alpha_m-1}, \quad \vec{r} >\mu\gamma\label{eq:truncated_gamma_dist}  
\end{equation}
where $\mu>0$, $ \smash{\vec{\alpha} = (\alpha_0,\dots,\alpha_{M-1})\ge 0}$, $\gamma\ge0$, and ${\Gamma(\alpha_m,\gamma)}$ is the upper incomplete gamma function. Note that when $\gamma=0$, \eqref{eq:truncated_gamma_dist} is only defined for $\vec{\alpha} > 0$.

Let $\rvec{r}\sim\trG(\mu,\vec{\alpha},\gamma)$. Then, the truncated $M$-dimensional distribution $\SGtr(\mu, \vec{\alpha},\gamma)$ is the distribution of a real vector $\rvec{s}=\sqrt{\rvec{r}}$ with \ac{pdf}
\begin{equation}
    f_{\rvec{s}}(\vec{s}) = f_{\rvec{r}}(\vec{s}^2)\prod_{m=0}^{M-1} 2s_m,
\label{eq:real_sqruare_root_truncated_gamma}
\end{equation}
where $f_{\rvec{r}}(\cdot)$ is defined in \eqref{eq:truncated_gamma_dist}.

The truncated $M$-dimensional distribution $\CSGtr(\mu,\vec{\alpha},\gamma)$ with $\mu>0$, $\vec{\alpha}\ge 0$, and $\gamma\ge0$ denotes the distribution of a complex circularly symmetric vector $\rvec{x}$ whose magnitude is distributed as $|\rvec{x}|=\rvec{s}\sim \SGtr(\mu,\vec{\alpha},\gamma)$ . The \ac{pdf} of $\rvec{x}$ is
\begin{align}
    f_{\rvec{x}}(\vec{x}) &= f_{\rvec{s}}(\abs{\vec{x}})\prod_{m=0}^{M-1}\frac{1}{2\pi|x_m|}\nonumber\\
     &= \frac{1}{\pi^M}f_{\rvec{r}}(\abs{\vec{x}}^2), \label{eq:complex_square_root_truncated_gamma}
\end{align}
where $f_{\rvec{r}}(\cdot)$ and $f_{\rvec{s}}(\cdot)$ are defined in \eqref{eq:truncated_gamma_dist} and \eqref{eq:real_sqruare_root_truncated_gamma}, respectively.  Note that if $\rvec{x}\sim \CSGtr(\mu,\vec{\alpha},\gamma)$, then $|\rvec{x}|\sim \SGtr(\mu,\vec{\alpha},\gamma)$ and $|\rvec{x}|^2\sim \trG(\mu,\vec{\alpha},\gamma)$.

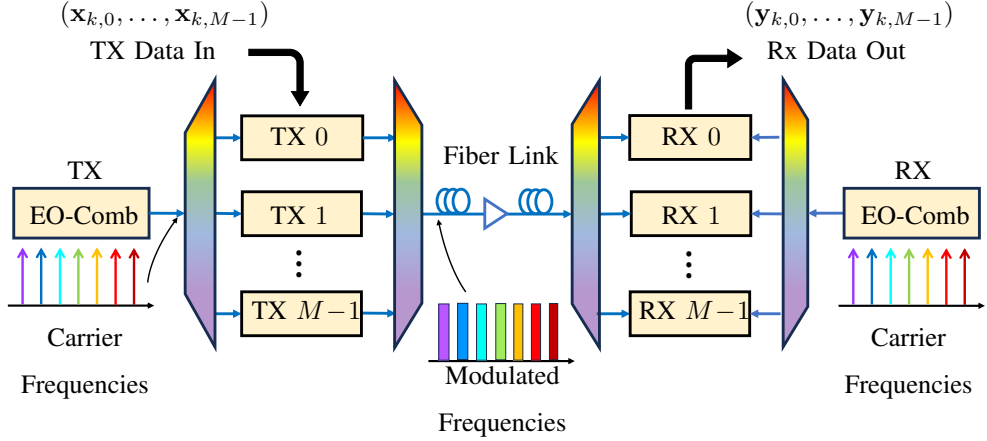
\begin{figure}[t]
    \centering
    \includegraphics[scale=0.36]{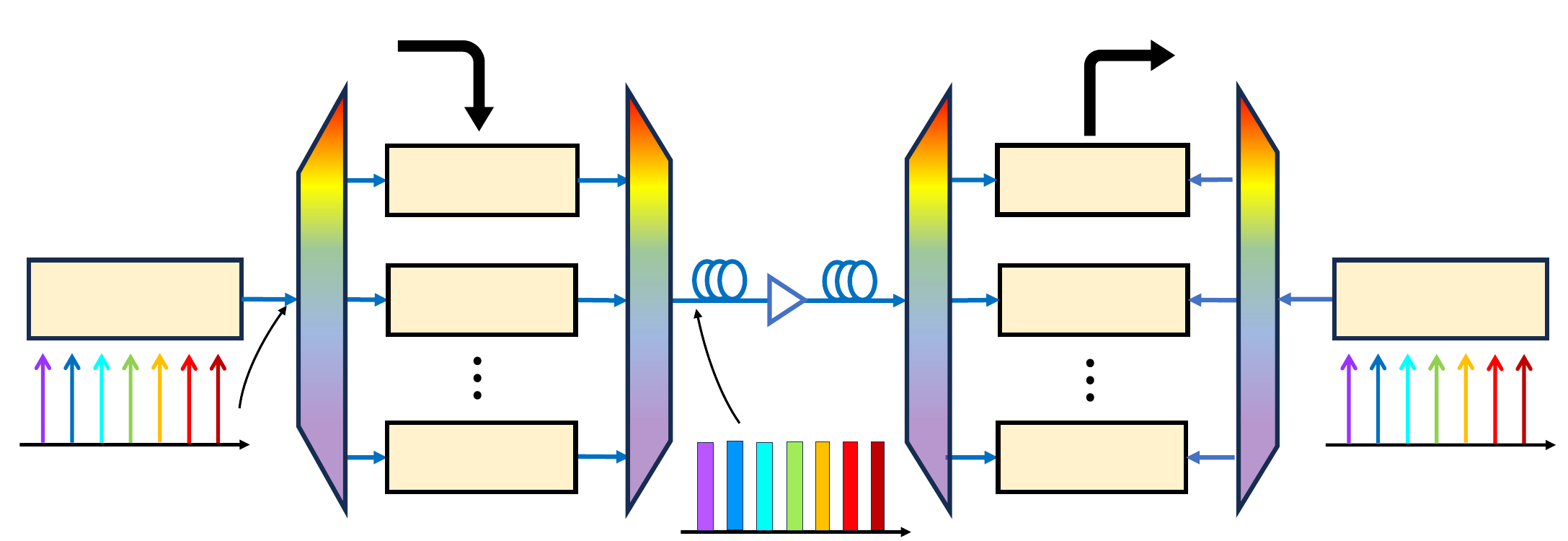}
    %---------- EO-Comb left ------------
    \put(-12.8cm,1.9cm){\rotatebox{0}{\small{EO-Comb}}}
    \put(-12.3cm,2.5cm){\rotatebox{0}{\small{TX}}}
    \put(-13.25cm,0pt){\rotatebox{0}{\small{ \begin{tabular}{c} Carrier \\ Frequencies \end{tabular}}}}
    %---------- Data In  ------------
    \put(-12cm,4.1cm){\rotatebox{0}{\small{TX Data In}}}
    \put(-12.4cm,4.6cm){\rotatebox{0}{\small{$(\mathbf{x}_{k,0},\dots,\mathbf{x}_{k,M-1})$}}}
    %---------- TX  ------------
    \put(-9.6cm,2.95cm){\rotatebox{0}{\small{TX $0$}}}
    \put(-9.6cm,1.9cm){\rotatebox{0}{\small{TX $1$}}}
    \put(-9.85cm,0.63cm){\rotatebox{0}{\small{TX $M\!-\!1$}}}
    %---------- midle part  ------------
    \put(-7.3cm,2.7cm){\rotatebox{0}{\small{Fiber Link}}}
    \put(-7.7cm,-0.5cm){\rotatebox{0}{\small{ \begin{tabular}{c} Modulated \\ Frequencies \end{tabular}}}}
    %---------- Data Out  ------------
    \put(-3cm,4.1cm){\rotatebox{0}{\small{Rx Data Out}}}
    \put(-3.3cm,4.6cm){\rotatebox{0}{\small{$(\mathbf{y}_{k,0},\dots,\mathbf{y}_{k,M-1})$}}}
     %---------- RX  ------------
    \put(-4.4cm,2.95cm){\rotatebox{0}{\small{RX $0$}}}
    \put(-4.4cm,1.9cm){\rotatebox{0}{\small{RX $1$}}}
    \put(-4.7cm,0.63cm){\rotatebox{0}{\small{RX $M\!-\!1$}}}
    %---------- EO-Comb right ------------
    \put(-2.3cm,0pt){\rotatebox{0}{\small{ \begin{tabular}{c} Carrier \\ Frequencies \end{tabular}}}}
    \put(-1.75cm,1.9cm){\rotatebox{0}{\small{EO-Comb}}}
    \put(-1.3cm,2.5cm){\rotatebox{0}{\small{RX}}}
    \caption{The structure of a comb-based \ac{WDM} link. The transmitter and receiver (local oscillator) \acp{EO} are free-running (uncorrelated).}
    \label{fig:EO_system}
\end{figure}

\subsection{\ac{EO} Phase-Noise Channel}\label{subsec: phase-noise-channel}
The \ac{EO} has gained popularity in experimental studies due to its simplicity in generation using standard components such as lasers, modulators, and \ac{RF} sources. Furthermore, it exhibits remarkable stability and consistency over extended periods of time.

As illustrated in Fig.~\ref{fig:EO_system}, the comb lines generated by an \ac{EO} can be used as carriers in a communication system by using a similar \ac{EO} at the receiver (but uncorrelated with the transmitter, i.e., \textit{free-running \acp{EO}}) \cite{lundberg2020phase}. Consequently, the collective phase noise encountered in a system utilizing \acp{EO} is the summation of both transmitter and receiver phase noises.  Typically, a comb-based optical system utilizes a large number of comb lines; thus, we are only interested in $M\ge2$ where using \acp{EO} is meaningful for communication purposes.

We consider a single-polarization $M$-dimensional transmission affected by the \ac{CW} laser and \ac{RF} oscillator phase noises, and \ac{ASE} noise at the receiver. We also assume that nonlinearities and equalization-enhanced phase noise are negligible and chromatic dispersion is compensated. %We also %Along with an optical link, several \ac{PDL}-inducing in-line components (e.g., amplifiers, isolators, couplers, multiplexers) are in place resulting in cascaded \ac{PDL} components. 

The \ac{EO} channel model can be expressed as 
\begin{equation}
    \rvec{y}_{k} = e^{j\vec{\thetab}_{k}}\circ\rvec{x}_{k}+\rvec{w}_{k},\label{eq:EO_chan_model}
\end{equation}
where  $\rvec{x}_k = \{\mbf{x}_{k,m}\}_{m=0}^{M-1}$ and $\rvec{y}_k = \{\mbf{y}_{k,m}\}_{m=0}^{M-1}$ denote the $M$-dimensional input and output vectors at discrete time $k$, respectively. The $M$-dimensional vector $\vec{\thetab}_k = \{\thetab_{k,m}\}_{m=0}^{M-1}$ denotes the phase-noise process. Moreover, the additive noise $\rvec{w}_k = \{\mbf{w}_{k,m}\}_{m=0}^{M-1}$ is $\mathcal{CN}(\vec{0},\I_M)$ and independent for all $k$ and $m$.

Note that the channel \eqref{eq:EO_chan_model} can describe various wireless \ac{MIMO} links as well as optical \ac{MIMO} channels. For instance, $\vec{\thetab}_k = \thetab_k \cdot \mbf{1}_{M}$ with $\thetab_k$ modeled as Wiener process denotes the case of common phase noise studied in \cite{durisi_MIMO_Cap:2013}. Moreover, assuming $\vec{\thetab}_k = (\thetab_{k,0}, \dots, \thetab_{k,M-1})$ with independent and stationary $\thetab_{k,m}$ can describe the model studied in \cite[Model B3]{yang:2017} when only the receiver phase noise is present and the channel matrix is identity. In the following, we introduce the \ac{EO} phase-noise model such that two independent phase-noise sources are available and the phase noises between \added{sub}channels are correlated. Thus, the \ac{EO} phase-noise model falls in the region between \cite{durisi_MIMO_Cap:2013} and \cite{yang:2017}.

\begin{figure}[t]
    \centering
    \includegraphics[scale=0.8]{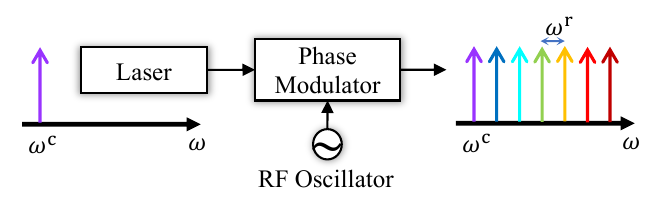}
    \put(-240pt,70pt){\footnotesize${\Ec(t) \sim e^{j(\omegac t + \thetacb{}(t))}}$}
    \put(-170pt,-5pt){\footnotesize${\Er(t) \sim \sin\fparen{\omegar t + \thetarb{}(t)}}$}
    \caption{ A typical configuration of an \ac{EO}. In this setup, a \ac{CW} laser at frequency $\omegac$ with phase noise $\thetacb{}(t)$ is fed to a phase modulator that is driven by an \ac{RF} oscillator operating at frequency $\omegar$ with phase noise $\thetarb{}(t)$. This modulation process results in the generation of $M$ comb lines characterized by a central frequency $\omegac$ and a frequency spacing $\omegar$.}
    \label{fig:EO_comb}
\end{figure}

The \ac{EO} phase noise of \added{sub}channel $m\in\{0,\dots,M-1\}$ at time $k\in \{0,1,\dots\}$ is modeled as\footnote{Note that $\theta\modplus\phi$ denotes $\wrap(\theta+\phi)$ where $\wrap(\theta)= \!\mod\!_{2\pi}(\theta+\pi)-\pi$. } \cite{ishizawa_pn_comb:2013}
\begin{equation}
    \thetab_{k,m} = \thetacb{k} \modplus m\thetarb{k},\label{def:theta_km_Eo_comb}
\end{equation}
where $\thetacb{k}$ and  $\thetarb{k}$ are the combined (transmitter and receiver) phase noise induced by \ac{CW} lasers and \ac{RF} sources, respectively. To understand where the \ac{CW} laser and \ac{RF} oscillator phase noises come from, we need to take the \ac{EO} configuration displayed in Fig.~\ref{fig:EO_comb} into account.  An \ac{EO} is created by employing a laser source that oscillates at the frequency $\omegac$ with phase noise of $\thetacb{}(t)$. This laser is then coupled with a phase modulator that is driven by an \ac{RF} source operating at frequency $\omegar$ and phase noise of $\thetarb{}(t)$ \cite{lundberg_comb:2018,metcalf:2013}. Note that $\thetacb{k}$ and $\thetarb{k}$ denote the discrete time samples of $\thetacb{}(t)$ and $\thetarb{}(t)$, respectively. 

We introduce the convention $\mathrm{c/r}$ to prevent repeating the same equations twice. The phase-noise sources are modeled as
\begin{align}
    \thetab^{\mathrm{c/r}}_{k} &=\Deltab^{\mathrm{c/r}}_{k}\modplus\thetab^{\mathrm{c/r}}_{k-1},  &&\text{if}\quad k=1,2,\dots\nonumber\\
    \thetab^{\mathrm{c/r}}_{k}&\sim \Uniform[-\pi,\pi), &&\text{if}\quad k=0,\label{eq:EO_chan_model_c}
\end{align}
where $\Deltab^{\mathrm{c/r}}_{k}\sim\mathcal{WN}(0,\sigma^{2}_\mathrm{c/r})$ independent with $\sigma^{2}_\mathrm{c/r} = 2\pi B_\mathrm{c/r}/R_\mathrm{s}$. Moreover, $R_{\mathrm{s}}$ is the symbol rate and $B_\mathrm{c/r}>0$ are the \ac{CW} laser and \ac{RF} oscillator linewidths, respectively. 
Since the initial phases are $\thetab_{0}^{\mathrm{c/r}}\sim \Uniform[-\pi,\pi)$, the processes $\{\thetab_{k}^{\mathrm{c/r}}\}$ and $\{\thetab_{k,m}\}$ are hence stationary. The \ac{iid} assumption on $\{\Deltab^{\mathrm{c/r}}_{k}\}$ implies that $\{\thetab^{\mathrm{c/r}}_{k}\}$ are Markov processes known as Wiener processes \cite{durisi_MIMO_Cap:2013} which makes the process $\{\vec{\thetab}_k\}$ a multivariate Wiener process.
Note that the assumption of $\sigma^2_\mathrm{c/r}>0$ is crucial for the validity of the results presented in this study, as it ensures that the differential entropy rate of $h(\bs{\Delta}_k^\mathrm{c/r})>-\infty$.

\subsection{Channel Capacity}
The capacity of the unitary \added{\ac{EO}}\label{r2:c1_d} phase-noise channel \eqref{eq:EO_chan_model}--\eqref{eq:EO_chan_model_c} %, which is equivalent to \eqref{eq:chModel}, 
is given by 
\begin{equation}\label{eq:capacity}
    C(\rho) = \lim_{n\to \infty} \frac{1}{n}\sup I\paren{\rvec{x}_{1}^n;\rvec{y}_{1}^n},
\end{equation}
where the supremum is over all probability distributions on $\rvec{x}_{1}^n$ that satisfy the average-power constraint
\begin{equation}\label{eq:pw_cons}
    \sum_{k=1}^{n}\E\left[{\norm{\rvec{x}_k}^2}\right]\le n\rho,
\end{equation}
where $\rho$ denotes the maximum available transmission power.

\section{Main Results}\label{sec:main_results}
This section summarizes the main results of this work.

\subsection{Capacity Bounds}\label{subsec:capacity-bounds}
In the following theorems, we derive upper and lower bounds on the capacity of the \ac{EO} phase-noise channel \eqref{eq:EO_chan_model}--\eqref{eq:EO_chan_model_c}, where the phase noises are correlated across \added{sub}channels. Specifically, Theorems~\ref{th:upper_bound} and \ref{th:lower_bound} characterize upper and lower bounds on the capacity $C(\rho)$ of the channel defined in \eqref{eq:EO_chan_model}--\eqref{eq:EO_chan_model_c}. The proofs are located in Appendix~\ref{app:sec:proof_ub_lb}.

\begin{theorem}\label{th:upper_bound}
For any $\lambda\ge0$, $\vec{\alpha}=(\alpha_0,\dots,\alpha_{M-1})>0$, and $M\ge 2$, the capacity of the channel \eqref{eq:EO_chan_model}--\eqref{eq:EO_chan_model_c} with power constraint \eqref{eq:pw_cons} can be upper-bounded as $C(\rho)\le U(\rho)$, where
\begin{align}
    U(\rho)&=\alpha_{\Sigma} \log \fparen{\frac{\rho+M}{ \alpha_{\Sigma}}}+2\log(2\pi) +\lambda-(M-2)\log e+\sum_{m=0}^{M-1}\log\Gamma(\alpha_m)\nonumber\\&\quad+\max_{\vec{s}\ge 0}\big\{R_{\lambda,\vec{\alpha}}(\rho,\vec{s})+F(M,\vec{s},\Deltacb{},\Deltarb{})\big\}.\label{eq:th_upper_bound}
\end{align}
Here, \label{r1:c1_a}
\begin{equation}
    \alpha_{\Sigma} = \sum_{m=0}^{M-1}\alpha_m,\label{eq:alpha_s}
\end{equation} 
\begin{align}
R_{\lambda,\vec{\alpha}}(\rho,\added{\rvec{s}}) &= (\alpha_{\Sigma}\log e-\lambda)\bigg(\frac{\added{\E[\norm{\rvec{s}}^2}]+M}{\rho+M}\bigg)+\!\!\sum_{m=0}^{M-1} (1-\alpha_m)\added{\fparen{\Eb{\log\fparen{\mbf{s}_{m}^2} +\mathrm{E}_1\fparen{\mbf{s}_{m}^2}}}}\nonumber\\&\quad-h\paren{|\added{\mbf{s}_{0}}+\mbf{z}_{0}|^2\bgiven \added{\mbf{s}_{0}}}-h\paren{|\added{\mbf{s}_{1}}+\mbf{z}_{1}|^2\bgiven \added{\mbf{s}_{1}}},\label{eq:Rs}
\end{align}
and 
\begin{align}
F(M,\rvec{s},\Deltacb{},\Deltarb{})= \begin{cases}
\!-h\!\fparen{\brace{\Deltacb{}\!\modplus\! m\Deltarb{}\!\modplus\!\phase{\mbf{s}_{m}+\mbf{z}_{m}}}_{0}^{1}\bgiven \rvec{s},|\rvec{s}+\rvec{z}|}.& \!\!M = 2\\ \\ \!-h(\Deltarb{})-h\big(\Deltacb{}\!\modplus\!\phase{\norm{\rvec{s}}+\mbf{v}}\bgiven\rvec{s},\abs{\norm{\rvec{s}}+\mbf{v}}\big).&\!\!M>2
    \end{cases}\label{eq:Fs}
\end{align}
For future needs, \added{$R_{\lambda,\vec{\alpha}}(\rho,\rvec{s})$ and} $ F(M,\rvec{s},\Deltacb{},\Deltarb{})$ \added{are} defined for a random $\rvec{s}$, although $\vec{s}$ is deterministic in Theorem~\ref{th:upper_bound}. Moreover, $\Deltacb{}\sim \mathcal{WN}(0,\sigma^2_\mathrm{c})$ and $\Deltarb{}\sim \mathcal{WN}(0,\sigma^2_\mathrm{r})$ are independent; the scalar $\mbf{v}\sim \CN(0,1)$ and the elements of the vector $\rvec{z}=( \mbf{z}_{0},\dots,\mbf{z}_{M-1})$ are \ac{iid} and $\CN(0,1)$ and independent of $\Deltacb{}$, $\Deltarb{}$, and $\mbf{v}$.
\qed
\end{theorem}

Theorem~\ref{th:upper_bound} is obtained by extending the method used in \cite{durisi_MIMO_Cap:2013} to derive an upper bound on the capacity of the \ac{MIMO} channel with a common phase noise between the \added{sub}channels. The proof is located in Appendix~\ref{proof:th_upper_bound}.

Theorem \ref{th:upper_bound} gives a family of upper bounds that can be tightened by minimizing \eqref{eq:th_upper_bound} over  $\lambda$ and $\vec{\alpha}$. As previously mentioned, we are interested in $M\ge2$ in this work. For the special case of $M=1$, we refer the reader to \cite[Theorem 1]{durisi_MIMO_Cap:2013}.  

\begin{theorem}\label{th:lower_bound}
For any real random vector $\rvec{s}=(\mbf{s}_0,\dots,\allowbreak\mbf{s}_{M-1})\ge0$ with independent elements that satisfies the power constraint $\E[\norm{\rvec{s}}^2]\le \rho$, the capacity of the channel \eqref{eq:EO_chan_model}--\eqref{eq:EO_chan_model_c} with power constraint \eqref{eq:pw_cons} can be lower-bounded as $C(\rho)\ge L(\rho)$, where
\begin{align}
    L(\rho) &=\log(2\pi)-(M-1)\log e+h(\Deltacb{},\Deltarb{})-2h\big(\fbrace{\Deltacb{}\!\modplus \!m\Deltarb{}\modplus\phase{\mbf{s}_m+\mbf{z}_{m}}}_{0}^{1}\given \rvec{s},\fbrace{\abs{\mbf{s}_m+\mbf{z}_{m}}}_{0}^{1}\!\big)\nonumber\\
    &\quad+h\paren{\rvec{s}^2}-\frac{1}{2}\E\brack{\log\fparen{1+2 \mbf{s}_{0}^2}}-\frac{1}{2} \E\brack{\log\fparen{1+2 \mbf{s}_{1}^2}} -\!\sum_{m=2}^{M-1}\E[g(m,\rvec{s})].
    %--------------------------------------
  \label{eq:th_lower_bound}
\end{align}
Here, $\Deltacb{}\sim \mathcal{WN}(0,\sigma^2_\mathrm{c})$ and $\Deltarb{}\sim \mathcal{WN}(0,\sigma^2_\mathrm{r})$.
The elements $\{\mbf{z}_{0},\dots,\mbf{z}_{M-1}\}$ are \ac{iid} $\CN(0,1)$ and independent of all other random quantities. Moreover, \label{r1:c2_a}
\begin{equation}
   g(m,\vec{s}) = \added{\min\fparen{\log(2\pi), \frac{1}{2} \log\fparen{2\pi e\phi(m,\vec{s})}}} 
   %h\fparen{\phi(m,\vec{s})\bgiven \vec{s},\fbrace{\abs{{s}_{i}+\mbf{z}_{i}}}_{i=0}^{m}}
   -h\fparen{\phase{{s}_{m}+\mbf{z}_{m}}\given {s}_{m},\abs{{s}_{m}+\mbf{z}_{m}}},\label{eq:g_func}
\end{equation}
where 
% \begin{align}
%      \phi(m,\vec{s})&\!=\!\!\begin{cases}
%      \phase{{s}_{2}+\mbf{z}_{2}}\modsub2\phase{{s}_{1}+\mbf{z}_{1}}\modplus\phase{{s}_{0}+\mbf{z}_{0}},& m=2.
%      \\\\
% \phase{{s}_{m}+\mbf{z}_{m}}\modsub\phase{{s}_{m-1}+\mbf{z}_{m-1}}
% \modsub\!\phase{{s}_{m-2}+\mbf{z}_{m-2}}\modplus\phase{{s}_{m-3}+\mbf{z}_{m-3}},& m>2.
%     \end{cases}\label{eq:Phi_func}
% \end{align}
\added{
\begin{align}
     \phi(m,\vec{s})&\!=\!\!\begin{cases}
     \frac{1}{s_2^2} +\frac{4}{s_1^2} +\frac{1}{s_2^2},& m=2.
     \\
     \sum_{i=m-3}^{m} \frac{1}{s_2^i} ,& m>2.
    \end{cases}\label{eq:Phi_func}
\end{align}}
\qed
\end{theorem}

The bound presented in Theorem~\ref{th:lower_bound} is a valid lower bound on the capacity for any real random vector $\rvec{s}\ge 0$; hence, it can be maximized over the distribution of $\rvec{s}$. The proof is located in Appendix~\ref{sec:proof_th_lb}. 

The reason behind selecting \added{sub}channels $m = 0$ and $m = 1$ in \eqref{eq:Rs} and \eqref{eq:th_lower_bound} is the requirement for two \added{sub}channels to acquire information about the two unknown phase noises. With the channel model exhibiting symmetry, any two adjacent \added{sub}channels could be chosen. For simplicity's sake, we opted for $m=0$ and $m=1$.

\subsection{High-\ac{SNR} Capacity Bounds}\label{subsec:high_snr_cap}

The following two theorems characterize the high-\ac{SNR} behavior of the capacity of the channel \eqref{eq:EO_chan_model}--\eqref{eq:EO_chan_model_c}.   In the high-\ac{SNR} regime, neglecting the additive noise, the output can be assumed as a rotated version of the input, i.e., $\rvec{y}_k\approx e^{j\vec{\thetab}_k}\circ\rvec{x}_k$; thus, we chose circularly symmetric input and output distributions to derive the high-\ac{SNR} bounds. The proofs are located in Appendix~\ref{app:sec:proof_asym_ub_lb}.  

\begin{theorem}\label{th:asym_upper_bound}
In the high-\ac{SNR} regime, the capacity of the channel \eqref{eq:EO_chan_model}--\eqref{eq:EO_chan_model_c} with power constraint \eqref{eq:pw_cons} behaves as $C(\rho)\le U_{\mathrm{hsnr}}(\rho)$, where 
\begin{equation}
U_{\mathrm{hsnr}}(\rho)= \!\paren{M-1}\log\fparen{\frac{\rho}{M-1}}+2\log\pi-h\paren{\Deltacb{},\Deltarb{}}+o(1), \qquad \rho \to \infty.\label{eq:th_asym_upper_bound}
\end{equation}
Here, $\Deltacb{}\sim \mathcal{WN}(0,\sigma^2_\mathrm{c})$ and $\Deltarb{}\sim \mathcal{WN}(0,\sigma^2_\mathrm{r})$ and $o(1)$ indicates a function of $\rho$ that vanishes in the limit $\rho\to\infty$. 
\qed
\end{theorem}

Theorem \ref{th:asym_upper_bound} can be interpreted as follows: at high \acp{SNR} $\rho$, the capacity of the $M$-dimensional channel \eqref{eq:EO_chan_model}--\eqref{eq:EO_chan_model_c} is upper-bounded by the capacity of an $(M-1)$-dimensional \ac{AWGN} channel plus a correction term that accounts for the memory in the channel and does not depend on the \ac{SNR} $\rho$. 

\begin{theorem}\label{th:asym_lower_bound}
In the high-\ac{SNR} regime, the capacity of the channel \eqref{eq:EO_chan_model}--\eqref{eq:EO_chan_model_c} with power constraint \eqref{eq:pw_cons}  behaves as $C(\rho) \ge L_\mathrm{hsnr}(\rho)$, where
\begin{equation}
    L_{\mathrm{hsnr}}(\rho)= \fparen{M-1}\log \fparen{\frac{\rho}{M-1}}\!+\!2\log \pi-h\paren{\Deltacb{},\Deltarb{}}\!-\!\sum_{m=2}^{M-1}g_{\mathrm{hsnr}}(m)+o(1), \quad \rho \to \infty,\label{eq:th_asymp_lb}
\end{equation}
where $\rvec{u}\sim \SGtr(1,\vec{\alpha}^{*},0)$, and 
\begin{align}
    {\alpha}^{\ast}_m =\begin{cases}
    \frac{1}{2}, &  m \in \{0,1\}. \\
    1, & m\in \{2,\dots,M-1\}.
    \end{cases}\label{eq:alpha_star}
\end{align}
Finally,
\begin{align}
  g_{\mathrm{hsnr}}(m) \!=\!\begin{cases}
      \E\brack{\log\fparen{1+4\frac{\mbf{u}_{2}^2}{\mbf{u}_{1}^2}+\frac{\mbf{u}_{2}^2}{\mbf{u}_{0}^2}}},& \!\!\!m=2.\\\\
      \E\brack{\log\fparen{1+\frac{\mbf{u}_{m}^2}{\mbf{u}_{m-1}^2}+\frac{\mbf{u}_{m}^2}{\mbf{u}_{m-2}^2}+\frac{\mbf{u}_{m}^2}{\mbf{u}_{m-3}^2}}},& \!\!\!m>2.
  \end{cases}\label{eq:g_asym_th}  
\end{align}
\qed
\end{theorem}

The term $g_{\mathrm{hsnr}}(m)$ in Theorem~\ref{th:asym_lower_bound} is independent of the \ac{SNR} $\rho$, and determines the gap between the high-\ac{SNR} lower and upper bounds. To the best of our knowledge, there are no closed-form expressions for the expectations in \eqref{eq:g_asym_th}; however, it is rather straightforward to compute them numerically.

At high \acp{SNR}, the gap between the upper and lower bounds is
\begin{equation}
    U_{\mathrm{hsnr}}(\rho)-L_{\mathrm{hsnr}}(\rho) = \sum_{m=2}^{M-1} g_{\mathrm{hsnr}}(m) + o(1), \quad \rho\to\infty,
\end{equation}
where $g_{\mathrm{hsnr}}(m)$ defined as in \eqref{eq:g_asym_th}.
For the special case of $M=2$, we have $U_\mathrm{hsnr}(\rho) - L_\mathrm{hsnr}(\rho) = o(1)$ resulting in the characterization of the high-\ac{SNR} capacity for $M=2$ as
    \begin{equation}
        C(\rho) = \log \rho +2\log \pi-h\paren{\Deltacb{},\Deltarb{}}+o(1), \qquad \rho\to\infty, \label{cor:hsnr_cap}
    \end{equation}
where $o(1)$ indicates a function of $\rho$ that vanishes as $\rho\to \infty$. 

From Theorems~\ref{th:asym_upper_bound} and \ref{th:asym_lower_bound} it can be deduced that at high \acp{SNR}, one has to give up two real \added{sub}channels (one complex \added{sub}channel) to get full knowledge about the unknown phase noises. Thus, a feasible transmission strategy might involve utilizing the phase information from two \added{sub}channels (which convey no data) to estimate and then eliminate the phase noises from the remaining \added{sub}channels, as done in \cite{lundberg_joint:2017,farsi_ecoc:2023}.

%=========================== Section VI =================================

\section{Numerical Results}\label{sec:numerical_res}
In this section, we numerically evaluated the upper bound $U(\rho)$ and lower bound $L(\rho)$ in \eqref{eq:th_upper_bound} and \eqref{eq:th_lower_bound} and compared them with their high-\ac{SNR} expressions $U_\mathrm{hsnr}(\rho)$  and $L_\mathrm{hsnr}(\rho)$---the $o(1)$ terms are neglected---in \eqref{eq:th_asym_upper_bound} and \eqref{eq:th_asymp_lb}. For every $\rho$, we minimized the upper bound \eqref{eq:th_upper_bound} over\footnote{Our analysis was confined to the intervals of $0 < \vec{\alpha} \le 10$ and $0 \leq \lambda \leq 2M\alpha_{\Sigma}$, with $\alpha_{\Sigma}$ defined in \eqref{eq:alpha_s}.} $\lambda$ and $\vec{\alpha}$. For the evaluation of the lower bound \eqref{eq:th_lower_bound}, we chose $\rvec{s}\sim\mathcal{G}_\mathrm{tr}(\mu, \vec{\alpha}, \gamma)$ where $\mu= \rho/(M-1)$ and $\gamma >0$. Then, we maximized the lower bound over\footnote{We restricted our analysis to the range of $0 < \vec{\alpha} \leq 5$ and $0 \leq \gamma \leq 2/(e^{2M/(M-2)}-1)$.} $\gamma$ and $\vec{\alpha}$ such that %$\E[\norm{\rvec{s}}^2]\le \rho$ 
\begin{equation}
    \E[\norm{\rvec{s}}^2] = \mu \sum_{m=0}^{M-1}\frac{\Gamma(1+\alpha_m,\gamma)}{\Gamma(\alpha_m,\gamma)} \le \rho,\label{eq:pww}
\end{equation}
is satisfied. We optimized parameters using the Nelder--Mead simplex algorithm \cite{nelder_simplex:1965} accompanied by the Lagrange multiplier method to handle the power constraint \eqref{eq:pww}. We also employed the toolbox in \cite{szabo_ITE:2014} for the numerical evaluation of differential entropy terms in both upper and lower bounds.

We also considered the capacity 
\begin{equation}
    C_\mathrm{awgn}(\rho)=M\log\fparen{1+\frac{\rho}{M}}\label{eq:c_awgn}
\end{equation}
of an \ac{AWGN} channel with per-\added{sub}channel \ac{SNR} equal to $\rho/M$, which is intuitively a good upper bound on $C(\rho)$ of the channel \eqref{eq:EO_chan_model}--\eqref{eq:EO_chan_model_c} at low \ac{SNR} as the additive noise is the dominant source of impairment. The \acp{AIR} using $64$-\text{QAM} and $1024$-\text{QAM}, which is a lower bound on $C(\rho)$, are denoted by $L_\text{64-\text{QAM}}(\rho)$ and $L_\text{1024-\text{QAM}}(\rho)$, respectively. We evaluated these rates using the algorithm proposed in \cite{barletta:2012} for computation of the information rates for finite-state channels. Specifically, we used $512$ levels for the discretization of the phase-noise process and a block of $2000$ channel uses. Additionally, we visually represented a shaded region denoting the ``Capacity Area''. This area spans between the minimum value among all upper bounds, i.e., $\min(U(\rho), C_\mathrm{awgn}(\rho))$, and the maximum value among the lower bounds, i.e., $\max\fparen{L_\text{64-\ac{QAM}}(\rho), L_\text{1024-\ac{QAM}}(\rho), L(\rho)}$.

\begin{figure}
    \centering
    \input{Figs/M21_sigmac_5e-5}
    \caption{Upper and lower bounds on the capacity of the channel \eqref{eq:EO_chan_model}--\eqref{eq:EO_chan_model_c} and $M=21$, $v_\mathrm{c} = 5\cdot 10^{-6}$, and $v_\mathrm{r} = 5\cdot 10^{-9}$, i.e., $\sigma^2_\mathrm{c} = \pi \cdot10^{-5}$ and $\sigma^2_\mathrm{r} = \pi\cdot10^{-8}$.}
    \label{fig:M_5_practical}
\end{figure}
We define the normalized linewidth $v_{\mathrm{c/r}} = B_{\mathrm{c/r}}/R_\mathrm{s}$ (linewidth divided by symbol rate). In practice, the symbol rate $R_\mathrm{s}$ is in the range $0.1$--$100$ Gbaud, and the number of \added{sub}channels is typically high ($M>10$) in a system utilizing \acp{EO}. The laser
linewidth $B_\mathrm{c}$ and RF oscillator linewidth $B_\mathrm{r}$ are typically in
the range of $1-1000$ kHz and $1-1000$ Hz, respectively. Hence, the normalized linewidth of the \ac{CW} laser $v_\mathrm{c}$ may vary in the range of $10^{-8}-10^{-2}$ depending on the particular application and transmission scenario. Similarly, the normalized linewidth of the \ac{RF} oscillator $v_\mathrm{r}$ could fall within the range of $10^{-11}-10^{-5}$. 

\begin{figure}
    \centering
    \definecolor{mycolor1}{rgb}{0.75294,0.75294,0.75294}%
\definecolor{mycolor2}{rgb}{0.00000,0.39216,0.00000}%
\definecolor{mycolor3}{rgb}{0.00000,0.50196,0.00000}%
\newcommand\w{7.5}
\newcommand\h{6.566}
\newcommand\lineWidth{1.2}
\begin{tikzpicture}

\begin{axis}[%
width=\w cm,
height=\h cm,
at={(0cm,0cm)},
scale only axis,
xmin=0,
xmax=43,
grid=both,
xminorticks=true,
minor tick num=4,
xlabel style={font=\color{white!15!black}},
xlabel={SNR per \added{sub}channel (dB)},
ymin=0.01,
ymax=16,
ylabel style={font=\color{white!15!black}},
ylabel={bits per \added{sub}channel},
axis background/.style={fill=white},
title style={font=\bfseries},
xminorgrids,
yminorgrids,
grid style={dotted,white!70!black},
legend style={at={(0.03,0.97)}, anchor=north west, font=\scriptsize,legend cell align=left, align=left, draw=white!15!black}
]
\pgfplotsset{every tick label/.append style={font=\footnotesize},}
\addplot[area legend, draw=none, fill=mycolor1]
table[row sep=crcr] {%
x	y\\
-3.01029995663981	0.458374154871375\\
-1.01029995663981	0.684046427385777\\
0.989700043360188	0.967786326885898\\
2.98970004336019	1.31353648209027\\
4.98970004336019	1.70283290760232\\
6.98970004336019	2.1327404235839\\
8.98970004336019	2.57866135823533\\
10.9897000433602	3.03940734803073\\
12.9897000433602	3.50081923592318\\
14.9897000433602	3.95675549081705\\
16.9897000433602	4.40076683285218\\
18.9897000433602	4.83331806228322\\
20.9897000433602	5.23707274165628\\
22.9897000433602	5.63351517324072\\
24.9897000433602	6.01479319542638\\
26.9897000433602	6.38483674132647\\
28.9897000433602	6.90591716714878\\
30.9897000433602	7.54076569903137\\
32.9897000433602	8.1515949602787\\
34.9897000433602	8.73790886829041\\
36.9897000433602	9.29585795622371\\
38.9897000433602	9.82503901110362\\
40.9897000433602	10.3258683981572\\
42.9897000433602	10.7998967736642\\
44.9897000433602	11.2490578650495\\
46.9897000433602	11.6773097959582\\
46.9897000433602	12.0904350435382\\
44.9897000433602	11.7651134595001\\
42.9897000433602	11.4293743586527\\
40.9897000433602	11.0961040727924\\
38.9897000433602	10.7635194708674\\
36.9897000433602	10.4221566313418\\
34.9897000433602	10.0977215852962\\
32.9897000433602	9.75877491981213\\
30.9897000433602	9.29352336954657\\
28.9897000433602	8.75317098833452\\
26.9897000433602	8.21560732474469\\
24.9897000433602	7.66008356285243\\
22.9897000433602	7.10145610607267\\
20.9897000433602	6.54601188953204\\
18.9897000433602	5.97982425580195\\
16.9897000433602	5.41012097428575\\
14.9897000433602	4.82731980646834\\
12.9897000433602	4.19791499492918\\
10.9897000433602	3.59314442875816\\
8.98970004336019	3.06887165928912\\
6.98970004336019	2.44601991378832\\
4.98970004336019	2.02230495533734\\
2.98970004336019	1.58040401380984\\
0.989700043360188	1.17373075418175\\
-1.01029995663981	0.841930136220867\\
-3.01029995663981	0.584962500721156\\
}--cycle;\addlegendentry{Capacity Area}

\addplot [color=black, line width=1.2pt]
  table[row sep=crcr]{%
-3.01029995663981	0.584962500721156\\
-1.01029995663981	0.841930136220867\\
0.989700043360188	1.17373075418175\\
2.98970004336019	1.58040401380984\\
4.98970004336019	2.05477442105536\\
6.98970004336019	2.58496250072116\\
8.98970004336019	3.15776584123307\\
10.9897000433602	3.76122485738499\\
12.9897000433602	4.38580088075509\\
14.9897000433602	5.0244911053703\\
16.9897000433602	5.6724253419715\\
18.9897000433602	6.32633347725687\\
20.9897000433602	6.98406888349338\\
22.9897000433602	7.64424267406752\\
24.9897000433602	8.30596446829023\\
26.9897000433602	8.96866679319521\\
28.9897000433602	9.63198931397432\\
30.9897000433602	10.2957037600375\\
32.9897000433602	10.9596657368193\\
34.9897000433602	11.6237839916194\\
36.9897000433602	12.2880008897076\\
38.9897000433602	12.9522800428473\\
40.9897000433602	13.6165984823797\\
42.9897000433602	14.2809417123833\\
44.9897000433602	14.9453005850856\\
46.9897000433602	15.6096693280491\\
};\addlegendentry{$C_\mathrm{awgn}(\rho)$ \eqref{eq:c_awgn}}

\addplot [color=red, dashdotted, line width=1.2pt]
  table[row sep=crcr]{%
-3.01029995663981	3.76154462188656\\
-1.01029995663981	4.0937374313753\\
0.989700043360188	4.42593024086403\\
2.98970004336019	4.75812305035277\\
4.98970004336019	5.09031585984151\\
6.98970004336019	5.42250866933024\\
8.98970004336019	5.75470147881898\\
10.9897000433602	6.08689428830772\\
12.9897000433602	6.41908709779645\\
14.9897000433602	6.75127990728519\\
16.9897000433602	7.08347271677392\\
18.9897000433602	7.41566552626266\\
20.9897000433602	7.7478583357514\\
22.9897000433602	8.08005114524013\\
24.9897000433602	8.41224395472887\\
26.9897000433602	8.74443676421761\\
28.9897000433602	9.07662957370634\\
30.9897000433602	9.40882238319508\\
32.9897000433602	9.74101519268381\\
34.9897000433602	10.0732080021725\\
36.9897000433602	10.4054008116613\\
38.9897000433602	10.73759362115\\
40.9897000433602	11.0697864306388\\
42.9897000433602	11.4019792401275\\
44.9897000433602	11.7341720496162\\
46.9897000433602	12.066364859105\\
};\addlegendentry{$U_{\mathrm{hsnr}}(\rho)$ \eqref{eq:th_asym_upper_bound}, $L_{\mathrm{hsnr}}(\rho)$  \eqref{eq:th_asymp_lb}}

\addplot [color=red, line width=1.2pt]
 table[row sep=crcr]{%
-3.01029995663981	0.621997197997496\\
-1.01029995663981	0.885445578325207\\
0.989700043360188	1.2257297918959\\
2.98970004336019	1.60174824206557\\
4.98970004336019	2.02230495533734\\
6.98970004336019	2.44601991378832\\
8.98970004336019	3.06887165928912\\
10.9897000433602	3.59314442875816\\
12.9897000433602	4.19791499492918\\
14.9897000433602	4.82731980646834\\
16.9897000433602	5.41012097428575\\
18.9897000433602	5.97982425580195\\
20.9897000433602	6.54601188953204\\
22.9897000433602	7.10145610607267\\
24.9897000433602	7.66008356285243\\
26.9897000433602	8.21560732474469\\
28.9897000433602	8.75317098833452\\
30.9897000433602	9.29352336954657\\
32.9897000433602	9.75877491981213\\
34.9897000433602	10.0977215852962\\
36.9897000433602	10.4221566313418\\
38.9897000433602	10.7635194708674\\
40.9897000433602	11.0961040727924\\
42.9897000433602	11.4293743586527\\
44.9897000433602	11.7651134595001\\
46.9897000433602	12.0904350435382\\
};\addlegendentry{$U(\rho)$ \eqref{eq:th_upper_bound}}

\addplot [color=blue, line width=1.2pt]
table[row sep=crcr]{%
-3.01029995663981	-6.53492253237752\\
-1.01029995663981	-5.58241800015568\\
0.989700043360188	-4.62993958764867\\
2.98970004336019	-3.67871265074529\\
4.98970004336019	-2.72625812630954\\
6.98970004336019	-1.77428014656304\\
8.98970004336019	-0.836988784814595\\
10.9897000433602	0.0798415815515212\\
12.9897000433602	0.962187807846464\\
14.9897000433602	1.81118889013992\\
16.9897000433602	2.62339822676552\\
18.9897000433602	3.40239300212079\\
20.9897000433602	4.15104521562977\\
22.9897000433602	4.8728590399942\\
24.9897000433602	5.5721640064351\\
26.9897000433602	6.24939388387648\\
28.9897000433602	6.90591716714878\\
30.9897000433602	7.54076569903137\\
32.9897000433602	8.1515949602787\\
34.9897000433602	8.73790886829041\\
36.9897000433602	9.29585795622371\\
38.9897000433602	9.82503901110362\\
40.9897000433602	10.3258683981572\\
42.9897000433602	10.7998967736642\\
44.9897000433602	11.2490578650495\\
46.9897000433602	11.6773097959582\\
};\addlegendentry{$L(\rho)$ \eqref{eq:th_lower_bound}}

\addplot [color=mycolor3, dashed, line width=1.2pt, mark=+, mark options={solid, mycolor3}]
table[row sep=crcr]{%
-3.01029995663981	0.459282811630734\\
-1.01029995663981	0.680504721960291\\
0.989700043360188	0.969952753880413\\
2.98970004336019	1.3107415860063\\
4.98970004336019	1.70061002786073\\
6.98970004336019	2.12401592630257\\
8.98970004336019	2.5703840421003\\
10.9897000433602	3.02432008540281\\
12.9897000433602	3.48346698272829\\
14.9897000433602	3.93197013620038\\
16.9897000433602	4.3609603666028\\
18.9897000433602	4.75804215359739\\
20.9897000433602	5.10024222965351\\
22.9897000433602	5.3605742562552\\
24.9897000433602	5.54767110389216\\
26.9897000433602	5.66937811313419\\
28.9897000433602	5.74194884362008\\
30.9897000433602	5.78126542860795\\
32.9897000433602	5.80059303091039\\
34.9897000433602	5.80790763126229\\
36.9897000433602	5.81031076312659\\
38.9897000433602	5.81181942413942\\
40.9897000433602	5.81140079025879\\
42.9897000433602	5.81409777520394\\
44.9897000433602	5.81172273559258\\
46.9897000433602	5.8044089244475\\
};\addlegendentry{$L_\text{64-QAM}(\rho)$}

\addplot [color=mycolor3, dashed, line width=1.2pt, mark size=2.0pt, mark=asterisk, mark options={solid, mycolor3}]
table[row sep=crcr]{%
-3.01029995663981	0.458374154871375\\
-1.01029995663981	0.684046427385777\\
0.989700043360188	0.967786326885898\\
2.98970004336019	1.31353648209027\\
4.98970004336019	1.70283290760232\\
6.98970004336019	2.1327404235839\\
8.98970004336019	2.57866135823533\\
10.9897000433602	3.03940734803073\\
12.9897000433602	3.50081923592318\\
14.9897000433602	3.95675549081705\\
16.9897000433602	4.40076683285218\\
18.9897000433602	4.83331806228322\\
20.9897000433602	5.23707274165628\\
22.9897000433602	5.63351517324072\\
24.9897000433602	6.01479319542638\\
26.9897000433602	6.38483674132647\\
28.9897000433602	6.74079472829791\\
30.9897000433602	7.08698051246854\\
32.9897000433602	7.42049031650242\\
34.9897000433602	7.73057363936699\\
36.9897000433602	8.02382125410065\\
38.9897000433602	8.2939664485261\\
40.9897000433602	8.54020052019245\\
42.9897000433602	8.7539372751339\\
44.9897000433602	8.91798093378268\\
46.9897000433602	9.01078923995261\\
};\addlegendentry{$L_\text{1024-QAM}(\rho)$}

\node[rotate=0](p1) at(40,6){};
\node[rotate=0](p2) at(40,4.75){\scriptsize \textcolor{black}{$5.81$}};
\draw[->,line width=1 pt] (p2)--(p1);
\node[rotate=0](y1) at(40,8.4){};
\node[rotate=0](y2) at(40,7.15){\scriptsize \textcolor{black}{$8.41$}};
\draw[->,line width=1 pt] (y2)--(y1);
\end{axis}

\begin{axis}[%
width=0in,
height=0in,
at={(0in,0in)},
scale only axis,
xmin=0,
xmax=1,
ymin=0,
ymax=1,
axis line style={draw=none},
ticks=none,
axis x line*=bottom,
axis y line*=left
]
\end{axis}
\end{tikzpicture}%
    \caption{Upper and lower bounds on the capacity of the channel  \eqref{eq:EO_chan_model}--\eqref{eq:EO_chan_model_c}  and  $M=2$, $v_\mathrm{c} = 5\cdot 10^{-3}$, and $v_\mathrm{r} = 5\cdot 10^{-5}$, i.e., $\sigma^2_\mathrm{c} = \pi \cdot10^{-2}$ and $\sigma^2_\mathrm{r} = \pi\cdot10^{-4}$.}
    \label{fig:M_2}
\end{figure}
Here, we utilized real-world channel parameters by setting\footnote{As an example one can set the symbol rate $R_\mathrm{s} = 20$ Gbaud, $B_\mathrm{c} = 100$ kHz, and $B_\mathrm{r} = 100$ Hz.}  $v_\mathrm{c} = \added{5\cdot10^{-6}}$ and $v_\mathrm{r} = 5\cdot10^{-9}$ %$R_\mathrm{s} = 28 ~\mathrm{Gbaud}$, $B_\mathrm{c} = 100~\mathrm{kHz}$, and $B_\mathrm{r} = 100~\mathrm{Hz}$
, corresponding to
$\sigma^2_\mathrm{c}= \pi\cdot10^{-5}$ and $\sigma^2_\mathrm{r}= \pi\cdot 10^{-8}$. Fig.~\ref{fig:M_5_practical} shows the bounds for $M=21$, where it can be seen that $C_\mathrm{awgn}(\rho)$ is a tighter upper bound than our bound throughout the studied \ac{SNR} range as it performs closer to the lower bound from the \acp{QAM} and $L(\rho)$. This is expected, since with the selected channel parameters, the phase noise is extremely low. As a result, the additive Gaussian noise emerges as the dominant impairment within the shown \ac{SNR} range. Consequently, the capacity is expected to closely resemble that of the \ac{AWGN} channel. Furthermore,  the converging nature of $U(\rho)$ towards $C_\mathrm{awgn}(\rho)$ curve is evident, and they are projected to intersect at higher \ac{SNR} levels. However, such high \ac{SNR} values fall beyond practical relevance. From the analysis, we can conclude that in scenarios with extremely low phase noises---typical in practical \ac{EO} applications---the \ac{AWGN} capacity serves as a sufficiently stringent upper bound.

In Fig.~\ref{fig:M_2}, we show the results for $M=2$ with %$B_\mathrm{c} = 100~\mathrm{MHz}$, and $B_\mathrm{r} = 1~\mathrm{MHz}$, 
$v_\mathrm{c} = 5\cdot 10^{-3}$, and $v_\mathrm{r} = 5\cdot 10^{-5}$,
which correspond to
$\sigma^2_\mathrm{c}= \pi\cdot10^{-2}$ and $\sigma^2_\mathrm{r}= \pi\cdot 10^{-4}$. It can be seen that $U(\rho)$ is a tighter upper bound than $C_\mathrm{awgn}(\rho)$ throughout the studied \ac{SNR} range. The bound $L_\text{1024-\ac{QAM}}(\rho)$ is tighter than $L(\rho)$ up to about $28$ dB, above which $L(\rho)$ is tighter. This behavior is expected, as the input distribution for the $L(\rho)$ is chosen such that it achieves the capacity at high \acp{SNR}. 
The lack of saturation in $L_\text{1024-\ac{QAM}}$ to its designated nominal point of $10$ bits per \added{sub}channel can be attributed to the average constellation rotation induced by phase noise, which consistently surpasses the maximum tolerable rotation for the $1024\text{-\ac{QAM}}$ constellation. The bounds approach the high-\ac{SNR} expressions as the \ac{SNR} increases, confirming that the high-\ac{SNR} capacity for $M=2$ follows Theorem~\ref{th:asym_upper_bound}. 
It is important to emphasize that the case where $M=2$ lacks practical relevance in systems employing \acp{EO} as a light source. Nevertheless, from an information-theoretic perspective, this scenario holds significance as it serves to illustrate the capacity achieved at high \ac{SNR}, as expressed in \eqref{cor:hsnr_cap}.

\label{r1:c3_start}\added{In Fig.~\ref{fig:M_2}, we observe that the gap between the lower bounds of the \ac{QAM} constellations and their corresponding nominal values remains significant, particularly at moderate to high \ac{SNR}. This raises the question of whether alternative input distributions could yield tighter lower bounds. One promising approach is to optimize ring constellations that maximize \eqref{eq:th_lower_bound}, where symbols are drawn from a distribution with a uniformly distributed phase and discrete magnitudes \cite{Krishnan2013}. Since the capacity-achieving distribution is circularly symmetric, a sufficiently large number of rings can approximate any circularly symmetric input distribution, potentially leading to tighter lower bounds. However, optimizing ring constellations for finite \acp{SNR} introduces substantial complexity. The number of rings, their respective probability mass functions, and their magnitude values must be optimized under the average power constraint \eqref{eq:pw_cons}. This optimization can quickly become intractable, particularly if the optimal number of rings scales exponentially with capacity. Given these challenges, we evaluated the lower bounds for the truncated gamma distribution $\CSGtr(\cdot)$ and \ac{QAM} constellations, which offer a balance between analytical tractability and practical relevance. Although optimizing ring constellations could potentially improve lower bounds, their optimization is beyond the scope of this work and is left for future investigation.}\label{r1:c3_end}

\begin{figure}
    \centering
    \input{Figs/M21_sigmac_5e-3}
    \caption{Upper and lower bounds on the capacity of the channel \eqref{eq:EO_chan_model}--\eqref{eq:EO_chan_model_c} and $M=21$, $v_\mathrm{c} = 5\cdot 10^{-3}$, and $v_\mathrm{r} = 5\cdot 10^{-5}$, i.e., $\sigma^2_\mathrm{c} = \pi \cdot10^{-2}$ and $\sigma^2_\mathrm{r} = \pi\cdot10^{-4}$.}
    \label{fig:M_5}
\end{figure}
%In Fig. \ref{fig:M_2}, the curves for $M=2$ are displayed. 

In Fig.~\ref{fig:M_5}, results are shown for $M=21$. The upper bound $C_\mathrm{awgn}(\rho)$ is tighter up to $28$ dB (see the magnified window on the figure); then, $U(\rho)$ becomes tighter. This behavior mainly comes from the $F(\cdot)$ function defined in \eqref{eq:Fs}, where we used a loose upper bound for $M>2$. The bounds approach the high-\ac{SNR} expressions as the \ac{SNR} increases. Moreover, at high \ac{SNR}, the gap between the lower bound $L(\rho)$ and the upper bound $U(\rho)$ approaches a constant gap (approximately $1.19$ bits per \added{sub}channel) as expressed in \eqref{eq:g_asym_th}. 

While the linewidth of the studied \ac{CW} laser and \ac{RF} oscillator, as depicted in Figs.~\ref{fig:M_2} and \ref{fig:M_5}, may not currently hold practical significance, the ongoing trend in optical communication systems is geared towards developing more cost-effective and accessible solutions. This trajectory may lead to the utilization of lower-cost lasers and oscillators characterized by higher linewidths. Furthermore, in specific applications like space communications, exceptionally low symbol rates may be employed, resulting in elevated normalized linewidth values. Consequently, the mentioned figures can offer valuable insights into assessing the influence of the linewidth of the \ac{CW} laser and \ac{RF} oscillator on the capacity of the \ac{EO} channel. This is especially relevant in the context of emerging cost-conscious and accessible system designs or potential applications in specialized niches.

%====================== CONCLUSION ========================
\section{Discussion and Conclusions}\label{sec:conclusion}
We obtained lemmas that establish upper and lower bounds for the generic parallel \label{r2:c1_e} channel under the influence of multivariate Wiener phase noise. These lemmas can serve as a foundation for deriving capacity bounds across various phase-noise models. Then, we studied the capacity of a parallel channel affected by correlated phase noises originating from \acp{EO}. Specifically, the phase noise of each \added{sub}channel (comb line) is a combination of two independent Wiener phase-noise sources: the \ac{CW} laser phase noise, which uniformly affects all \added{sub}channels, and \ac{RF} oscillator phase noises that increase linearly with the \added{sub}channel number.
We derived lower and upper bounds on channel capacity, illustrating the capacity's behavior for various values of \ac{SNR} and phase-noise parameters. Additionally, high-\ac{SNR} capacity upper and lower bounds were derived, revealing a pre-log of $M-1$, where $M$ represents the number of \added{sub}channels. A physical intuition to the loss of one complex \added{sub}channel (equivalent to two signal space degrees of freedom) is attributed to the sacrifice of two real dimensions to account for the impact of the two phase noise sources. Hence, a viable transmission scheme could be using the phase of two \added{sub}channels (which carry no data) to estimate the phase noises and remove them from the rest of the \added{sub}channels. The same intuition can be employed to justify the high-\ac{SNR} capacity bound derived in \cite{durisi_MIMO_Cap:2013}, wherein the pre-log becomes $M-1/2$ due to the presence of only one unknown phase noise source.

Numerical evaluations indicated that in scenarios with extremely low phase noises---typical in practical optical applications---the \ac{AWGN} channel capacity serves as a sufficiently accurate upper bound.

While the capacity bounds presented in this paper are initially derived for a specific \ac{EO} phase-noise channel, the majority of techniques and derivations hold a general applicability that can be adapted for various phase-noise channels. More specifically:
\begin{itemize}
    \item The capacity upper bound (\ref{eq:cap_ub}) from Lemma~\ref{lemma:general_ub} applies not only to the \ac{EO} phase-noise model (\ref{def:theta_km_Eo_comb}--\ref{eq:EO_chan_model_c}) but also to any phase noise modeled as a multivariate Wiener process. Hence, altering the upper bounds for a different channel requires revisiting \eqref{eq:duality_bound} from Lemma~\ref{lemma:duality_bound}, which utilizes the duality bound, and \eqref{eq:lemm_I_y_theta} from Lemma~\ref{lemma:I_y_theta}, which exploits the memory and correlations. Note that the techniques in the proofs of the aforementioned lemmas are general to some extent and could be reused.
    \item The capacity lower bound \eqref{eq:cap_lb} in Lemma~\ref{lemma:general_lb} is not limited to \ac{EO} phase-noise model \eqref{def:theta_km_Eo_comb}--\eqref{eq:EO_chan_model_c} and remains valid even if the phase-noise vector is modeled as a multivariate Wiener process. Therefore, adjusting the lower bound for a different channel involves deriving the \ac{RHS} terms in \eqref{eq:cap_lb}. The choice of the input distribution and its parameters depends on the channel parameters and requires careful consideration.
    \item The results can also potentially be extended to account for more than two phase-noise sources. One example would be the soliton microcombs channels where phase noise arises from three different sources, namely, \ac{CW} laser, pump laser, and shot noise \cite{lei_microcomb:2022}. Our hypothesis is that the pre-log term might alter to $M-d/2$ where $d$ is the number of independent phase-noise sources. The rationale behind our hypothesis is that one needs to sacrifice $d$ \added{real} \added{sub}channels to gain the full knowledge of $d$ unknown phases.
\end{itemize}
 
An interesting open question is to establish tighter upper bounds in the low-\ac{SNR} regime for $M>2$. We believe the lack of tightness in the upper bound originates from \eqref{eq:lemm_I_y_theta} in Lemma~\ref{lemma:I_y_theta}, where the derived bound on mutual information is loose and could potentially be refined. Another area for future exploration is refining the lower bound for the low-\ac{SNR} regime, as Theorem~\ref{th:lower_bound} presents loose bounds for this regime.

%============================== APPENDIX ===================

\section{Acknowledgments}
The authors express gratitude to Prof. Luca Barletta for generously providing the source codes used to compute the information rates of \ac{QAM} inputs.

%\appendix
\appendices

\section{Mathematical Preliminaries}\label{app:sec:preliminaries}

\begin{definition}\label{def:circularly_symmetric}
    A vector random process $\{\vec{\bs{\omega}}_k\}$ is said to be \textit{circularly symmetric} if 
    \begin{equation}
        \{\vec{\bs{\omega}}_k\} \sim \{e^{j\bs{\Theta}_k}\vec{\bs{\omega}}_k\},
    \end{equation}
    where the process $\{\bs{\Theta}_k\}$ is $\Uniform[-\pi,\pi)$ and independent of $\{\vec{\bs{\omega}}_k\}$. 
\end{definition}

\begin{lemma}\label{prop:1}
The input process $\{\rvec{x}_k\}$ that achieves the capacity of the channel~\eqref{eq:EO_chan_model}--\eqref{eq:EO_chan_model_c} is circularly symmetric. 
\end{lemma}
\begin{IEEEproof}
The proof follows the same steps as the one in \cite[Prop. 7]{Moser_fading_number:2009} and relies on the fact that $e^{j\bs{\Theta}_k}\rvec{w}_k\sim\rvec{w}_k$ for any circularly symmetric random variable $\rvec{w}_k$.
\end{IEEEproof}

\begin{lemma}[escape-to-infinity]\label{lemma:escape_to_infty}
Fix a real-valued scalar $\xi>0$. % and let %$\mathcal{K}(\xi)=\{\vec{x}\in \mathbb{C}^M:|x_i|\ge\xi\, \forall i\}$. 
Denote by $C^{(\xi)}(\rho)$ the capacity of the channel \eqref{eq:EO_chan_model}--\eqref{eq:EO_chan_model_c} when the input signal is subject to the average-power constraint \eqref{eq:pw_cons} and to the additional constraint that $\abs{\rvec{x}_{k}}\ge \xi$ almost surely for all $k$. Then,
\begin{equation}
    C(\rho) = C^{(\xi)}(\rho) + o(1), \quad \rho \to \infty,
\end{equation}
with $C(\rho)$ obtained in \eqref{eq:capacity}.
\begin{IEEEproof}
The lemma follows directly from \cite[Def. 4.11, Thm. 4.12]{lapidoth_duality:2003}.
\end{IEEEproof}
\end{lemma}

\begin{lemma}[\hspace{-.03cm}{\cite[Lemma 6.9]{lapidoth_duality:2003}}]\label{lemma:a}
Let $\bs{\upsilon}$ and $\bs{\tau}$ be two independent real random variables satisfying $\Eb{\abs{\bs{\upsilon}}}< \infty$ and $\Eb{\abs{\bs{\tau}}}< \infty$. Then,
\begin{equation}
    \lim_{\varepsilon \to 0} h(\bs{\upsilon}+\varepsilon\bs{\tau}) = h(\bs{\upsilon}).
\end{equation}
\end{lemma}

\begin{lemma}[\hspace{-.03cm}{\cite[Lemma 6.15]{lapidoth_duality:2003}}]\label{lemma:b}
For any real random variable  $\bs{\tau}\ge 0$,
\begin{equation}
   h(\log \bs{\tau}) = h(\bs{\tau})-\E[\log \bs{\tau}],
\end{equation}
and
\begin{equation}
   h\paren{\bs{\tau}^2} = h(\bs{\tau})+\E\brack{\log \bs{\tau}} +\log 2.\label{eq:lemma_b}
\end{equation}
\end{lemma}

%See Lemma 6.15 in \cite{lapidoth_duality:2003}.

\begin{lemma}[\hspace{-.03cm}{\cite[Lemmas 3 and 4]{yang:2017}}]\label{lemma:polar_entropy}
For any $m$-dimensional complex random vector $\vec{\bs{\omega}}$ with $h(\vec{\bs{\omega}})>-\infty$, we have 
\begin{equation}
    h(\vec{\bs{\omega}})= h\left(|\vec{\bs{\omega}}|^2\right)+ h\fparen{\phase{\vec{\bs{\omega}}}\bgiven|\vec{\bs{\omega}}|}-m\log 2.\label{eq:lemma2_a}
\end{equation}
In particular, if the elements of $\vec{\bs{\omega}}$ are circularly symmetric with independent phases, then
\begin{equation}
    h(\vec{\bs{\omega}}) = h\left(|\vec{\bs{\omega}}|^2\right) +m\log\pi.\label{eq:lemma2_b}
\end{equation}
\end{lemma}

% \begin{lemma}\label{lemma:4}
% Let $\bs{\omega}\sim\CN(0,1)$ and for a given real $\beta$, define $\bs{\tau}=|\beta+\bs{\omega}|^2$. Then,
% \label{r1:c1_lemma6}\added{
% \begin{align}
%        h\paren{\bs{\tau}} & \le  \frac{1}{2}\log (4\pi e (1+\beta^2)),\\ 
%   \Eb{\log \bs{\tau}} &= \log \beta^2 - \mathrm{Ei}(-\beta^2),
% \end{align}
% where $\mathrm{Ei}(\cdot)$ denotes the exponential integral function \cite{} [8, Sec, 8.21]
% \begin{equation}
%     \mathrm{Ei}(-x) = - \int_{x}^{\inf} \frac{e^-t}{t}\mathrm{d}t, \quad x>0.
% \end{equation}}
% % where the correction term $o(1)$ vanishes when $\beta\to \infty$.
% \begin{IEEEproof}
% Substitute $\nu\to 2$, $\mbf{T}\to 2\bs{\tau}$, $\mbf{Z} \to \sqrt{2}e^{j\bs{\Theta}}\bs{\omega}$, $x \to \sqrt{2}\beta$ in \cite[Eqs. (6), (9), and (14)]{lapidoth:2002}. Note that in \cite{lapidoth:2002}, $\bs{Z}$ is zero mean with variance $2$, therefore we modified the equation to match our assumption that $\bs{\omega}\sim \CN(0,1)$.
% \end{IEEEproof}
% \end{lemma}
% \end{align}
% 
\begin{lemma}\label{lemma:4}
    Let $\bs{\omega}\sim\CN(0,1)$ and for a given real $\beta$, define $\bs{\tau}=|\beta+\bs{\omega}|^2$. Then,
    \begin{align}
        h\paren{\bs{\tau}} & = \frac{1}{2}\log\beta^2+
        \frac{1}{2}(\log4\pi e)+o(1), \label{eq:chi_square_entropy}                                \\
                           &\added{\Eb{\log \bs{\tau}} = \log \beta^2 + \mathrm{E}_{1}(\beta^2),}\label{eq:chi_square_elog}
    \end{align}
    where the correction term $o(1)$ vanishes when $\beta\to \infty$ \added{ and $\mathrm{E}_1(x) =-\mathrm{Ei}(-x)$ is defined in Section~\ref{subsec:notation}}.
    \begin{IEEEproof}
        \added{For \eqref{eq:chi_square_entropy},} substitute $\nu\to 2$, $\mbf{T}\to 2\bs{\tau}$, $\mbf{Z} \to \sqrt{2}e^{j\bs{\Theta}}\bs{\omega}$, $x \to \sqrt{2}\beta$ in \cite[Eqs. (6), (9), and (14)]{lapidoth:2002}. Note that in \cite{lapidoth:2002}, $\bs{Z}$ is zero mean with variance $2$, therefore we modified the equation to match our assumption that $\bs{\omega}\sim \CN(0,1)$. \added{Moreover, \eqref{eq:chi_square_elog} follows directly from \cite[Definition 3, Eq.~(35)]{moser2020expected} and that $\mathrm{E}_1(x) =-\mathrm{Ei}(-x)$ for real values of $x>0$.}
    \end{IEEEproof}
\end{lemma}

\begin{lemma}\label{lemma:upper_bound_on_squared_rv}
Let $\bs{\omega}\sim\CN(0,1)$ and for any real random variable $\bs{\tau}$, we can write
\begin{equation}
    h(\abs{\bs{\tau}+\bs{\omega}}^2 \given \bs{\tau}) \le \frac{1}{2}\Eb{\log(2\pi e(1+2\bs{\tau}^2))}.
\end{equation}
\begin{IEEEproof}
    We can upper-bound $h(|\bs{\tau}+\bs{\omega}|^2\given \bs{\tau})$ by the entropy of a Gaussian distribution with the same variance. Note that $\mathrm{var}(|\bs{\tau}+\bs{\omega}|^2\given \bs{\tau}) = 1+2\bs{\tau}^2$. The expectation emerges because $h(|\bs{\tau}+\bs{\omega}|^2\given \bs{\tau}) = \Eb{ h\paren{|\tau+\bs{\omega}|^2\given \bs{\tau}= \tau}}$
\end{IEEEproof}
\end{lemma}

\begin{lemma}\label{lemma:upper_bound_on_pn_var}
Let $\bs{\Theta}\in [-\pi,\pi)$ with $h(\bs{\Theta})>-\infty$ and independent of $\bs{\omega}\sim \CN(0,1)$; then, for any real random variable $\bs{\tau}$
\begin{equation}
    h(e^{j\bs{\Theta}}\bs{\tau}+\bs{\omega}|\bs{\tau}) \le  \frac{1}{2}\Eb{\log\fparen{1+2\bs{\tau}^2}} + h\fparen{\bs{\Theta}\modplus\phase{\bs{\tau}+\bs{\omega}}\bgiven\bs{\tau}} +\frac{1}{2}\log(\pi e)-\frac{1}{2}\log2.\label{eq:lemma5}
\end{equation}
\begin{IEEEproof}
\begin{align}
    h\paren{e^{j\bs{\Theta}}\bs{\tau}+\bs{\omega}\given \bs{\tau}}&\note{a}{=}h\paren{e^{j\bs{\Theta}}(\bs{\tau}+\bs{\omega})\given \bs{\tau}}\nonumber\\
    &\note{b}{=}h\fparen{\big|\bs{\tau}+\bs{\omega}\big|^2\given \bs{\tau}}
    +h\fparen{\bs{\Theta}\modplus\phase{\bs{\tau}+\bs{\omega}}\bgiven \bs{\tau},\abs{\bs{\tau}+\bs{\omega}}}-\log 2\nonumber\\
    &\note{c}{\le} \frac{1}{2} \E\brack{\log\fparen{2\pi e\brack{1+2\bs{\tau}^2}}}+h\paren{\bs{\Theta}\modplus\phase{\bs{\tau}+\bs{\omega}}\given \bs{\tau}}-\log 2 ,\label{eq:lemma5_temp}
\end{align}
where (a) follows because $\bs{\omega}$ is circularly symmetric; (b) is a consequence of Lemma \ref{lemma:polar_entropy}; and finally, in (c), we applied Lemma~\ref{lemma:upper_bound_on_squared_rv} and that \added{conditioning does not increase} entropy and $h(|\bs{\tau}+\bs{\omega}|^2\given \bs{\tau})$. Simplifying the \ac{RHS} of \eqref{eq:lemma5_temp} gives \eqref{eq:lemma5}. 
\end{IEEEproof}
\end{lemma}

\begin{lemma}\label{lemma:von_mises}
Let $\bs{\omega}\sim\CN(0,1)$. For a given real $\beta>0$ and $\tau>0$, the conditional distribution of $\phase{\beta+\bs{\omega}}$ given $\abs{\beta+\bs{\omega}} = \tau$ is $\mathcal{VM}(0,2\beta\tau)$.
% \begin{align}
%    %f_{\bs{\Theta}\given \bs{\tau}}(\Theta\given \tau)\sim \mathcal{VM}(0,2\beta\tau).
%    (\bs{\Theta}\given \bs{\tau})\sim \mathcal{VM}(0,2\beta\bs{\tau}).
% \end{align} 
%\begin{IEEEproof}
\begin{IEEEproof}
Substitute $\epsilon(k) \to \phase{\beta+\bs{\omega}}$, $r(k) \to \tau$, and $\sigma^2/(2A) \to 1$ in \cite[Eq. (12)]{fu_von_mises:2008}.
\end{IEEEproof}
\end{lemma}

\begin{lemma}\label{lemma:von_mises_ent}
Let $f_{\thetab}^\mathrm{VM} (\theta;\mu,1/\sigma^2)$ and $f_{\thetab}^\mathrm{WN}(\theta;\mu,\sigma^2)$ denote the \acp{pdf} of the von Mises distribution $\mathcal{VM}(\mu,1/\sigma^2)$ and the wrapped normal distribution $\mathcal{WN}(\mu,\sigma^2)$, respectively. For small $\sigma^2$, the distribution of $\bs{\theta}$ tends to a normal distribution with zero mean $\mu=0$ and variance of $\sigma^2$ such that 
\begin{equation}
 f_{\thetab}^\mathrm{VM} (\theta;0,1/\sigma^2)-f_{\thetab}^\mathrm{WN}\fparen{\theta;0,\sigma^2} = O(\sigma), \quad \sigma \to 0.
\end{equation} 
\end{lemma}
\begin{IEEEproof}
See \cite{kent:1978} and \cite[Eq. 3.5.24]{mardia_directional:2000}.
\end{IEEEproof}

\begin{lemma}\label{lemma:igamma_lowerbound_helper}
For all $0 \le \alpha \le 1$ and $0<x\le{x}_0$,
\begin{equation} \label{e:zeta}
(2+x)^\alpha - x^\alpha \ge 2\alpha ,
\end{equation}
where ${x}_0 \approx 0.1770$ is the unique solution $x>0$ of
\begin{equation} \label{e:xbar}
(2+x)\log_e(2+x) - x\log_e x - 2 = 0 .
\end{equation}
\end{lemma}

\begin{IEEEproof}
We define for any fixed $0<x\le x_0$
\begin{equation} 
\zeta(\alpha) = (2+x)^\alpha - x^\alpha - 2\alpha
\end{equation}
and calculate its derivatives
\begin{align}
\zeta'(\alpha) &= (2+x)^\alpha \log_e(2+x) - x^\alpha \log_e x - 2 , \\
\zeta''(\alpha) &= (2+x)^\alpha (\log_e(2+x))^2 - x^\alpha (\log_e x)^2 ,
\end{align}
which are continuous and differentiable for $x>0$.
The equation $\zeta''(\alpha) = 0$ has a unique solution
\begin{equation}
\alpha = \bar\alpha = 2 \frac{ \log_e \left( \frac{-\log_e x}{\log_e(2+x)} \right) } {\log_e \left( \frac{2+x}{x} \right) } < 1 ,
\end{equation}
where the inequality follows because $\zeta''(\alpha)$ is increasing and $\zeta''(1)>0$. Hence, $\zeta''(\alpha) \le 0$ for $0 \le \alpha \le \bar\alpha$ and $\zeta''(\alpha) \ge 0$ for $\bar\alpha \le \alpha \le 1$. These properties will now be used to prove that $\zeta(\alpha) \ge 0$ in both intervals.

We first consider $\alpha=1$. Here, $\zeta(1) = 0$ and
\begin{equation}
\zeta'(1) = (2+x)\log_e(2+x)-x\log_e x-2 .
\end{equation}
This function increases monotonically for all $x>0$ and equals zero for $x={x}_0$ by \eqref{e:xbar}. Hence, $\zeta'(1)\le 0$ for $0<x\le {x}_0$.

We next consider $\bar\alpha \le \alpha \le 1$. Since $\zeta(\alpha)$ is convex in this interval, it is not less than its tangent at $\alpha = 1$, i.e.,
\begin{align}
\zeta(\alpha) &\ge \zeta(1) + \zeta'(1)(\alpha-1) \\
& \ge 0. \label{e:zetatangent}
\end{align}

In $0 \le \alpha \le \bar\alpha$, finally, $\zeta(\alpha)$ is concave and by Jensen's equality not less than its secant, i.e.,
\begin{align}
\zeta(\alpha) &\ge \frac{\bar\alpha-\alpha}{\bar\alpha} \zeta(0) + \frac{\alpha}{\bar\alpha} \zeta(\bar\alpha) \\
& \ge 0, \label{e:zetasecant}
\end{align}
because $\zeta(0)=0$ and $\zeta(\bar\alpha) \ge 0$ by \eqref{e:zetatangent}.

Together, \eqref{e:zetatangent} and \eqref{e:zetasecant} prove \eqref{e:zeta} for all $0\le\alpha\le 1$.
\end{IEEEproof}

\begin{lemma} \label{lemma:igamma_lb}
For all $0<\alpha\le1 $ and $0<x\le x_0$,
\begin{equation}
\Gamma(\alpha,x) \ge \Gamma(1,x),
\end{equation}
where $x_0\approx 0.1770$ was defined in Lemma \ref{lemma:igamma_lowerbound_helper}.
\end{lemma}

\begin{IEEEproof}
For all $0 < \alpha \le 1$ and $0<x\le x_0$, from \cite[Ch.~8, Eq.~(8.10.10)]{olver2010nist} and that $\Gamma(1,x) = e^{-x}$ we have
\begin{equation}
    x^{1-\alpha}e^{x} \Gamma(\alpha,x) \ge \frac{x}{2\alpha}\fparen{\fparen{1+\frac{2}{x}}^\alpha-1}.\label{eq:igamma_inequality}
\end{equation}
With some basic mathematical operations, we can rewrite \eqref{eq:igamma_inequality} as
\begin{align}
    \Gamma(\alpha,x) &\ge  e^{-x}\frac{(x+2)^\alpha -x^{\alpha}}{2\alpha}\nonumber\\
     &\note{a}{\ge} e^{-x}\nonumber\\
     &=\Gamma(1,x),\label{eq:igamma_inequality_b}
\end{align}
where in (a) we utilized Lemma~\ref{lemma:igamma_lowerbound_helper}.
\end{IEEEproof}

\begin{lemma}\label{lemma:truncated_mean}
Let the random scalar $\mbf{r} \sim \trG(\mu,\alpha,\gamma)$ where $\mu>0$, $\alpha>0$, and $\gamma> 0$. Also, define 
\begin{equation}
    J(\alpha,\gamma) = \alpha + \frac{e^{-\gamma}\gamma^{\alpha}}{\Gamma(\alpha,\gamma)}.\label{eq:def_J}
\end{equation}
Then,
\begin{align}
    \E[\mbf{r}] &= \mu\frac{\Gamma(\alpha+1,\gamma)}{\Gamma(\alpha,\gamma)}\nonumber\\
    &=\mu J(\alpha,\gamma).\label{eq:lemma_truncated_mean}
\end{align}
\begin{IEEEproof}
    Using \eqref{eq:truncated_gamma_dist} and setting $M=1$ to get the \ac{pdf} for a scalar random variable, we can write
    \begin{align}
        \E[\mbf{r}] &= \frac{1}{\Gamma(\alpha,\gamma)} \int_{\mu\gamma}^{\infty} e^{-\frac{r}{\mu}} \mu^{-\alpha} r^{\alpha} \mathrm{d}r\nonumber\\
        &\note{a}{=}  \frac{\mu}{\Gamma(\alpha,\gamma)} \int_{\gamma}^{\infty} e^{-u} u^{\alpha} \mathrm{d}u,\nonumber\\
        &\note{b}{=}\mu\frac{\Gamma(\alpha+1,\gamma)}{\Gamma(\alpha,\gamma)}\nonumber\\
        &\note{c} = \mu \fparen{\alpha+ \frac{e^{-\gamma} \gamma^{\alpha}}{\Gamma(\alpha,\gamma)}}\nonumber\\
        & = \mu J(\alpha,\gamma),
    \end{align}
    where in (a) we employed the change of variable technique by defining $u = r/\mu$, in (b) we used that ${\Gamma(\alpha+1, \gamma)}=\int_{\gamma}^{\infty} e^{-u} u^{\alpha} \mathrm{d}u$,  and in (c) we used that $\Gamma(\alpha+1,\gamma) = \alpha \Gamma(\alpha,\gamma) + e^{-\gamma} \gamma^{\alpha}$.
    \end{IEEEproof}
\end{lemma}

\begin{lemma}\label{lemma:gamma_pow_c}
For any $m\in\{0,\dots,M-1\}$ and any $0\le x\le x_\mr{max}$, let 
\begin{equation}
    c_m(x)  = \frac{W_\mr{L}(x^{\alpha_m^*}\log_{e}x)}{\log_{e}x} = \frac{x^{\alpha_m^*}}{e^{W_\mr{L}(x^{\alpha_m^*}\log_e x)}},\label{eq:c_gamma}
\end{equation}
where $\alpha_m^*$ defined in \eqref{eq:alpha_star}. Moreover, $W_\mr{L}(x)$ is the principal branch of the Lambert W function, which is defined for any $x\ge -e^{-1}$ by $e^{W_L(x)}W_L(x)=x$ and $W_L(x)\ge -1$ %the Lambert W function\footnote{\hlight{When $-1/e\le x <0$, there are two possible real values of Lambert W function. Here, we denote $W_\mr{L}(x)$ as the branch satisfying $-1 \le W_\mr{L}(x)$.}} with the definition $e^{W_\mr{L}(x)} W_\mr{L}(x) = x$
\cite{corless1996lambert}, and $x_\mr{max} \approx 0.00471$ is the smallest $x>0$ for which
\begin{equation}
    x^{\alpha_m^*}\log_{e}x \ge -\frac{1}{e}.
\end{equation}
Then, we have 
\begin{equation}
    x^{c_m(x)} = \frac{x^{\alpha_m^*}}{c_m(x)}.\label{eq:lemma_gamma_pow_c}
\end{equation}
\end{lemma}
\begin{IEEEproof}
\begin{align}
    x^{c_m(x)} &= e^{\log_{e}(x)c_m(x)} \nonumber\\
    &= e^{W_\mr{L}(x^{\alpha_m^*}\log_{e}(x))} \nonumber\\
    & \note{a}{=} \frac{x^{\alpha_m^*} \log_{e}(x)}{W_\mr{L}(x^{\alpha_m^*}\log_{e}(x))} \nonumber\\
    & = \frac{x^{\alpha_m^*}}{c_m(x)}, 
\end{align}
where (a) follows from the definition of the Lambert W function.
\end{IEEEproof}

\begin{lemma}\label{lemma:von_mises_entropy}
%Let %$\thetab \sim \mathcal{VM}(0,1/\sigma^2)$. 
Let $\thetab \in [-\pi,\pi)$ be a random variable with \ac{pdf}
\begin{equation}
 g_{\thetab}(\theta)= f_{\thetab}^\mathrm{WN}\fparen{\theta;0,\sigma^2} + O(\sigma), \quad \sigma \to 0,
\end{equation} 
where $f_{\thetab}^\mathrm{WN}({\theta;0,\sigma^2})$ denotes the zero-mean wrapped normal distribution with variance $\sigma^2$. 
Then,
  \begin{equation}
      h(\thetab) = \frac{1}{2}\log(2\pi e \sigma^2)+O(\sigma), \quad \sigma \to 0.
  \end{equation}
\end{lemma}
\begin{IEEEproof}
The \ac{pdf} of a zero-mean wrapped normal distribution with variance $\sigma^2$ is defined as
\begin{equation}
    f^\mr{WN}_{\thetab}(\theta;0,\sigma^2) =\frac{1}{\sqrt{2\pi\sigma^2}} \sum_{l=-\infty}^{\infty} \exp\fparen{-\frac{(\theta-2\pi l)^2}{2\sigma^2}},
\end{equation}
where $\theta \in [-\pi,\pi)$. %Then, based on Lemma~\ref{lemma:von_mises_ent}, we can write 
% \begin{align}
%     g(\theta) =f(\theta)+ O(\sigma), \qquad \sigma \to 0.\label{app:g_approx}
% \end{align}
Now let
\begin{equation}
    f_\mathrm{G}(\theta) =\frac{1}{\sqrt{2\pi\sigma^2}}\exp\fparen{-\frac{\theta^2}{2\sigma^2}}, \qquad -\infty\le \theta \le \infty.
\end{equation}
For any $\theta \in [-\pi,\pi)$,
\begin{align}
     f^\mr{WN}_{\thetab}(\theta;0,\sigma^2)- f_\mathrm{G}(\theta) &= \frac{1}{\sqrt{2\pi\sigma^2}} \sum_{\substack{l=-\infty\\ l \neq 0}}^{\infty} \exp\fparen{-\frac{(\theta-2\pi l)^2}{2\sigma^2}} \nonumber\\
    &\note{a}{\le} \frac{2}{\sqrt{2\pi\sigma^2}} \sum_{\substack{l=1}}^{\infty} \exp\fparen{-\frac{\pi^2 l}{2\sigma^2}}\nonumber\\
    &= O\fparen{\frac{1}{\sigma}\exp\fparen{-\frac{\pi^2}{2\sigma^2}}\!},
\end{align}
where (a) follows since $(\theta-2\pi l)^2 \ge \pi^2 l^2 \ge \pi^2 l$ for $\abs{\theta} \le \pi$ and $l \ge1$. Now for $\theta\in [-\pi,\pi)$, we can write 
\begin{align}
       g_{\thetab}(\theta) &= f_\mathrm{G}(\theta)+ O\fparen{\frac{1}{\sigma}\exp\fparen{-\frac{\pi^2}{2\sigma^2}}\!}+O(\sigma)\nonumber\\
        &= f_\mathrm{G}(\theta)+ O(\sigma),\label{eq:app_g}
\end{align}
where the last equality holds since $\exp\paren{-\pi^2/(2\sigma^2)}/\sigma$ has faster decay than $\sigma$ as $\sigma \to 0$. 

Defining the entropy of $\thetab$ as 
\begin{equation}
    h(\thetab) = -\int_{-\pi}^{\pi} g_{\thetab}(\theta)\log g_{\thetab}(\theta) d\theta,
\end{equation}
we can write
\begin{align}
    h(\thetab) &= -\!\!\int_{-\pi}^{\pi} \!\!\!\!\paren{f_\mathrm{G}(\theta)+ O(\sigma)}\log g_{\thetab}(\theta) d\theta\nonumber\\ 
    &\note{a}{=}-\!\!\int_{-\pi}^{\pi}\!\!\!\! f_\mathrm{G}(\theta)\log g_{\thetab}(\theta) d\theta+ O(\sigma)\nonumber\\
    &= -\!\!\int_{-\pi}^{\pi}\!\!\!f_\mathrm{G}(\theta)\!\fparen{\log f_\mathrm{G}(\theta)\! +\! \log\frac{g_{\thetab}(\theta)}{f_\mathrm{G}(\theta)}} d\theta + O(\sigma)\nonumber\\
   & = -\!\!\int_{-\pi}^{\pi}\!\!\!f_\mathrm{G}(\theta)\log f_\mathrm{G}(\theta) d\theta\nonumber\\
   &\quad-\!\!\int_{-\pi}^{\pi}\!\!\!\!f_\mathrm{G}(\theta) \log\frac{g_{\thetab}(\theta)}{f_\mathrm{G}(\theta)}d\theta + O(\sigma),\label{eq:app_h_vm}
    %& = \frac{1}{2}\log(2\pi e\sigma^2)+O(\sigma),\label{eq:app_h_vm}
\end{align}
where in (a) we used the fact that $f_{G}(\theta)>0$ for all $\theta \in [-\pi,\pi)$ and $\sigma\ge0$. 

The first term on the \ac{RHS} of \eqref{eq:app_h_vm} can be evaluated as
\begin{align}
    -\int_{-\pi}^{\pi}\!\!\!\!f_\mathrm{G}(\theta)\log f_\mathrm{G}(\theta) d\theta
    & = \! \fparen{\!1-\mathrm{erfc}\fparen{\frac{\pi}{\sqrt{2\sigma^2}}}\!\!} \frac{\log(2\pi e\sigma^2)}{2}\! -\sqrt{\frac{\pi}{2\sigma^2}}\exp\!\fparen{-\frac{\pi^2}{2\sigma^2}}\nonumber\\
    & = \frac{1}{2}\log(2\pi e\sigma^2)+O\fparen{\frac{1}{\sigma}\exp\fparen{-\frac{\pi^2}{2\sigma^2}}\!}\label{eq:app_h_vm_a}.
\end{align}

%To deal with the second term on the \ac{RHS} of \eqref{eq:app_h_vm} we utilize we that $\log(x) \ge 1-1/x$ for $x> 0$.  
The second term on the \ac{RHS} of \eqref{eq:app_h_vm} can be written as
\begin{align}
    -\int_{-\pi}^{\pi}\!\!\!\!f_\mathrm{G}(\theta) \log\fparen{\frac{g_{\thetab}(\theta)}{f_\mathrm{G}(\theta)}}d\theta &\note{a}{\le} \int_{-\pi}^{\pi}\!\!\!\!f_\mathrm{G}(\theta) \!\fparen{1-\frac{g_{\thetab}(\theta)}{f_\mathrm{G}(\theta)}\!}d\theta\nonumber\\
    &=\int_{-\pi}^{\pi}\!\!\!\!\fparen{f_\mathrm{G}(\theta) -g_{\thetab}(\theta)} d\theta\nonumber\\
    & \note{b}{=} \int_{-\pi}^{\pi}\!\!\!\!O(\sigma) d\theta\nonumber\\
    & = O(\sigma),\label{eq:app_h_vm_b}
\end{align}
where in (a) we utilize that $-\log(x) \le 1-x$ for $x> 0$. We can do this because both $g_{\thetab}(\theta)>0$ and $f_\mathrm{G}(\theta)>0$; (b) directly follows from \eqref{eq:app_g}.

Combining \eqref{eq:app_h_vm_b} and \eqref{eq:app_h_vm_a} into \eqref{eq:app_h_vm} we get
\begin{equation}
    h(\thetab) =   \frac{1}{2}\log(2\pi e\sigma^2)+ O(\sigma),
\end{equation}
and the proof is complete.
\end{IEEEproof}

\section{Proof of Capacity Upper and Lower Bounds}\label{app:sec:proof_ub_lb}
This section is dedicated to proving capacity upper and lower bounds in Theorems~\ref{th:upper_bound} and~\ref{th:lower_bound}. 

To establish the upper bound in Theorem~\ref{th:upper_bound}, we adopt the duality approach, leveraging that any output distribution provides an upper bound on the capacity. Specifically, we focus on a family of circularly symmetric distributions where the squared magnitude of the outputs follows the gamma distribution $\trG(\mu>0,\vec{\alpha}>0,0)$.

On the other hand, for the lower bound in Theorem~\ref{th:lower_bound}, we rely on the insight that any input distribution provides a lower bound on the capacity. Consequently, we introduce a circularly symmetric input distribution with the squared magnitude of the inputs following a truncated gamma distribution $\trG(\mu>0,\vec{\alpha}>0,\gamma>0)$.

Both theorems are proven under the assumption of circularly symmetric input distributions, as Lemma~\ref{prop:1} establishes that the capacity-achieving distribution for \added{\ac{EO}}\label{r2:c1_f} phase-noise channels \eqref{eq:EO_chan_model}--\eqref{eq:EO_chan_model_c} is indeed circularly symmetric.

\subsection{Proof of Theorem~\ref{th:upper_bound} (Upper Bound)}\label{proof:th_upper_bound}
The following lemma characterizes an upper bound on the capacity of the channel \eqref{eq:EO_chan_model}--\eqref{eq:EO_chan_model_c} and serves as starting points for formulating an upper bound on the capacity of a \added{\ac{EO}}\label{r2:c1_g} phase-noise channel, whether the phase noises are independent or correlated. 
\begin{lemma}\label{lemma:general_ub}
The capacity of the channel \eqref{eq:EO_chan_model}--\eqref{eq:EO_chan_model_c} under the power constraint \eqref{eq:pw_cons} can be upper-bounded as
\begin{equation}
    C(\rho) \le \sup_{\mathcal{Q}_{\rvec{x}_1}}\left\{ 
    I(\rvec{x}_1;\rvec{y}_1)+I(\rvec{y}_1;\vec{\thetab}_0\given\rvec{x}_1)
    \right\},\label{eq:cap_ub}
\end{equation}
where the supremum is over all probability distributions $\mathcal{Q}_{\rvec{x}_1}$ on $\rvec{x}_1$ that satisfy the power constraint $\E[\norm{\rvec{x}_1}^2]\le \rho$. 
%Here, $I(\rvec{x}_1;\rvec{y}_1)$ is equivalent to the mutual information of a memoryless channel with correlated uniformly distributed phase noises over the channels. 
\end{lemma}
\begin{IEEEproof}
    See Appendix~\ref{sec:proof_lemma_general_ub}.
\end{IEEEproof}

We start by upper-bounding each term on the \ac{RHS} of \eqref{eq:cap_ub}. Thanks to Lemma~\ref{prop:1}, our focus can be narrowed down to input processes with circular symmetry. Specifically, we will examine $\rvec{x}_k$ whose amplitude $\abs{\rvec{x}_k}$ and phase $\phase{\rvec{x}}_k$ are independent of each other.

In the next lemma, we utilize the duality approach \cite[Theorem. 5.1]{lapidoth_duality:2003} to upper-bound $I(\rvec{x}_1;\rvec{y}_1)$. %the first term in the \ac{RHS} of \eqref{eq:cap_ub}.
\begin{lemma}\label{lemma:duality_bound}
For channel \eqref{eq:EO_chan_model}--\eqref{eq:EO_chan_model_c} and for any circularly symmetric distribution $\mathcal{Q}_{\rvec{x}_1}$ on $\rvec{x}_1$ that satisfies $\E[\norm{\rvec{x}_1}^{\added{2}}] \le \rho$, $\vec{\alpha}=(\alpha_0,\dots,\alpha_{M-1}) > 0$, and $ \lambda \ge 0$, we have%the first term on the \ac{RHS} of \eqref{eq:cap_ub} is
\begin{equation}
       I(\rvec{x}_1;\rvec{y}_1) \le\alpha_{\Sigma}\log\left(\frac{\rho+M}{\alpha_{\Sigma}}\right)+d_{\lambda,\vec{\alpha}}+\E\brack{R_{\lambda,\vec{\alpha}}(\rho,\abs{\rvec{x}_1})},\label{eq:duality_bound}
\end{equation}
where 
\begin{equation}
    d_{\lambda,\vec{\alpha}} = \lambda-(M-2)\log e +\sum_{m=0}^{M-1}\log\Gamma(\alpha_m),\label{eq:d_lambda}
\end{equation}
and $\alpha_{\Sigma}$ and $R_{\lambda,\vec{\alpha}}(\cdot)$ are defined in \eqref{eq:alpha_s} and \eqref{eq:Rs}, respectively. 
\end{lemma}
\begin{IEEEproof}
%Proof that exploits the duality approach is provided in Appendix \ref{appef:A}.
See Appendix \ref{app:lemma_duality_bound}.
\end{IEEEproof}

It is rather challenging to characterize $I(\rvec{y}_1;\vec{\thetab}\bgiven \rvec{x}_1)$ for $\smash{M>2}$.  In the following lemma, we present a precise characterization of this term specifically for the case of $M=2$ and provide a looser upper bound when $M>2$.
\begin{lemma}\label{lemma:I_y_theta}
For channel \eqref{eq:EO_chan_model}--\eqref{eq:EO_chan_model_c} and for any circularly symmetric $\mathcal{Q}_{\rvec{x}_1}$ on $\rvec{x}_1$ that satisfies $\E[\norm{\rvec{x}_1}^{\added{2}}] \le \rho$, the second term on the \ac{RHS} of \eqref{eq:cap_ub} is
\begin{equation}
    I(\rvec{y}_1;\vec{\thetab}_0\given\rvec{x}_1) \le 2\log(2\pi)+F(M,\abs{\rvec{x}_1},\Deltacb{1},\Deltarb{1}),\label{eq:lemm_I_y_theta}
\end{equation}
where $F(\cdot)$ is defined in \eqref{eq:Fs}. Here, $\Deltacb{1} \sim \mathcal{WN}(0,\sigma^2_\mathrm{c})$ and $\Deltarb{1}\sim \mathcal{WN}(0,\sigma^2_\mathrm{r})$ are independent.
\begin{IEEEproof}
     See Appendix~\ref{app:lemma:I_y_theta}.
\end{IEEEproof}
\end{lemma}

Substituting \eqref{eq:lemm_I_y_theta} and \eqref{eq:duality_bound} into \eqref{eq:cap_ub}, we obtain
\begin{equation}
       C(\rho)\le \alpha_{\Sigma}\log\fparen{\frac{\rho+M}{\alpha_{\Sigma}}}+2\log(2\pi)+d_{\lambda,\vec{\alpha}} +\sup_{\mathcal{Q}_{\abs{\rvec{x}_1}}}\!\big\{\E\brack{R_{\lambda,\vec{\alpha}}(\rho,\abs{\rvec{x}_1})}+F(M,\abs{\rvec{x}_1},\Deltacb{1},\Deltarb{1})\big\},
        \label{eq:mi_up_B_temp}
\end{equation}
where the supremum over $\mathcal{Q}_{\rvec{x}_1}$ is replaced with supremum over $\mathcal{Q}_{\abs{\rvec{x}_1}}$ that satisfies the power constraint $\E[\norm{\rvec{x}_1}^2]\le \rho$. %Next, the supremum over all $\mathcal{Q}_{\abs{\rvec{x}_1}}$ is upper-bounded by removing the power constraint and maximizing the \ac{RHS} of \eqref{eq:mi_up_B_temp} over all deterministic $\vec{s}\ge 0$

Defining a real deterministic vector $\vec{s}\ge 0$, we can write
\begin{equation}
       C(\rho)\le \alpha_{\Sigma}\log \fparen{\frac{\rho+M}{\alpha_{\Sigma}}}+2\log(2\pi)+d_{\lambda,\vec{\alpha}}+\max_{\vec{s}\ge 0}\big\{R_{\lambda,\vec{\alpha}}(\rho,\vec{s})+F(M,\vec{s},\Deltacb{},\Deltarb{})\big\},
        \label{eq:mi_up_B}
\end{equation}
which follows as the supremum over all $\mathcal{Q}_{\abs{\rvec{x}_1}}$ is upper-bounded by removing the power constraint and maximizing over all deterministic $\vec{s}\ge 0$. For convenience, the time index is dropped and all the random quantities are replaced by timeless random variables with the same distributions, i.e., $\mbf{v}_1\to \mbf{v}$, $\Deltacb{1}\to \Deltacb{}$, $\Deltarb{1}\to \Deltarb{}$ and $\mbf{z}_{1,m}\to \mbf{z}_{m} $ for all $m$. Replacing \eqref{eq:d_lambda} into \eqref{eq:mi_up_B} concludes the proof of Theorem \ref{th:upper_bound}.

\subsection{Proof of Theorem~\ref{th:lower_bound} (Lower Bound)}\label{sec:proof_th_lb}
The following lemma characterizes a lower bound on the capacity of the channel \eqref{eq:EO_chan_model}--\eqref{eq:EO_chan_model_c} and it holds whether the phase noises are independent or correlated. 
\begin{lemma}\label{lemma:general_lb}
The capacity of the channel \eqref{eq:EO_chan_model}--\eqref{eq:EO_chan_model_c} under the power constraint \eqref{eq:pw_cons} can be lower-bounded as
\begin{equation}
    C(\rho)\ge I\paren{\rvec{x}_2;\rvec{y}_2\given \vec{\thetab}_{1}} - I\paren{\rvec{x}_2;\vec{\thetab}_{1}\given \rvec{x}_{1},\rvec{y}_{1},\rvec{y}_2},\label{eq:cap_lb}
\end{equation}
for any distribution on $\rvec{x}_1$ and $\rvec{x}_2$ that fulfill $\E[\norm{\rvec{x}_1}^2]\le \rho$ and $\E[\norm{\rvec{x}_2}^2]\le \rho$.
%Here, $I(\rvec{x}_1;\rvec{y}_1)$ is equivalent to the mutual information of a memoryless channel with correlated uniformly distributed phase noises over the channels. 
\end{lemma}
\begin{IEEEproof}
    See Appendix~\ref{sec:proof_lemma_general_lb}.
\end{IEEEproof}

To derive the lower bound, we use  Lemma~\ref{lemma:general_lb} and start by examining the two terms on the \ac{RHS} of \eqref{eq:cap_lb} separately. 

\subsubsection*{\added{\underline{Characterizing the first term on the \ac{RHS} of \eqref{eq:cap_lb}}}}
We can rewrite the first term as
\begin{equation}
    I\paren{\rvec{x}_2;\rvec{y}_2\given \vec{\thetab}_{1}} = h\paren{\rvec{y}_2\given \vec{\thetab}_{1}}-h\paren{\rvec{y}_2\given \rvec{x}_2,\vec{\thetab}_{1}}.\label{eq:I(y2;x2)}
\end{equation}

Lemma~\ref{prop:1} allows us to narrow our attention to input processes exhibiting circular symmetry. Here, we consider a circularly symmetric input vector with independent elements. Thus, the first term on the \ac{RHS} of \eqref{eq:I(y2;x2)} can be bounded as 
\begin{align}
    h\paren{\rvec{y}_2\given \vec{\thetab}_{1}}&\ge h\paren{\rvec{y}_2\given \rvec{w}_2,\vec{\thetab}_{1}}\nonumber\\
    &=h\paren{e^{j\vec{\thetab}_2}\circ\rvec{x}_2\given \vec{\thetab}_{1}}\nonumber\\
    &\note{a}{=} h(\rvec{x}_2)\nonumber\\
    &\note{b}{=} \sum_{m=0}^{M-1} \left(h(\abs{\mbf{x}_{2,m}}^2)+h(\phase{\mbf{x}}_{2,m}\bgiven\abs{\mbf{x}_{2,m}})-\log 2\right)\nonumber\\
    &\note{c}{=}M\log\pi+\sum_{m=0}^{M-1} h(\abs{\mbf{x}_{2,m}}^2)\\
    &= M\log\pi+h\paren{\abs{\rvec{x}_2}^2},\label{eq:h(y2)}
\end{align}
where (a) holds since $\rvec{x}_2$ is circularly symmetric and rotation does not change its distribution, i.e., $e^{j\vec{\thetab}_2}\circ\rvec{x}_2 \sim \rvec{x}_2$; in (b) we applied Lemma~\ref{lemma:polar_entropy} and that the squared amplitudes $\abs{\mbf{x}_{2,m}}^2$ are \ac{iid} for any $m$; in (c) we used that the phases $\phase{\mbf{x}}_{2,m}\sim \Uniform[-\pi,\pi)$ independently of the amplitudes $\abs{\mbf{x}_{2,m}}$.

Continuing with the chain rule, we can express the second term on the \ac{RHS} of \eqref{eq:I(y2;x2)} as
\begin{equation}
    h\fparen{\rvec{y}_{2}\given \rvec{x}_2,\vec{\thetab}_1}\le h\fparen{\brace{\mbf{y}_{2,m}}_{0}^{1}\bgiven \rvec{x}_2,\vec{\thetab}_{1}}+h\fparen{\{\mbf{y}_{2,m}\}_{2}^{M-1}\bgiven \rvec{x}_2,\brace{\mbf{y}_{2,m}}_{0}^{1}},
    \label{eq:lb_h(y2|s2)}
\end{equation}
where the inequality follows as \added{conditioning does not increase} entropy.
% \begin{lemma}\label{lemma:lb_h(y2|s2)_a}
% The first term on the \ac{RHS} of \eqref{eq:lb_h(y2|s2)} can be upper-bounded as
% \begin{align}
%     h\paren{\brace{\mbf{y}_{2,m}}_{0}^{1}\bgiven \rvec{x}_2,\vec{\thetab}_{1}}
%     &\le  \log(\pi e)-\log 2+\frac{1}{2} \E\brack{\log\fparen{1+2 \abs{\mbf{x}_{2,0}}^2}}+\frac{1}{2} \E\brack{\log\fparen{1+2 \abs{\mbf{x}_{2,1}}^2}}\nonumber\\
%     &\quad\! \!+h\!\fparen{\!\brace{\Deltacb{2}\!\modplus\! m\Deltarb{2}\!\modplus\!\phase{\abs{\mbf{x}_{2,m}}+\mbf{z}_{2,m}}}_{0}^{1}\bgiven\abs{\rvec{x}_2},\!\fbrace{\abs{\mbf{y}_{2,m}}}_{0}^{1}}\!,\label{eq:lb_h(y2|s2)_a_final}
% \end{align}
% where $|\mbf{y}_{2,m}|\!=\!\big||\mbf{x}_{2,m}|+\mbf{z}_{2,m}\big|$ and $\mbf{z}_{2,m}\sim \CN(0,1)$.
% \end{lemma}

% \begin{IEEEproof}
%     See Appendix~\ref{app:lb_h(y2|s2)_a}.
% \end{IEEEproof}

\label{r2:c3_a}\moved{The first term on the \ac{RHS} of \eqref{eq:lb_h(y2|s2)} can be written as
\begin{equation}
    h\paren{\!\brace{\mbf{y}_{2,m}}_{0}^{1}\!\bgiven \rvec{x}_2,\vec{\thetab}_{1}} \note{a}{=} h\fparen{\brace{|\mbf{y}_{2,m}|^2}_{0}^{1}\bgiven \rvec{x}_2,\vec{\thetab}_{1}}-2\log2  + h\paren{\brace{\phase{\mbf{y}}_{2,m}}_{0}^{1}\bgiven \rvec{x}_2,\vec{\thetab}_{1},\!\brace{|\mbf{y}_{2,m}|}_{0}^{1}},
    \label{eq:lb_h(y2|x2)_a}
\end{equation}
where (a) follows by utilizing Lemma~\ref{lemma:polar_entropy}. %Here, we can use Lemma~\ref{lemma:polar_entropy} because it is clear from \eqref{eq:app:abs_ykm} that $|\mbf{y}_{2,m}|$ are independent for all $m$ when $\rvec{x}_2$ is given. 
We continue by upper-bounding each entropy term in \eqref{eq:lb_h(y2|x2)_a}. The first term can be upper-bounded as 
\begin{align}
    h\!\fparen{\!\brace{|\mbf{y}_{2,m}|^2}_{0}^{1}\bgiven \rvec{x}_2,\vec{\thetab}_{1}}
    &\!\note{a}{=} h\fparen{\brace{|\mbf{y}_{2,m}|^2}_{0}^{1}\given\! \abs{\rvec{x}_2},\vec{\thetab}_{1}}\nonumber\\
    &\!\note{b}{=} h\fparen{|\mbf{y}_{2,0}|^2 \given\!\abs{\mbf{x}_{2,0}}}
    \!+\!h\paren{|\mbf{y}_{2,1}|^2 \given\! \abs{\mbf{x}_{2,1}}}\nonumber\\
    & \note{c}{\le} \log(2\pi e)+\frac{1}{2} \E\brack{\log\fparen{1+2 \abs{\mbf{x}_{2,0}}^2}}+\frac{1}{2} \E\brack{\log\fparen{1+2 \abs{\mbf{x}_{2,1}}^2}},\label{eq:lb_h(y2|x2)_a_1} 
\end{align}
where (a) holds since $\mbf{y}_{2,m}$ are independent of $\phase{\rvec{x}}_2$. Moreover,  (b) follows as $|\mbf{y}_{2,0}|$ and  $|\mbf{y}_{2,1}|$ in \eqref{eq:app:abs_ykm} are independent of $\vec{\thetab}_1$ and each other; the inequality in (c) is obtained by recalling \eqref{eq:app:abs_ykm} and applying Lemma~\ref{lemma:upper_bound_on_squared_rv} on both entropy terms.}

\moved{The second term on the \ac{RHS} of \eqref{eq:lb_h(y2|x2)_a} can be upper-bounded as 
\begin{align}
     h\fparen{\fbrace{\phase{\mbf{y}}_{2,m}}_{0}^{1}\bgiven \rvec{x}_2,\vec{\thetab}_{1},\brace{|\mbf{y}_{2,m}|}_{0}^{1}}
    &=  h\fparen{\fbrace{\phase{\mbf{y}}_{2,m}\modsub\phase{\mbf{x}}_{2,m}\modsub\thetab_{1,m}}_{0}^{1}\bgiven \rvec{x}_2,\vec{\thetab}_{1},\brace{|\mbf{y}_{2,m}|}_{0}^{1}}\nonumber\\   
    &\note{a}{=} h\fparen{\brace{\Deltacb{2}\!\modplus m\Deltarb{2}\!\modplus\!\phase{\abs{\mbf{x}_{2,m}}+\mbf{z}_{2,m}}}_{0}^{1}\bgiven\abs{\rvec{x}_2},\!\brace{|\mbf{y}_{2,m}|}_{0}^{1}},
    \label{eq:lb_h(y2|x2)_a_2} 
\end{align}
where %(a) follows by removing the contribution of $\{\thetab_{1,m}\}_{0}^{1}$ from the ; 
in (a) we used from \eqref{eq:app:phase_ykm} that $\phase{\mbf{y}}_{2,m} = \thetab_{2,m}\modplus\phase{\mbf{x}}_{2,m}\modplus\phase{\abs{\mbf{x}_{2,m}}+\mbf{z}_{2,m}}$ and that $\thetab_{2,m}\modsub\thetab_{1,m} =\Deltacb{2}\modplus m\Deltarb{2}$. Note that the elements of $\phase{\rvec{x}}_2$ are irrelevant to the entropy. Hence, they are omitted from the conditions.} \added{Substituting \eqref{eq:lb_h(y2|x2)_a_2} and \eqref{eq:lb_h(y2|x2)_a_1} into \eqref{eq:lb_h(y2|x2)_a} gives} \moved{
\begin{align}
    h\paren{\brace{\mbf{y}_{2,m}}_{0}^{1}\bgiven \rvec{x}_2,\vec{\thetab}_{1}}
    &\le  \log(\pi e)-\log 2+\frac{1}{2} \E\brack{\log\fparen{1+2 \abs{\mbf{x}_{2,0}}^2}}+\frac{1}{2} \E\brack{\log\fparen{1+2 \abs{\mbf{x}_{2,1}}^2}}\nonumber\\
    &\quad\! \!+h\!\fparen{\!\brace{\Deltacb{2}\!\modplus\! m\Deltarb{2}\!\modplus\!\phase{\abs{\mbf{x}_{2,m}}+\mbf{z}_{2,m}}}_{0}^{1}\bgiven\abs{\rvec{x}_2},\!\fbrace{\abs{\mbf{y}_{2,m}}}_{0}^{1}}\!.\label{eq:lb_h(y2|s2)_a_final}
\end{align}}

\moved{We can characterize the second term on the \ac{RHS} of \eqref{eq:lb_h(y2|s2)} as follows
\begin{align}
    h&\fparen{\{\mbf{y}_{2,m}\}_{2}^{M-1}\given \rvec{x}_2,\big\{\mbf{y}_{2,m}\big\}_{0}^{1}}\nonumber\\
    &=\sum_{m=2}^{M-1} h\fparen{\mbf{y}_{2,m}\given \rvec{x}_2,\big\{\mbf{y}_{2,i}\big\}_{i=0}^{m-1}} \nonumber\\
    &\note{a}{=} \sum_{m=2}^{M-1} I\fparen{\mbf{y}_{2,m};\vec{\thetab}_2\given \rvec{x}_2,\big\{\mbf{y}_{2,i}\big\}_{i=0}^{m-1}} +h\fparen{\mbf{y}_{2,m}\given \rvec{x}_2,\vec{\thetab}_2,\big\{\mbf{y}_{2,i}\big\}_{i=0}^{m-1}}\nonumber\\
    & \note{b}{=} (M-2)\log(\pi e)+\sum_{m=2}^{M-1} I\fparen{\phase{\mbf{y}}_{2,m};\vec{\thetab}_2\given \rvec{x}_2,\big\{\phase{\mbf{y}}_{2,i}\big\}_{i=0}^{m-1},\big\{{|\mbf{y}_{2,i}|}\big\}_{i=0}^{m}}\nonumber\\
    & = (M-2)\log(\pi e)\nonumber\\&\quad+\sum_{m=2}^{M-1} \bigg(h\fparen{\phase{\mbf{y}}_{2,m}\given \rvec{x}_2,\big\{\phase{\mbf{y}}_{2,i}\big\}_{i=0}^{m-1},\big\{|\mbf{y}_{2,i}|\big\}_{i=0}^{m}} - h\fparen{\phase{\mbf{y}}_{2,m}\given \rvec{x}_2,\vec{\thetab}_2,\big\{\phase{\mbf{y}}_{2,i}\big\}_{i=0}^{m-1},\big\{|\mbf{y}_{2,i}|\big\}_{i=0}^{m}}\!\bigg)
   \label{eq:app_lb_h(y2|x2)}
\end{align}
where (a) follows from the definition of mutual information and (b) holds since it is evident from \eqref{eq:app:ykm}--\eqref{eq:app:abs_ykm} that $|\mbf{y}_{2,m}|$ and $\vec{\thetab}_2$ are independent, and $(\mbf{y}_{2,m}\given\mbf{x}_{2,m},\vec{\thetab}_2)\sim\CN(e^{j\thetab_{2,m}}\mbf{x}_{2,m},1)$ for all $m\in\{2,\dots,M-1\}$.} 

\moved{Next, we upper-bound each entropy term in the first summation term on the \ac{RHS} of \eqref{eq:app_lb_h(y2|x2)}. We consider the cases of $m=2$ and $m>2$ separately.}

\moved{We first define
\begin{equation}
        \mbf{y}'_{k,m} = e^{-j\phase{\mbf{x}}_{k,m}} \mbf{y}_{k,m},
\end{equation}
which implies
\begin{align}
    \phase{\mbf{y}}'_{k,m} &=     \phase{\mbf{y}}_{k,m} \modsub \phase{\mbf{x}}_{k,m}\nonumber\\
    & = \thetab_{k,m} \modplus \phase{\mbf{s}_{k,m} + \mbf{z}_{k,m}}.\label{eq:app:phase_ykm'}
\end{align}
Then, starting with the case $m=2$, we obtain
\begin{align}
      h&\fparen{\phase{\mbf{y}}_{2,2}\given \rvec{x}_2,\{\phase{\mbf{y}}_{2,i}\}_{i=0}^{1},\brace{|\mbf{y}_{2,i}|}_{i=0}^{2}} \nonumber\\
      &= h\!\fparen{\phase{\mbf{y}}_{2,2} \!\modsub\! \phase{\mbf{x}}_{2,2} \given \phase{\rvec{x}}_2, \abs{\rvec{x}_2},\{\phase{\mbf{y}}_{2,i}\}_{i=0}^{1},\brace{|\mbf{y}_{2,i}|}_{i=0}^{2}} \nonumber\\
       & \note{a}{=} h\fparen{\phase{\mbf{y}}'_{2,2}\given \abs{\rvec{x}_2},\{\phase{\mbf{y}}'_{2,i}\}_{i=0}^{1},\brace{|\mbf{y}_{2,i}|}_{i=0}^{2}} \nonumber\\
      &\note{b}{=}h\Big(\phase{\mbf{y}}'_{2,2}\!\modsub\!2\phase{\mbf{y}}'_{2,1}\!\modplus\!\phase{\mbf{y}}'_{2,0}\given \abs{\rvec{x}_2},\brace{\phase{\mbf{y}}'_{2,i}}_{i=0}^{1},\brace{|\mbf{y}_{2,i}|}_{i=0}^{2}\Big)\nonumber\\
      &\note{c}{\le}h\Big(\phase{\mbf{y}}'_{2,2}\modsub2\phase{\mbf{y}}'_{2,1}\modplus\phase{\mbf{y}}'_{2,0}\given \abs{\rvec{x}_2}\Big)\nonumber\\
    &\note{d}{=}h\Big(\phase{\abs{\mbf{x}_{2,2}}+\mbf{z}_{2,2}}\!\modsub\!2\phase{\abs{\mbf{x}_{2,1}}+\mbf{z}_{2,1}}\!\modplus\!\phase{\abs{\mbf{x}_{2,0}}+\mbf{z}_{2,0}}\bgiven \abs{\rvec{x}_2}\Big),
    % & \note{e}{=}h\fparen{\phi(2,\abs{\rvec{x}_2})\given \abs{\rvec{x}_2},\brace{|\mbf{y}_{2,i}|}_{i=0}^{2}},\nonumber \\
    \label{eq:app_lb_h(y2|x2)_tmp}
\end{align}
where (a) follows directly from \eqref{eq:app:phase_ykm'} and that $\phase{\rvec{x}}_2$ is irrelevant for $\phase{\mbf{y}}'_{2,2}$, (b) follows as we added or subtracted given phases which does not change the entropy, and (c) holds as \added{conditioning does not increase} entropy. \added{Finally}, in (d) we used that according to \eqref{def:theta_km_Eo_comb} for $m=2$ we have $\thetab_{2,2} = 2\thetab_{2,1}\modsub\thetab_{2,0}$.}

\label{r1:c2_b}\added{ 
Since $\phase{\abs{\mbf{x}_{2,2}}+\mbf{z}_{2,2}}\!\modsub\!2\phase{\abs{\mbf{x}_{2,1}}+\mbf{z}_{2,1}}\!\modplus\!\phase{\abs{\mbf{x}_{2,0}}+\mbf{z}_{2,0}}$ is supported on $[-\pi,\pi)$, we can apply the maximum entropy theorem as in \cite[Eq. (22)]{barletta_ub:2017} to upper-bound the conditional entropy term in \eqref{eq:app_lb_h(y2|x2)_tmp} by the entropy of a wrapped Gaussian random variable with the same variance, yielding}
\begin{align}
   \moved{h}&\moved{\Big(\phase{\abs{\mbf{x}_{2,2}}+\mbf{z}_{2,2}}\!\modsub\!2\phase{\abs{\mbf{x}_{2,1}}+\mbf{z}_{2,1}}\!\modplus\!\phase{\abs{\mbf{x}_{2,0}}+\mbf{z}_{2,0}}\bgiven \abs{\rvec{x}_2}\Big)} \nonumber\\
    & \added{\note{a}{\le} \min \left(\log(2\pi), \frac{1}{2} \Eb{\log\fparen{\pi e \fparen{ \frac{1}{|\mbf{x}_{2,2}|^2} +\frac{4}{ |\mbf{x}_{2,1}|^2}  + \frac{1}{|\mbf{x}_{2,0}|^2}}}}\right)} \nonumber\\ 
    & \added{\note{b}{=} \min \left(\log\fparen{2\pi}, \frac{1}{2} \E\left[\log\left( \pi e \phi(2,\abs{\rvec{x}_2}\right)\right]\right),}
     \label{eq:app_lb_h(y2|x2)_a}
\end{align}
\added{where in (a), we used the bound $
\mathsf{Var}\fparen{\phase{\abs{\mbf{x}_{2,2}}+\mbf{z}_{2,2}}\!-\!2\phase{\abs{\mbf{x}_{2,1}}+\mbf{z}_{2,1}}\!+\!\phase{\abs{\mbf{x}_{2,0}}+\mbf{z}_{2,0}} \bgiven \abs{\rvec{x}_2}} \le 1/(2|\mbf{x}_{2,2}|^2) + 2/(|\mbf{x}_{2,1}|^2) + 1/(2|\mbf{x}_{2,0}|^2),$  
and in (b), we used the function of $\phi(\cdot)$ defined in \eqref{eq:Phi_func}.  
}

\moved{For the case $m>2$, we follow similar footsteps as in \eqref{eq:app_lb_h(y2|x2)_a} and write}
\begin{align}
  \moved{h}&\moved{\fparen{\phase{\mbf{y}}_{2,m}\given \rvec{x}_2,\brace{\phase{\mbf{y}}_{2,i}}_{i=0}^{m-1},\brace{|\mbf{y}_{2,i}|}_{i=0}^{m}}} \nonumber\\
&\moved{= h\fparen{\phase{\mbf{y}}'_{2,m}\given \abs{\rvec{x}_2},\brace{\phase{\mbf{y}}'_{2,i}}_{i=0}^{m-1},\brace{|\mbf{y}_{2,i}|}_{i=0}^{m}}} \nonumber\\
  &\moved{=h\Big( \phase{\mbf{y}}'_{2,m}\modsub\phase{\mbf{y}}'_{2,m-1}\modsub\phase{\mbf{y}}'_{2,m-2}\modplus\phase{\mbf{y}}'_{2,m-3}\bgiven \abs{\rvec{x}_2},\brace{\phase{\mbf{y}}'_{2,i}}_{i=0}^{m-1},\brace{|\mbf{y}'_{2,i}|}_{i=0}^{m}\Big)}\nonumber\\
  & \added{\note{a}{\le} h\Big( \phase{\mbf{y}}'_{2,m}\modsub\phase{\mbf{y}}'_{2,m-1}\modsub\phase{\mbf{y}}'_{2,m-2}\modplus\phase{\mbf{y}}'_{2,m-3}\bgiven \abs{\rvec{x}_2}\Big)}\nonumber \\
  &\added{\note{b}{\le }\min \left(\log\fparen{2\pi}, \frac{1}{2} \E\left[\log\left( \pi e \phi(m,\abs{\rvec{x}_m}\right)\right]\right),}
  \label{eq:app_lb_h(y2|x2)_b}
\end{align}
\moved{where \added{ in (a)}, we used that \added{conditioning does not increase} entropy and that $\thetab_{2,m}= \thetab_{2,m-1}\modplus\thetab_{2,m-2}\modsub\thetab_{2,{m-3}}$, for $m>2$\added{; in (b), we upper-bounded the entropy term using the maximum entropy theorem, as in \eqref{eq:app_lb_h(y2|x2)_a}.}  Note that we could possibly use $\thetab_{2,m} = 2\thetab_{2,m-1}\modsub\thetab_{2,m-2}$ as in \eqref{eq:app_lb_h(y2|x2)_a} but it would lead to a looser bound. Finally, combining \eqref{eq:app_lb_h(y2|x2)_a} and \eqref{eq:app_lb_h(y2|x2)_b} we obtain for $m\ge2$} 
\begin{equation}
\moved{h\fparen{\phase{\mbf{y}}_{2,m}\given \rvec{x}_2,\brace{\phase{\mbf{y}}_{2,i}}_{i=0}^{m-1},\brace{\abs{\mbf{y}_{2,i}}}_{i=0}^{m}}} \le \added{\min \left(\log\fparen{2\pi}, \frac{1}{2} \E\left[\log\left( \pi e \phi(m,\abs{\rvec{x}_m}\right)\right]\right)}.\label{eq:app_lb_h(y2|x2)_ab_final} 
\end{equation}

\moved{For the second entropy term on the \ac{RHS} of \eqref{eq:app_lb_h(y2|x2)},
\begin{align}
h\fparen{\phase{\mbf{y}}_{2,m}\given \rvec{x}_2,\vec{\thetab}_2,\big\{\phase{\mbf{y}}_{2,i}\big\}_{i=0}^{m-1},\big\{\abs{\mbf{y}_{2,i}}\big\}_{i=0}^{m}} 
    &= h\fparen{\phase{\mbf{y}}_{2,m}-\phase{\mbf{x}}_{2,m}\given \rvec{x}_2,\vec{\thetab}_2,\big\{\phase{\mbf{y}}_{2,i}\big\}_{i=0}^{m-1},\big\{\abs{\mbf{y}_{2,i}}\big\}_{i=0}^{m}} \nonumber\\
    &\note{a}{=}  h\fparen{\thetab_{2,m}\modplus\phase{\abs{\mbf{x}_{2,m}}+\mbf{z}_{2,m}}\bgiven \abs{\mbf{x}_{2,m}},\thetab_{2,m},\abs{\mbf{y}_{2,m}}}\nonumber\\
    &\note{b}{=} h\paren{\phase{\abs{\mbf{x}_{2,m}}+\mbf{z}_{2,m}}\bgiven \abs{\mbf{x}_{2,m}},\abs{\mbf{y}_{2,m}}},\label{eq:app_lb_h(y2|x2)_c}
\end{align}
where (a) follows as $\brace{\phase{\mbf{y}}_{2,i}}_{i=0}^{m-1}$ are irrelevant given ${\thetab}_{2,m}$ and (b) follows since $\thetab_{2,m}$ is independent of $\phase{\abs{\mbf{x}_{2,m}}+\mbf{z}_{2,m}}$.} 
%\vspace{0.1cm}

\moved{Substituting \eqref{eq:app_lb_h(y2|x2)_ab_final} and \eqref{eq:app_lb_h(y2|x2)_c} into \eqref{eq:app_lb_h(y2|x2)} gives}
\begin{align}
     \moved{h}&\moved{\fparen{\brace{\mbf{y}_{2,m}}_{m=2}^{M-1}\bgiven \rvec{x}_2,\mbf{y}_{2,0},\mbf{y}_{2,1}}}\nonumber\\
     &\moved{\le (M-2)\log(\pi e)} \nonumber \\ & \quad + \moved{\sum_{m=2}^{M-1} \bigg(\!\added{\min \left(\log\fparen{2\pi}, \frac{1}{2} \E\left[\log\left( \pi e \phi(m,\abs{\rvec{x}_m}\right)\right]\right)} -h\fparen{\phase{\abs{\mbf{x}_{2,m}}+\mbf{z}_{2,m}}\bgiven \abs{\mbf{x}_{2,m}},|\mbf{y}_{2,m}|}\bigg)}\nonumber\\
     &\moved{=(M-2)\log(\pi e)+\sum_{m=2}^{M-1}\E[g(m,\abs{\rvec{x}_2})],}\label{eq:lb_h(y2|s2)_b_final}
\end{align}
\moved{where $g(\cdot)$ is defined in \eqref{eq:g_func}.} 
\label{r1:c2_b_end}

% Which gives
% \begin{equation}
%         h\fparen{\{\mbf{y}_{2,m}\}_{2}^{M-1}\bgiven \rvec{x}_2,\brace{\mbf{y}_{2,m}}_{0}^{1}}\le (M-2)\log(\pi e)+\sum_{m=2}^{M-1}\E[g(m,\abs{\rvec{x}_2})],\label{eq:lb_h(y2|s2)_b_final}
% \end{equation}
%where the last equality follows as $\mbf{s}_{2,m}$ and $\mbf{z_{2,m}}$ are identically distributed for $m\ge2$. 
\moved{Substituting \eqref{eq:lb_h(y2|s2)_b_final} and \eqref{eq:lb_h(y2|s2)_a_final} into \eqref{eq:lb_h(y2|s2)} gives}
\begin{align}
  \moved{h\paren{\rvec{y}_{2}\given \rvec{x}_2,\vec{\thetab}_1}} & \le \moved{(M-1)\log(\pi e)-\log 2+\frac{1}{2}\E\brack{\log\fparen{1+2 \abs{\mbf{x}_{2,0}}^2}}+\frac{1}{2} \E\brack{\log\fparen{1+2 \abs{\mbf{x}_{2,1}}^2}}}\nonumber\\
    &\quad +\moved{h\fparen{\!\brace{\Deltacb{2}\modplus m\Deltarb{2}\!\modplus\!\phase{\abs{\mbf{x}_{2,m}}+\mbf{z}_{2,m}}}_{0}^{1}\bgiven\abs{\rvec{x}_2},\fbrace{\abs{\mbf{y}_{2,m}}}_{0}^{1}}\!+\!\sum_{m=2}^{M-1}\E[g(m,\abs{\rvec{x}_2})].}\label{eq:lb_h(y2|x2)_final}
\end{align}
\moved{Substituting \eqref{eq:lb_h(y2|x2)_final} and \eqref{eq:h(y2)} in \eqref{eq:I(y2;x2)} gives}
\begin{align}
   \moved{I\paren{\rvec{x}_2;\rvec{y}_2\given \vec{\thetab}_{1}}}
   &\ge \moved{\log(2\pi)-(M-1)\log e+h\paren{\abs{\rvec{x}_2}^2}}\nonumber\\&\quad-\moved{\frac{1}{2}\E\brack{\log\fparen{1+2 \abs{\mbf{x}_{2,0}}^2}}-\frac{1}{2} \E\brack{\log\fparen{1+2 \abs{\mbf{x}_{2,1}}^2}}}\nonumber\\
    &\quad -\moved{h\!\fparen{\brace{\Deltacb{2}\!\modplus m\Deltarb{2}\!\modplus\!\phase{\abs{\mbf{x}_{2,m}}+\mbf{z}_{2,m}}}_{0}^{1}\bgiven\abs{\rvec{x}_2},\fbrace{\abs{\mbf{y}_{2,m}}}_{0}^{1}}\!-\!\sum_{m=2}^{M-1}\E[g(m,\abs{\rvec{x}_2})].}\label{eq:I(y2;x2)_final}
\end{align}

\moved{To this point, \eqref{eq:I(y2;x2)_final} provides a lower bound for the first term on the \ac{RHS} of \eqref{eq:cap_lb}. To finalize the derivation of the lower bound, \added{the next part provides} an upper bound on the second term on the \ac{RHS} of \eqref{eq:cap_lb}.}

\subsubsection*{\added{\underline{Characterizing the second term on the \ac{RHS} of \eqref{eq:cap_lb}}}}
% \begin{lemma}\label{lemma:epsilon2}

% The second term on the \ac{RHS} of \eqref{eq:cap_lb} can be upper-bounded as
% \begin{equation}
%    I\paren{\rvec{x}_2;\vec{\thetab}_{1}\given \rvec{x}_{1},\rvec{y}_{1},\rvec{y}_2} \le h\big(\brace{\Deltacb{2}\modplus m\Deltarb{2}\!\modplus\!\phase{\abs{\mbf{x}_{2,m}}+\mbf{z}_{2,m}}}_{0}^{1}\!\bgiven\abs{\rvec{x}_2},\fbrace{\abs{\mbf{y}_{2,m}}}_{0}^{1}\big)-h\paren{\Deltacb{2},\Deltarb{2}},
%    %&\note{a}{\triangleq} 
%   % h\big(\{\Deltacb{2}\modplus m\Deltarb{2}\!\modplus\!\phase{\xi+\mbf{z}_{2,m}}\}_{0}^{1}\bgiven\!\{\abs{\xi+\mbf{z}_{2,m}}\}_{0}^{1}\big)\nonumber\\
%   %  &\quad-h\paren{\Deltacb{2},\Deltarb{2}}
% \label{eq:epsilon_final}
% \end{equation} 
% where $|\mbf{y}_{2,m}| = \big||\mbf{x}_{2,m}|+\mbf{z}_{2,m}\big|$ and $\mbf{z}_{2,m} \sim\CN(0,1)$. Moreover, $\Deltacb{2}\sim \mathcal{WN}(0,\sigma^2_\mathrm{c})$ and $\Deltarb{2}\sim \mathcal{WN}(0,\sigma^2_\mathrm{r})$ are independent of all the other random variables. 
% \begin{IEEEproof}
%      See Appendix~\ref{app:epsilon2}.
% \end{IEEEproof}
% \end{lemma}

\moved{Following similar steps as in \cite[App. IX]{lapidoth_duality:2003}, we can write %\eqref{eq:epsilon_final} as
\begin{align}
I\paren{\rvec{x}_2;\vec{\thetab}_1\given \rvec{x}_{1},\rvec{y}_{1},\rvec{y}_2}
   &= h\paren{\vec{\thetab}_1\given \rvec{x}_{1},\rvec{y}_{1},\rvec{y}_2}-h\paren{\vec{\thetab}_1\given \rvec{x}_{1},\rvec{x}_{2},\rvec{y}_{1},\rvec{y}_2}\nonumber\\
   &\note{a}{\le} h\paren{\vec{\thetab}_1\given\rvec{x}_{1},\rvec{y}_{1}}-h\paren{\vec{\thetab}_1\given \rvec{x}_{1},\rvec{x}_{2},\rvec{y}_{1},\rvec{y}_2,\vec{\thetab}_2}\nonumber\\
   &\note{b}{=} h\paren{\vec{\thetab}_1\given\rvec{x}_{1}, \rvec{y}_{1}}-h\paren{\vec{\thetab}_1\given \rvec{x}_{1},\rvec{y}_{1},\vec{\thetab}_2}\nonumber\\
   &= I\paren{\vec{\thetab}_2;\vec{\thetab}_1\bgiven\rvec{x}_{1},\rvec{y}_{1}}\nonumber\\
   &\note{c}{=} I\paren{\{\thetab_{2,m}\}_{0}^{1};\vec{\thetab}_1\bgiven\rvec{x}_{1},\rvec{y}_{1}}\nonumber\\
   &= h\paren{\{\thetab_{2,m}\}_{0}^{1}\bgiven\rvec{x}_{1},\rvec{y}_{1}}-h\paren{\{\thetab_{2,m}\}_{0}^{1}\bgiven\rvec{x}_{1},\rvec{y}_{1},\vec{\thetab}_1}\nonumber\\
   &\note{d}{=} h\paren{\{\thetab_{2,m}\}_{0}^{1}\bgiven\rvec{x}_{1},\rvec{y}_{1}}-h\paren{\{\thetab_{2,m}\}_{0}^{1}\bgiven \vec{\thetab}_1},
   \label{eq:epsilon_proof}
\end{align}
where (a) holds because \added{conditioning does not increase} the entropy; (b) follows since given $\thetacb{2}$, $\thetacb{1}$ is independent of the pair $(\rvec{y}_2,\rvec{x}_2)$; (c) follows as $\{\thetab_{2,m}\}_{0}^{1}$ is sufficient statistics for $\vec{\thetab}_2$; (d) holds since $\rvec{\thetab}_2$ and $(\rvec{y}_1,\rvec{x}_1)$ are independent given $\rvec{\thetab}_1$.}

\moved{The first term on the \ac{RHS} of \eqref{eq:epsilon_proof} can be upper-bounded as
\begin{align}
h\paren{\brace{\thetab_{2,m}}_{0}^{1}\bgiven \rvec{x}_{1},\rvec{y}_{1}}
   &=h\fparen{\brace{\thetab_{2,m}}_{0}^{1}\bgiven \abs{\rvec{x}_{1}},\phase{\rvec{x}}_{1}, |\rvec{y}_{1}|, \phase{\rvec{y}}_{1}}\nonumber\\
   &=h\fparen{\brace{\thetab_{2,m}\modsub\phase{\mbf{y}}_{1,m} \modplus \phase{\mbf{x}}_{1,m}}_{0}^{1}\bgiven \abs{\rvec{x}_{1}},\phase{\rvec{x}}_{1}, |\rvec{y}_{1}|, \phase{\rvec{y}}_{1}}\nonumber\\
   &\note{a}\le h\paren{\brace{\thetab_{2,m}\modsub\phase{\mbf{y}}_{1,m}\modplus \phase{\mbf{x}}_{1,m}}_{0}^{1}\bgiven \abs{\rvec{x}_1},\brace{|\mbf{y}_{1,m}|}_{0}^{1}}\nonumber\\
   &\note{b}{=}h\big(\brace{\thetab_{2,m}\modsub\thetab_{1,m} \modsub\phase{\abs{\mbf{x}_{1,m}}+\mbf{z}_{1,m}}}_{0}^{1}\bgiven|\rvec{x}_1|,\brace{|\mbf{y}_{1,m}|}_{0}^{1}\big)\nonumber\\
   &\note{c}{=}h\big(\brace{\Deltacb{2}\modplus m\Deltarb{2}\modsub\phase{\abs{\mbf{x}_{1,m}}+\mbf{z}_{1,m}}}_{0}^{1}\bgiven|\rvec{x}_1|,\brace{|\mbf{y}_{1,m}|}_{0}^{1}\big)\nonumber\\
    &\note{d}{=} 
   h\big(\brace{\Deltacb{2}\modplus m\Deltarb{2}\modplus\phase{\abs{\mbf{x}_{2,m}}+\mbf{z}_{2,m}}}_{0}^{1}\bgiven|\rvec{x}_2|,\brace{|\mbf{y}_{2,m}|}_{0}^{1}\big)\label{eq:epsilon_a_final}
\end{align}
where (a) holds because \added{conditioning does not increase} entropy, (b) follows directly from \eqref{eq:app:phase_ykm} where $\phase{\mbf{y}}_{1,m} = \thetab_{1,m} \modplus \phase{\mbf{x}}_{1,m}\modplus \phase{|\mbf{x}_{1,m}| + \mbf{z}_{1,m}}$, and (c) holds since $\thetab_{2,m} \modsub \thetab_{1,m} = \Deltacb{2}\modplus m\Deltarb{2} $from \eqref{def:theta_km_Eo_comb}--\eqref{eq:EO_chan_model_c}; finally, in (d) we use that $-\phase{|\mbf{x}_{1,m}|+\mbf{z}_{1,m}}$ has the same distribution as $\phase{|\mbf{x}_{2,m}|+\mbf{z}_{2,m}}$, because the phase function is symmetric and the pairs $(|\rvec{x}_{1}|,|\rvec{x}_{2}|)$ and $(\mbf{z}_{1,m},\mbf{z}_{2,m})$ are exchangeable since they have the same distribution due to the \ac{iid} assumption of the input distribution.} 

\moved{We continue with the second term on the \ac{RHS} of \eqref{eq:epsilon_proof} 
\begin{align}
   h\paren{\{\thetab_{2,m}\}_{0}^{1}\given \vec{\thetab}_1} &\note{a}{=}h\paren{\thetacb{2},\thetacb{2}\modplus\thetarb{2}\given \thetacb{1},\thetarb{1}}\nonumber\\
   & \note{b}{=} h\paren{\thetacb{2},\thetarb{2}\given \thetacb{1},\thetarb{1}}\nonumber\\
   & = h(\Deltacb{2},\Deltarb{2}),\label{eq:epsilon_b_final}
\end{align}
where (a) follows since the pair $(\thetacb{1},\thetarb{1})$ is sufficient statistic for $\vec{\thetab}_1$ and (b) holds since the pair $(\thetacb{2},\thetarb{2})$ is sufficient statistics for the pair $(\thetacb{2},\thetacb{2}\modplus\thetarb{2})$.} 

\moved{Substituting \eqref{eq:epsilon_a_final} and \eqref{eq:epsilon_b_final} into \eqref{eq:epsilon_proof} gives
\begin{equation}
   I\paren{\rvec{x}_2;\vec{\thetab}_{1}\given \rvec{x}_{1},\rvec{y}_{1},\rvec{y}_2} \le h\big(\brace{\Deltacb{2}\modplus m\Deltarb{2}\!\modplus\!\phase{\abs{\mbf{x}_{2,m}}+\mbf{z}_{2,m}}}_{0}^{1}\!\bgiven\abs{\rvec{x}_2},\fbrace{\abs{\mbf{y}_{2,m}}}_{0}^{1}\big)-h\paren{\Deltacb{2},\Deltarb{2}}.
\label{eq:epsilon_final}
\end{equation}} 
For sufficiently large $\abs{\rvec{x}_2}$, \eqref{eq:epsilon_final} can be made arbitrarily close to zero. This property will be used in Section \ref{sec:asym_lb} to obtain a high-\ac{SNR} lower bound on the capacity.

Finally, substituting \eqref{eq:epsilon_final} and  \eqref{eq:I(y2;x2)_final} into \eqref{eq:cap_lb} results in the desired lower bound \eqref{eq:th_lower_bound} on the capacity. Note that we defined $\rvec{s} = \abs{\rvec{x}_2}$ and replaced all the random quantities with timeless random variables with the same distributions (i.e., $\Deltacb{2}\to \Deltacb{}$, $\Deltarb{2}\to \Deltarb{}$, $\rvec{z}_{2}\to \rvec{z}$, $\abs{\rvec{x}_2}\to \rvec{s}$).  

%=========================== Section VI =================================
\section{High-\ac{SNR} Capacity Bounds}\label{app:sec:proof_asym_ub_lb}
This section is dedicated to the proof of the high-\ac{SNR} bounds in Section \ref{subsec:high_snr_cap}. For the upper bound, we exploit the insight in Lemma~\ref{lemma:escape_to_infty} that at high \acp{SNR}, the capacity can be achieved using an input distribution that escapes to infinity. To establish the lower bound, we choose a particular input distribution that escapes to infinity as the \ac{SNR} approaches infinity. 

\subsection{Proof of Theorem~\ref{th:asym_upper_bound} (High-\ac{SNR} Upper Bound)}\label{sec:asym_lb}
Based on Lemma \ref{lemma:escape_to_infty}, the high-\ac{SNR} behavior of $C(\rho)$ does not change under the additional constraint that the support of the amplitude of each element of the input vector lies outside an arbitrary radius. Hence,  we present the high-\ac{SNR} behavior of a modified version of the upper bound presented in Theorem \ref{th:upper_bound} under the additional constraint that $\abs{\rvec{x}_k}\ge\xi$ for all $k$ and any $0<\xi<\sqrt{\rho/M}$.  

Set $\lambda=\lambda^{\ast} = (M-1)\log e$ and $\vec{\alpha} = \vec{\alpha}^{\ast}$ defined in \eqref{eq:alpha_star}, which results in $\alpha_\Sigma = M-1$ and $d_{\lambda^{\ast},\vec{\alpha}^{\ast}} = \log(\pi e)$. We define $C^{(\xi)}(\rho)$ similar to \eqref{eq:capacity}--\eqref{eq:pw_cons} with the additional constraint that $\abs{\rvec{x}_k}\ge\xi$ for all $k$. Then, a similar relation to \eqref{eq:cap_ub} can be obtained where the supremum is over all distributions $\mathcal{Q}_{\rvec{x}}$ such that $\E[\norm{\rvec{x}_1}^2]\le \rho$ and $\abs{\rvec{x}_1} \ge \xi$, and from there by applying Lemma~\ref{lemma:duality_bound} and Lemma~\ref{lemma:I_y_theta}%following the same steps leading to \eqref{eq:mi_up_B}, %but considering that $\rvec{x} \in \mathcal{K}(\xi)$ and setting $\lambda=\lambda^{\ast} = (M-1)\log e$ and $\vec{\alpha} = \vec{\alpha}^{\ast}$ defined in \eqref{eq:alpha_star}, 
, we obtain $C^{(\xi)}(\rho)\le U^{(\xi)}(\rho)$, where 
\begin{equation}
    U^{(\xi)}(\rho) =\fparen{M-1}\log\fparen{\frac{\rho+M}{M-1}} +\log (\pi e)+2\log(2\pi)+\max_{\vec{s}\ge{\xi}}\left\{R_{\lambda^{\ast},\vec{\alpha}^{\ast}}(\rho,\vec{s})+F(M,\vec{s},\Deltacb{},\Deltarb{})\right\}.\label{eq:hsnr_ub}
\end{equation}
Here, the supremum over all $\mathcal{Q}_{\rvec{\rvec{x}_1}}$ satisfying $\abs{\rvec{x}_1} \ge \xi$ is upper-bounded by removing the power constraint and maximizing over all deterministic $\vec{s}\ge \xi$.

% The following lemma helps to characterize the high-\ac{SNR} behavior of \eqref{eq:hsnr_ub}.
% \begin{lemma}\label{lemma:asym_ub}
% % For any finite $\rho$
% % \begin{equation}
% %     \lim_{\substack{
% %     \vec{s}\to\infty}} R_{\lambda^{\ast},\vec{\alpha}^{\ast}}(\rho,\vec{s}) = -\log(4\pi e),\label{eq:r_asym}
% % \end{equation}
% and
% \begin{equation}
%      \lim_{\substack{\vec{s}\to \infty}}  F(M,\vec{s},\Deltacb{},\Deltarb{}) = -h\paren{\Deltacb{},\Deltarb{}},\label{eq:f_asym}
% \end{equation}
% where $\vec{s}\to \infty$ stands for $s_m \to \infty$ for all $m\in \{0,\dots,M-1\}$.
% \end{lemma}
% \begin{IEEEproof}
% See Appendix~\ref{app:lemma_asym_ub}.
% %We shall show that \eqref{eq:r_asym} and \eqref{eq:f_asym} holds by letting $\xi \to \infty$ which results in $\vec{s} \to \infty$. 

% \end{IEEEproof}

\moved{By setting $\lambda=\lambda^{\ast} = (M-1)\log e$ and $\vec{\alpha} = \vec{\alpha}^{\ast}$ in \eqref{eq:Rs} we get}
\begin{align}
\moved{R_{\lambda^{\ast},\vec{\alpha}^{\ast}}(\rho,\vec{s})} &=\moved{\frac{1}{2}}\added{\fparen{\log|s_0|^2 + \mathrm{E}_1\fparen{s_0^2}}}-\moved{h\paren{|s_0+\mbf{z}_0|^2\bgiven s_0}}\nonumber\\
&\quad+\moved{\frac{1}{2}}\added{\fparen{\log|s_1|^2 + \mathrm{E}_1\fparen{s_1^2}}}-\moved{h\paren{|s_1+\mbf{z}_1|^2\bgiven s_1}}.\label{eq:R_prime}
\end{align}
\label{r1:c1_lemma23}
\added{Given the inequality \cite[Eq. 
6.8.1]{NIST:DLMF}  
\begin{equation}
    \frac{1}{2} e^{-x} \log\left(1 + \frac{2}{x} \right) < \mathrm{E}_1(x) < e^{-x} \log\left(1 + \frac{2}{x} \right), \quad \forall x>0
\end{equation}
we observe that both the upper and lower bounds approach zero as $x \to \infty$. Therefore, by the squeeze theorem, we conclude that  
\begin{equation}
    \lim_{x\to\infty} \mathrm{E}_1(x) = 0. \label{eq:asym_E1}
\end{equation}
Thus, using \moved{\eqref{eq:asym_E1}} and Lemma~\ref{lemma:4} gives}
\begin{align}
   \moved{\lim_{s_0 \to \infty}} &\moved{\frac{1}{2}}\added{\fparen{\log \fparen{s_0^2}+\mathrm{E}_1(s_0^2)}}\!-\!\moved{h\paren{|s_0+\mbf{z}_0|^2\bgiven s_0}} =\!\moved{-\frac{1}{2}\log (4\pi e)},\label{eq:lim_log_a}\\
  \moved{\lim_{s_1 \to \infty}}&\moved{\frac{1}{2}}\added{\fparen{\log \fparen{s_1^2}+\mathrm{E}_1(s_1^2)}}\!-\!\moved{h\paren{|s_1+\mbf{z}_1|^2\bgiven s_1}=\!-\frac{1}{2}\log (4\pi e)}.\label{eq:lim_log_b}
\end{align}
\added{Thus,} \moved{for any finite $\rho$}, \added{we have}
\moved{\begin{equation}
    \lim_{\substack{
    \vec{s}\to\infty}} R_{\lambda^{\ast},\vec{\alpha}^{\ast}}(\rho,\vec{s}) = -\log(4\pi e),\label{eq:r_asym}
\end{equation}}

\moved{To characterize the asymptotic behavior of \eqref{eq:Fs}, we use Lemma \ref{lemma:a} and that the phases $\phase{s_0+\mbf{z}_0} \to 0$, $\phase{s_1+\mbf{z}_1} \to 0$, and $\phase{\norm{\vec{s}}+\mbf{v}} \to 0$ as $\vec{s}\to \infty$. Then,  
\begin{align}
\lim_{\substack{\vec{s} \to \infty}} h\!\fparen{\brace{\Deltacb{}\!\modplus\! m\Deltarb{}\!\modplus\!\phase{{s}_{m}+\mbf{z}_{m}}}_{0}^{1}\bgiven|\vec{s}+\rvec{z}|} = h(\Deltacb{},\Deltarb{})\label{eq:lim_F_a}
\end{align}
and
\begin{align}
    \lim_{\substack{\vec{s} \to \infty}} h(\Deltarb{})+h\fparen{\Deltacb{}\!\modplus\!\phase{\norm{\vec{s}}+\mbf{v}}\,\Big|\,\big|\!\norm{\vec{s}}+\mbf{v}\big|}= h(\Deltacb{},\Deltarb{}),\label{eq:lim_F_b}
\end{align}
which results in
\begin{equation}
     \lim_{\substack{\vec{s}\to \infty}}  F(M,\vec{s},\Deltacb{},\Deltarb{}) = -h\paren{\Deltacb{},\Deltarb{}}.\label{eq:f_asym}
\end{equation}}
\label{r2:c3_a_end}

Choosing\footnote{For instance, setting $\xi = \sqrt{\rho}/M$ satisfies the required conditions. Note that certain other functions will give the same result. %Any choice of $0<\xi<\sqrt{\rho/M}$ that ensures $\E[\norm{\rvec{x}_k}^2]\le \rho$ and leads to $\xi\to\infty$ as $\rho\to \infty$ leads to the same results.
} $\xi < \sqrt{\rho/M}$ as a function of $\rho$ %such that $\E[\norm{\rvec{x}_k}^2]\le \rho$ for all $k$ and 
such that $\xi \to \infty$ as $\rho \to \infty$, we obtain from \eqref{eq:hsnr_ub}, \added{\eqref{eq:r_asym}, and \eqref{eq:f_asym}}
\begin{equation}
 U^{(\xi)}(\rho) = U_\mathrm{hsnr}(\rho),\label{eq:lim_u_xi}
\end{equation}
where $U_\mathrm{hsnr}(\rho)$ is defined in Theorem~\ref{th:asym_upper_bound}. Finally, using \eqref{eq:lim_u_xi} and recalling from Lemma~\ref{lemma:escape_to_infty} that $ C^{(\xi)}(\rho) = C(\rho)+o(1)$, we can write
\begin{equation}
       C(\rho) \le  U_\mathrm{hsnr}(\rho),
\end{equation}
which completes the proof of Theorem~\ref{th:asym_upper_bound}.

\subsection{Proof of Theorem~\ref{th:asym_lower_bound} (High-\ac{SNR} Lower Bound)}
Here, we characterize the high-\ac{SNR} characteristics of the lower bound by selecting a particular distribution for $\rvec{s}$ in Theorem~\ref{th:lower_bound}. We opt for the truncated gamma distribution for $\rvec{s}$ based on the rationale that in the high-\ac{SNR} regime, the output can be considered as a rotated version of the input. Given that we employed the truncated gamma distribution for the squared amplitude of the output to derive the high-\ac{SNR} upper bound, it is reasonable to employ the same distribution for the squared amplitude of the input. Furthermore, the truncated gamma distribution serves as a versatile representation encompassing various well-known distributions.

Let $\mu =\rho/(M-1)$, $\gamma >0$, and  $\vec{\alpha}'(\gamma)\ge 0$ and take $\rvec{s} \sim \SGtr(\mu, \vec{\alpha}'(\gamma),\gamma)$. Using Lemma~\ref{lemma:truncated_mean} we can write
\begin{align}
    \E\big[\norm{\rvec{s}}^2\big] &= \sum_{m=0}^{M-1} \E\big[\abs{\mbf{s}_m}^2\big] \nonumber\\
    &=\mu \sum_{m = 0}^{M-1} J(\alpha'_m(\gamma),\gamma),\label{eq:pw_cons_truncated_a}
\end{align}
where $J(\cdot)$ is defined in \eqref{eq:def_J}. The distribution parameters $\gamma$ and $\vec{\alpha}'(\gamma)$ must be selected such that  the power constraint $\E[\norm{\rvec{s}}^2] \le \rho $ is satisfied.

The proof hinges on the choice of $\vec{\alpha}'(\gamma)$ and $\gamma$ as a function of $\rho$ such that \eqref{eq:pw_cons_truncated_a} is satisfied, and as $\rho\to\infty$, we have $\rho\gamma\to \infty$, $\gamma \to 0$, and $\vec{\alpha}'(\gamma)\to \vec{\alpha}^*$ where $\vec{\alpha}^{*}$ is defined in \eqref{eq:alpha_star}. Here, we select\footnote{Any other $\vec{\alpha}'(\gamma)$ that lead to $\E[\norm{\rvec{s}}^2] \le \rho$ and $\lim_{\gamma \to 0^+}\vec{\alpha}'(\gamma) = \vec{\alpha}^*$ will give the same result.}
\begin{equation}
    \alpha'_m(\gamma) = \alpha^*_m - c_m(\gamma),\label{eq:alpha_prime_def}
\end{equation}
where $\alpha^*_m$ and $c_m(\gamma)$  are defined in \eqref{eq:alpha_star} and \eqref{eq:c_gamma}, respectively. 
We also choose\footnote{Many other functions will give the same result.} 
\begin{align}
    \gamma = x_\mr{max}\cdot\begin{cases}
        1/\sqrt{\rho}, & \rho \ge 1,\\
        \rho, & 0< \rho <1,
    \end{cases}
\end{align}
which results in having $0<\gamma \le x_\mr{max}$ and as $\rho \to \infty$ we have $\rho\gamma \to \infty$ and $\gamma \to 0$.

\begin{lemma}\label{lemma:alpha_prime_power}
For $\vec{\alpha}'(\gamma)$ defined in \eqref{eq:alpha_prime_def} and for any $0\le\gamma\le x_\mr{max}$ where $x_\mr{max}\approx 0.00471$ is defined in Lemma~\ref{lemma:gamma_pow_c}, we have
\begin{equation}
    J(\alpha'_m(\gamma), \gamma) \le \alpha^*_m,\label{eq:lemma:j_upper_bound}
\end{equation}
and 
\begin{equation}
    \lim_{\gamma \to 0^{+}} \alpha_m'(\gamma) = \alpha_m^*. \label{eq:alpha_prime_limit}
\end{equation}
\end{lemma}
\begin{IEEEproof}
    See Appendix~\ref{app:lemma:alpha_prime}.
\end{IEEEproof}

Substituting \eqref{eq:lemma:j_upper_bound} into \eqref{eq:pw_cons_truncated_a} leads to $\E[\norm{\rvec{s}}^2] \le \rho$. Moreover, \eqref{eq:alpha_prime_limit} shows that $\vec{\alpha}'(\gamma)$ converges to $\vec{\alpha}^*$ element-wise as $\gamma \to 0$.

Note that with the chosen truncated distribution we have 
\begin{equation}
   |\rvec{s}|> \sqrt{\mu\gamma} = \sqrt{\rho\gamma/(M-1)}\label{eq:xi_prime}
\end{equation}
from \eqref{eq:truncated_gamma_dist}; thus,  $\mathrm{Pr}\{{\rvec{s}}\ge \sqrt{\mu\gamma}\} = 1$.

Using Lemma \ref{lemma:a} and that the phases $\phase{\mbf{s}_0+\mbf{z}_0} \to 0$ and $\phase{\mbf{s}_1+\mbf{z}_1} \to 0$, with probability $1$ as $\rho \to \infty$, we can write  
\begin{equation}
 h \big(\brace{\Deltacb{}\!\modplus\! m\Deltarb{} \modplus\phase{\mbf{s}_{m}+\mbf{z}_{m}}}_{0}^{1}\bgiven \rvec{s}, \{|\mbf{s}_{m}+\mbf{z}_{m}|\}_{0}^{1}\big) = h(\Deltacb{},\Deltarb{}) +o(1), \quad \rho \to \infty.\label{eq:lim_hdelta}
\end{equation}
Then, using Theorem~\ref{th:lower_bound} and substituting \eqref{eq:lim_hdelta} into \eqref{eq:th_lower_bound}, we obtain $C(\rho)\ge L^{(\xi)}(\rho)$, where 
\begin{align}
    L^{(\xi)}(\rho)&= \log(2\pi)-(M-1)\log(e)-h(\Deltacb{},\Deltarb{})\nonumber\\
    &\quad+h\paren{\rvec{s}^2}\!-\!\frac{1}{2}\E\brack{\log\fparen{1+2\mbf{s}_{0}^2}}\!-\!\frac{1}{2} \E\brack{\log\fparen{1+2\mbf{s}_{1}^2}} -\sum_{m=2}^{M-1}\E[g(m,\rvec{s})]+o(1).\label{eq:asym_lb_temp}
\end{align}
Next, we will characterize the high-\ac{SNR} behavior of each term on the \ac{RHS} of \eqref{eq:asym_lb_temp}. Recall that $\rvec{s}\sim \SGtr(\mu,\vec{\alpha}'(\gamma),\gamma)$ and that as $\rho\to\infty$, we have $\gamma \to 0$ and $\vec{\alpha}'(\gamma)\to\vec{\alpha}^{\ast}$. Thus, the \ac{pdf} $f_{\rvec{s}}(\vec{t})$ converges to $f_{\rvec{\tilde{s}}}(\vec{t})$ for every $\vec{t}$, where $\rvec{\tilde{s}}\sim \SGtr(\mu,\vec{\alpha}^{*},0)$, and we obtain
\begin{align}
    \lim_{\gamma\to 0}&h\fparen{\rvec{s}^2} = h\fparen{\rvec{\tilde{s}}^2},\label{eq:lb_asym_hu_rho}\\
    \lim_{\gamma\to 0}&\E\brack{\log \mbf{s}_m^2} =\E\brack{\log \mbf{\tilde{s}}_m^2}.\label{eq:lb_asym_hlog}
 \end{align}
%where $\rvec{u} \sim \SGtr(\mu,\vec{\alpha}^{*},0)$. 
Using that $\log(1+a) = \log a+o(1)$ for all real $a>1$, we can write
 \begin{equation}
%\lim_{\rho\to\infty}
\E\brack{\log\fparen{1+2\mbf{s}_{m}^2}}\!= \E\brack{\log \mbf{\tilde{s}}_m^2}+\log 2+o(1).\label{eq:log_u_gamma}
 \end{equation}

Moreover, since  $\E[\log\tilde{\mbf{s}}^2_m] = \log \mu + \psi(\alpha^*_m)$ \cite{durisi_MIMO_Cap:2013} where $\psi(\cdot)$ is the Euler’s digamma function, we can write 
\begin{align}
    \E[\log\tilde{\mbf{s}}^2_0] = \log \mu + \psi\fparen{\frac{1}{2}},\nonumber\\
    \E[\log\tilde{\mbf{s}}^2_1] = \log \mu + \psi\fparen{\frac{1}{2}}.\label{eq:Elog_s}
\end{align}
Note that for all $m$, the elements $\tilde{\mbf{s}}_m^2 \sim \trG(\mu,\vec{\alpha}^*, 0)$ and independent of each other. Thus, the entropy of $\rvec{\tilde{s}}^2$ can be written as \cite[Ch.~4.9]{michalowicz:diff_ent_handbook:2013}
\begin{align}
    h(\rvec{\tilde{s}}^2) &=\!\sum_{m=0}^{M-1}\!\!\paren{\alpha^*_m\log e+\log\mu+\log\Gamma(\alpha^*_m)+(1\!-\!\alpha^*_m)\psi(\alpha^*_m)}\nonumber\\
    &\note{a}{=} M\log\mu+\log(\pi)+(M-1)\log(e)+\psi\fparen{\frac{1}{2}},\label{eq:lb_asym_hu}
\end{align}
where (a) follows from the definition of $\vec{\alpha}^*$ in \eqref{eq:alpha_s} and that $\sum_{m=0}^{M-1} \alpha^*_m= M-1$. 

The following lemma helps to characterize the high-\ac{SNR} behavior of the \ac{RHS} of \eqref{eq:asym_lb_temp}. 

\begin{lemma}\label{lemma:g_asym}
For any deterministic vector $\vec{s}>\!\sqrt{\mu\gamma}$, we have
\begin{align}
   g(m,\vec{s})
    = \!\begin{cases}
       \frac{1}{2}\log\!\fparen{1\!+\!\frac{4{s}_2^2}{{s}^2_1}\!+\!\frac{{s}_2^2}{{s}^2_0}}+O\fparen{\frac{1}{\sqrt{\mu\gamma}}},& \!\!\!m=2,\\\\
     \frac{1}{2}\log\!\fparen{1+ \sum_{i=m-3}^{m-1} \frac{{s}_m^2}{{s}_i^2}}+O\fparen{\frac{1}{\sqrt{\mu\gamma}}},& \!\!\!m>2,
    \end{cases}\label{eq:gm_deterministic}
\end{align}
\begin{IEEEproof}
See Appendix~\ref{app:lemma_g_asym}. 
\end{IEEEproof}

\end{lemma}

Let $\tvec{\mbf{u}} = \rvec{s}/\sqrt{\mu}$, which leads to $\tvec{\mbf{u}} \sim \SGtr(1,\vec{\alpha}'(\gamma), \gamma)$. Then, using \eqref{eq:gm_deterministic} and recalling that as $\rho\to \infty$, we have $\mu\gamma \to\infty$, $\gamma \to 0$, and $\vec{\alpha}'(\gamma)\to\vec{\alpha}^{*}$, we obtain
\begin{align}
 \lim_{\rho\to\infty} \E[g(m,\rvec{s})] &= \lim_{\rho\to\infty} \E[g(m,\sqrt{\mu}\tvec{\mbf{u}})] \nonumber\\
 &\note{a}{=} g_{\mathrm{hsnr}}(m),\label{eq:g_hsnr}
\end{align}
where $g_{\mathrm{hsnr}}(\cdot)$ is defined in \eqref{eq:g_asym_th} and (a) follows because the \ac{pdf} $f_{\tvec{\mbf{u}}}(\vec{t})$ converges to $f_{\rvec{u}}(\vec{t})$ for every $\vec{t}$, where $\rvec{u}\sim \SGtr(1,\vec{\alpha}^{*},0)$.

Finally, recalling $\mu=\rho/(M-1)$ and substituting \eqref{eq:lb_asym_hu} into \eqref{eq:lb_asym_hu_rho}, \eqref{eq:Elog_s} into \eqref{eq:log_u_gamma}, and finally \eqref{eq:lb_asym_hu_rho}, \eqref{eq:log_u_gamma}, and \eqref{eq:g_hsnr} into \eqref{eq:asym_lb_temp} gives
\begin{equation}
    L^{(\xi)}(\rho) = L_\mathrm{hsnr}(\rho),
\end{equation}
where $L_\mathrm{hsnr}(\rho)$ is defined in \eqref{eq:th_asymp_lb}, which completes the proof of Theorem \ref{th:asym_lower_bound}.

\section{Proof of Lemmas Needed in Theorems~\ref{th:upper_bound}--\ref{th:asym_lower_bound}}\label{app:sec:proof_lemmas}
This section is dedicated to the proof of lemmas used to prove Theorems~\ref{th:upper_bound}--\ref{th:asym_lower_bound}. Before delving into the proofs, we define the following relations which will be used in some of the lemmas.

For all $m\in \{0,\dots,M-1\}$ and $k\in \{0,1,\dots\}$, we define
\begin{equation}
    \mbf{z}_{k,m} =e^{-j\fparen{\thetab_{k,m}+\phase{\mbf{x}}_{k,m}}}\mbf{w}_{k,m},\label{eq:app:z1}
\end{equation}
which gives $\mbf{z}_{k,m}\sim \CN(0,1)$ and
\begin{equation}
    \mbf{s}_{k,m} =\abs{\mbf{x}_{k,m}}, \label{eq:app:s1}
\end{equation}
which results in rewriting \eqref{eq:EO_chan_model} as
\begin{equation}
    \mbf{y}_{k,m} = e^{j\fparen{\mbf{\thetab}_{k,m}+ \phase{\mbf{x}}_{k,m}}}(\mbf{s}_{k,m}+\mbf{z}_{k,m}).\label{eq:app:ykm}
\end{equation}
Consequently, we have
\begin{align}
    \abs{\mbf{y}_{k,m}} &= \abs{\mbf{s}_{k,m}+\mbf{z}_{k,m}},\label{eq:app:abs_ykm} \\
    \phase{\mbf{y}}_{k,m} &= \thetab_{k,m} \modplus \phase{\mbf{x}}_{k,m} \modplus \phase{\mbf{s}_{k,m} + \mbf{z}_{k,m}}.\label{eq:app:phase_ykm}
\end{align}
% and define
% \begin{align}
%     \mbf{y}'_{k,m} &= e^{-j\phase{\mbf{x}}_{k,m}} \mbf{y}_{k,m}\\
%     \phase{\mbf{y}}'_{k,m} &=     \phase{\mbf{y}}_{k,m} \modsub \phase{\mbf{x}}_{k,m}\nonumber\\
%     & = \thetab_{k,m} \modplus \phase{\mbf{s}_{k,m} + \mbf{z}_{k,m}},\label{eq:app:phase_ykm'}
% \end{align}
%Also, recall from \eqref{def:theta_km_Eo_comb} that $\thetab_{1,m} = \thetacb{1} \modplus m\thetarb{1}$ and define 
%\begin{align}
 %   \mbf{y}''_{1,m} &=e^{-jm\thetarb{1}}  \mbf{y}_{1,m}\nonumber\\
 %   & = e^{j\thetacb{1}+ \phase{\mbf{x}}_{1,m}}(\mbf{s}_{1,m}+\mbf{z}_{1,m}).\label{eq:app:ykm_zegond}
%\end{align}
The vector forms of the abovementioned parameters are defined as $\rvec{z}_k = \{\mbf{z}_{k,m}\}_{m=0}^{M-1}$, $\rvec{s}_k = \{\mbf{s}_{k,m}\}_{m=0}^{M-1}$, and $\rvec{y}_k = \{\mbf{y}_{k,m}\}_{m=0}^{M-1}$.

In the following sections, the lemmas used in the proofs of Theorems~\ref{th:upper_bound}--\ref{th:asym_lower_bound} are presented. Specifically, Lemmas~\ref{lemma:general_ub}--\ref{lemma:I_y_theta} are used for Theorem~\ref{th:upper_bound}, Lemma~\ref{lemma:general_lb} for Theorem~\ref{th:lower_bound} for Theorem~\ref{th:asym_upper_bound}, and Lemmas~\ref{lemma:g_asym}--\ref{lemma:alpha_prime_power} for Theorem~\ref{th:asym_lower_bound}.

\subsection{Proof of Lemma~\ref{lemma:general_ub}}\label{sec:proof_lemma_general_ub}
Starting with the chain rule, we can write
\begin{equation}
    I(\rvec{x}_1^n;\rvec{y}_1^n) = \sum_{k=1}^{n} {I(\rvec{x}_1^n;\rvec{y}_k\given\rvec{y}_1^{k-1})}.\label{eq:mut_info}
\end{equation}
Following the footsteps of \cite{durisi_MIMO_Cap:2013}, we can upper-bound each term on the \ac{RHS} of \eqref{eq:mut_info} as
\begin{align}
    I(\rvec{x}_1^n;\rvec{y}_k\given\rvec{y}_1^{k-1}) & =  h(\rvec{y}_k\given\rvec{y}_1^{k-1})-h(\rvec{y}_k\given\rvec{y}_1^{k-1},\rvec{x}_1^n)\nonumber\\
    %---------------------------------
    & \overset{\text{(a)}}{\le}  h(\rvec{y}_k)-h(\rvec{y}_k\given\rvec{y}_1^{k-1},\rvec{x}_1^n)\nonumber\\
    %---------------------------------
    & \overset{\text{(b)}}{=}  h(\rvec{y}_k)-h(\rvec{y}_k\given\rvec{y}_1^{k-1},\rvec{x}_1^{k})\nonumber\\
    %--------------------------------
    &\overset{\text{(c)}}{\le} h(\rvec{y}_k) -h(\rvec{y}_k\given\rvec{y}_1^{k-1},\rvec{x}_1^{k-1},\rvec{x}_k,\vec{\thetab}_{k-1})\nonumber\\
    %---------------------------------
    & \overset{\text{(d)}}{=} h(\rvec{y}_k)-h(\rvec{y}_k\given\rvec{x}_k,\vec{\thetab}_{k-1})\nonumber\\
    %---------------------------------
    & = I(\rvec{y}_k;\rvec{x}_k,\vec{\thetab}_{k-1})\nonumber\\
    %---------------------------------
    & =  I(\rvec{x}_k;\rvec{y}_k)+I(\rvec{y}_k;\vec{\thetab}_{k-1}\given\rvec{x}_k)\nonumber\\
    & \note{e}{=}  I(\rvec{x}_1;\rvec{y}_1)+I(\rvec{y}_1;\vec{\thetab}_{0}\given\rvec{x}_1),
    \label{eq:mut_info_ub}
\end{align}
where (a) and (c) follow as \added{conditioning does not increase} entropy; (b) holds because based on \eqref{eq:EO_chan_model}--\eqref{eq:EO_chan_model_c}, given $\rvec{x}_{1}^{k}$ the output ${\rvec{y}_k}$ is independent of future inputs $\rvec{x}_{k+1}^{n}$; (d) follows since given $\vec{\thetab}_{k-1}$, the output $\rvec{y}_k$ is independent of the pair $(\rvec{y}_1^{k-1},\rvec{x}_1^{k-1})$. Finally, (e) follows because $\{\rvec{x}_k\}$,$\{\rvec{y}_k\}$, and $\{\vec{\thetab}_k\}$ are stationary processes. 

Note that \eqref{eq:mut_info_ub} holds for any channel in the form of \eqref{eq:EO_chan_model}--\eqref{eq:EO_chan_model_c} regardless of how the phase noises are correlated.

Finally, substituting \eqref{eq:mut_info_ub} into \eqref{eq:mut_info}, then \eqref{eq:mut_info} into \eqref{eq:capacity} gives \eqref{eq:cap_ub} and the proof is completed.

\subsection{Proof of Lemma \ref{lemma:duality_bound}}\label{app:lemma_duality_bound}
By duality, for every probability distribution $\mathcal{Q}_{\rvec{x}_1}$ on $\rvec{x}_1$ and any \ac{pdf} $f_{\rvec{y}_1}$ on $\rvec{y}_1$ we have \cite{lapidoth_duality:2003}
\begin{align}
    I(\rvec{x}_1;\rvec{y}_1) \le -\E[\log f_{\rvec{y}_1}(\rvec{y}_1)]-h(\rvec{y}_1|\rvec{x}_1).\label{eq:duality}
\end{align}
%Recall from \eqref{eq:ub_def_y1} that $\rvec{y}_1 = e^{j\fparen{\vec{\thetab}_1+\phase{\rvec{x}}_1}}\circ(\rvec{s}_1+\rvec{z}_1)$ where $\smash{\rvec{z}_1 = e^{-j\fparen{\vec{\thetab}_1+\phase{\rvec{x}}_1}}\circ\rvec{w}_1}$ and $\rvec{s}_1 = \abs{\rvec{x}_1}$.
For any probability distribution $\mathcal{Q}_{\rvec{x}_1}$ satisfying $\E[\norm{\rvec{x}_1}^2]\le \rho$, we have
\begin{align}
    1-\frac{\E[\norm{\rvec{x}_1}^2]+M}{\rho+M}\ge 0.
\end{align}
Now for any given $\lambda\ge 0$,
\begin{equation}
        I(\rvec{x}_1;\rvec{y}_1) \le -\E[\log f_{\rvec{y}_1}(\rvec{y}_1)]-h(\rvec{y}_1|\rvec{x}_1)+\lambda\bigg(1-\frac{\E[\norm{\rvec{x}_1}^2]+M}{\rho+M}\bigg)\label{eq:mi_dual}.
\end{equation}
To evaluate the first term on the \ac{RHS} of \eqref{eq:mi_dual}, we take $\rvec{y}_1 \sim \CSGtr(\mu,\vec{\alpha},0)$ and thus $f_{\rvec{y}_1}$ is defined according to \eqref{eq:complex_square_root_truncated_gamma}, where we set $\mu= (\rho+M)/\alpha_{\Sigma}$ and $\alpha_{\Sigma} = \sum_{m=0}^{M-1}\alpha_m$. 
%$\vec{\alpha}= \{\alpha_0,\dots,\alpha_{M-1}\}$ to be optimized later. We take $\alpha_{\Sigma} = \sum_{m=0}^{M-1}\alpha_m$ and set $\mu= \alpha_{\Sigma}/(\rho+M)$.
Essentially, we let each $\mbf{y}_{1,m}$ be independent and circularly symmetric with the squared magnitude $\abs{\mbf{y}_{1,m}}^2$ following a single-variate gamma distribution. 

Utilizing $f_{\rvec{y}_1}$ from \eqref{eq:complex_square_root_truncated_gamma} and that $\Gamma(\alpha,0) = \Gamma(\alpha)$, the first term in the \ac{RHS} of \eqref{eq:mi_dual} can be evaluated as
\begin{align}
    -\E[\log{f_{\rvec{y}_1}(\rvec{y}_1)}] &=  \alpha_{\Sigma}\log\frac{\rho+M}{\alpha_{\Sigma}}+\!\sum_{m=0}^{M-1}\!\log{\Gamma(\alpha_m)}+ M\log \pi\nonumber\\
    &\quad +\sum_{m=0}^{M-1}(1-\alpha_m) \Eb{\log|\mbf{y}_{1,m}|^2} +\frac{\alpha_{\Sigma}}{\rho+M}\Eb {\lVert{\rvec{y}_1}\rVert^2 }\cdot\log e.
    \label{eq:eqy}
\end{align}
% where (a) follows because 
% \begin{align}
%   \Eb{\norm{\rvec{y}_1}^2} &=\Eb{\norm{\rvec{s}_1}^2}+\Eb{\norm{\rvec{z}_1}^2}\nonumber\\
%   &=\Eb{\norm{\rvec{s}_1}^2}+M.
% \end{align}
Note that $\abs{\mbf{y}_{1,m}} = \abs{\mbf{s}_{1,m}+\mbf{z}_{1,m}}$ and $\E[\lVert\rvec{y}_1\rVert^2]= \E[\norm{\rvec{s}_1}^2]+M$.

To lower bound the second term of the \ac{RHS} of \eqref{eq:mi_dual}, we write the conditional differential entropy term as
\begin{align}
    h(\rvec{y}_1|\rvec{x}_1)&=h(\mbf{y}_{1,0}\given\rvec{x}_1)+h\left(\mbf{y}_{1,1}\given\rvec{x}_1,\mbf{y}_{1,0}\right)+\sum_{m=2}^{M-1}h\left(\mbf{y}_{1,m}|\rvec{x}_1,\brace{\mbf{y}_{1,i}}_{i=0}^{m-1}\right)\nonumber\\
    %Line 2
    %------------------------------------------
    &=h(\mbf{y}_{1,0}\given \mbf{x}_{1,0})+h(\mbf{y}_{1,1}\given \mbf{x}_{1,0},\mbf{x}_{1,1},\mbf{y}_{1,0})+\sum_{m=2}^{M-1}h\left(\mbf{y}_{m}|\brace{\mbf{x}_{1,i}}_{i=0}^{m},\brace{\mbf{y}_{1,i}}_{i=0}^{m-1}\right)\nonumber\\
    %Line 3
    %------------------------------------------
    &\note{a}{\ge}h(\mbf{y}_{1,0}\given \mbf{x}_{1,0})+h(\mbf{y}_{1,1}\given \mbf{x}_{1,0}, \mbf{x}_{1,1},\mbf{y}_{1,0},\thetacb{1})+\sum_{m=2}^{M-1}h\left(\mbf{y}_{1,m}|\brace{\mbf{x}_{1,i}}_{i=0}^{m},\brace{\mbf{y}_{1,i}}_{i=0}^{m-1},\thetacb{1},\thetarb{1}\right)\nonumber\\
    %Line 4
    %------------------------------------------
    &\note{b}{=}h(\mbf{y}_{1,0}\given \mbf{x}_{1,0})\!+\!h(\mbf{y}_{1,1}\given \mbf{x}_{1,1},\thetacb{1}\!+\!(M\!-\!2)\log(\pi e),
    \label{eq:hyx_final_temp}
    \end{align}
where (a) follows because \added{conditioning does not increase} the entropy; (b) holds since when $\thetacb{1}$ is given, $\mbf{y}_{1,1}$ is independent of the pair $(\mbf{x}_{1,0},\mbf{y}_{1,0})$ and given the pair $(\thetacb{1}, \thetarb{1})$, the last summation term becomes the summation of entropies of independent Gaussian variables, i.e., $(\mbf{y}_{1,m}|\mbf{x}_{1,m},\thetacb{1},\thetarb{1}) \sim \mathcal{CN}(e^{j(\thetacb{1}+m\thetarb{1})}\mbf{x}_{1,m},1)$.

Using \eqref{eq:app:ykm}, the first term on the \ac{RHS} of \eqref{eq:hyx_final_temp} is
\begin{align}
    h(\mbf{y}_{1,0}\given \mbf{x}_{1,0}) &=  h\left(e^{j(\thetacb{1}+\phase{\mbf{x}}_{1,0})}(\mbf{s}_{1,0}+\mbf{z}_{1,0})\given \mbf{s}_{1,0},\phase{\mbf{x}}_{1,0}\right) \nonumber\\
    &\note{a}{=}  h\left(e^{j\thetacb{1}}(\mbf{s}_{1,0}+\mbf{z}_{1,0})\given \mbf{s}_{1,0}\right) \nonumber\\
    &\note{b}{=} h\paren{|\mbf{s}_{1,0}+\mbf{z}_{1,0}|^2\given \mbf{s}_{1,0}}+\log\pi,\label{eq:app_c_hy0}
\end{align}
where (a) follows because $\mbf{z}_{1,0}$ defined in \eqref{eq:app:z1} is circularly symmetric and its distribution remains the same with rotation; to obtain (b) we applied Lemma~\ref{lemma:polar_entropy} as $e^{j\thetacb{1}}(\mbf{s}_{1,0}+\mbf{z}_{1,0})$ is circularly symmetric given $\mbf{s}_{1,0}$. Following the same steps leading to \eqref{eq:app_c_hy0}, the second term on the \ac{RHS} of \eqref{eq:hyx_final_temp} can be written as
\begin{align}
 h(\mbf{y}_{1,1}\given \mbf{x}_{1,1},\thetacb{1}) &= h\left(e^{j(\thetacb{1}+\thetarb{1})}(\mbf{s}_{1,1}+\mbf{z}_{1,1})\given \mbf{s}_{1,1},\thetacb{1}\right)\nonumber\\
 &= h\fparen{|\mbf{s}_{1,1}+\mbf{z}_{1,1}|^2\given \mbf{s}_{1,1}}+\log \pi. \label{eq:app_c_hy1}
\end{align}
Combining \eqref{eq:app_c_hy1} and \eqref{eq:app_c_hy0} into \eqref{eq:hyx_final_temp} gives 
\begin{equation}
    h(\rvec{y}_1\given\rvec{x}_1) \ge h\paren{|\mbf{s}_{1,0}+\mbf{z}_{1,0}|^2\given \mbf{s}_{1,0}}+h\paren{|\mbf{s}_{1,1}+\mbf{z}_{1,1}|^2\given \mbf{s}_{1,1}} +\log(\pi^M)+(M-2)\log e.\label{eq:app_c_hyx_final}
\end{equation}
Substituting \eqref{eq:eqy} and \eqref{eq:app_c_hyx_final} into \eqref{eq:mi_dual} \added{and that $\Eb{\log\fparen{\abs{\mbf{s}_{1,m} + \mbf{z}_{1,m}}^2}} = \Eb{\log\fparen{\mbf{s}_{1,m}^2} + \mathrm{E}_1\fparen{\mbf{s}_{1,m}^2}}$ from \eqref{eq:chi_square_elog}}\label{r1:c1_b} gives
\begin{align}
I(\rvec{x}_1;\rvec{y}_1) 
 &\le \alpha_{\Sigma}\log\frac{\rho+M}{\alpha_{\Sigma}}+ \lambda-(M-2)\log e\nonumber\\
 &\quad+\sum_{m=0}^{M-1}\fparen{\log\Gamma(\alpha_m)+ (1-\alpha_m)%\Eb{\log\fparen{|\mbf{s}_{1,m}+\mbf{z}_{1,m}|^2}}
 \added{\Eb{\log\fparen{\mbf{s}_{1,m}^2} +\mathrm{E}_1\fparen{\mbf{s}_{1,m}^2}}}
 }\nonumber\\
&\quad + (\alpha_{\Sigma}\log e-\lambda)\!\bigg(\frac{\E[\norm{\rvec{s}_1}^2]+M}{\rho+M}\bigg)\nonumber\\ 
&\quad-h\paren{|\mbf{s}_{1,0}+\mbf{z}_{1,0}|^2\bgiven \mbf{s}_{1,0}}-h\paren{|\mbf{s}_{1,1}+\mbf{z}_{1}|^2\bgiven \mbf{s}_{1,1}}\nonumber\\
& = \alpha_{\Sigma}\log\left(\frac{\rho+M}{\alpha_{\Sigma}}\right)\!+\!d_{\lambda,\vec{\alpha}}+\!R_{\lambda,\vec{\alpha}}(\rho,\rvec{s}_1),\label{eq:upper_bound_modified}
\end{align}
where $R_{\lambda,\vec{\alpha}}(\cdot)$ and $d_{\lambda,\vec{\alpha}}$ are defined in \eqref{eq:Rs} and \eqref{eq:d_lambda}, respectively.

Finally, replacing back $\rvec{s}_1$ with $\abs{\rvec{x}_1}$ gives \eqref{eq:duality_bound} and completes the proof.

\subsection{Proof of Lemma \ref{lemma:I_y_theta}}\label{app:lemma:I_y_theta}
The proof is divided into two parts. The first part simplifies the \ac{RHS} \eqref{eq:cap_ub} for $M=2$, and the second part provides an upper bound on the \ac{RHS} of \eqref{eq:cap_ub} for $M>2$.
\vspace{0.2cm}

\emph{The second term on the \ac{RHS} of \eqref{eq:cap_ub} for $M=2$:} \hfill
   
From \eqref{eq:app:z1}--\eqref{eq:app:ykm} we can write $\phase{\rvec{y}}_1 = \rvec{\thetab}_1 \modplus \phase{\rvec{x}}_1 \modplus \phase{\rvec{s}_1+\rvec{z}_1}$. Then, applying the chain rule on the second term on the \ac{RHS} of \eqref{eq:cap_ub} gives
\begin{align}
    I\paren{\rvec{y}_1;\vec{\thetab}_0\bgiven\rvec{x}_1}&\note{a}{=}I\fparen{\phase{\rvec{y}}_1;\vec{\thetab}_0\bgiven\rvec{x}_1,|\rvec{y}_1|}\nonumber\\
    & = I\fparen{\phase{\rvec{y}}_1\modsub\phase{\rvec{x}}_1;\vec{\thetab}_0\bgiven\rvec{x}_1,|\rvec{y}_1|}\nonumber\\
    & \note{b}{=} I\fparen{\vec{\thetab}_1\modplus\phase{\rvec{s}_1+\rvec{z}_1};\vec{\thetab}_0\bgiven\rvec{s}_1,|\rvec{y}_1|}\nonumber\\
    & = h\fparen{\vec{\thetab}_1\modplus\phase{\rvec{s}_1+\rvec{z}_1}\bgiven\rvec{s}_1,|\rvec{y}_1|} -h\fparen{\vec{\thetab}_1\modplus\phase{\rvec{s}_1+\rvec{z}_1}\bgiven\rvec{s}_1,|\rvec{y}_1|,\vec{\thetab}_0}\nonumber\\
     & \note{c}{=} 2\log(2\pi)-h\fparen{\vec{\thetab}_1\modplus\phase{\rvec{s}_1+\rvec{z}_1}\bgiven\rvec{s}_1,|\rvec{y}_1|,\vec{\thetab}_0}\nonumber\\
       & = 2\log(2\pi) -h\Big(\brace{\thetacb{1}\modplus m\thetarb{1}\modplus\phase{\mbf{s}_{1,m}+\mbf{z}_{1,m}}}_{0}^{1}\bgiven\rvec{s}_1,|\rvec{s}_1+\rvec{z}_1|,\vec{\thetab}_0\Big)\nonumber\\
    & = 2\log(2\pi)-h\Big(\brace{\Deltacb{1}\modplus m\Deltarb{1}\modplus\phase{\mbf{s}_{1,m}+\mbf{z}_{1,m}}}_{0}^{1}\bgiven\rvec{s}_1,|\rvec{s}_1+\rvec{z}_1|\Big)\nonumber\\
    \label{eq:I_y_theta_M2_final} 
\end{align}
where (a) holds since the pair $(|\rvec{y}_{1}|,\vec{\thetab}_{0})$ are independent; (b) follows since $\rvec{s}_1=\abs{\rvec{x}_1}$ and $\phase{\rvec{x}_1}$ are independent as $\rvec{x}_1$ is circularly symmetric; (c) holds since the two components of  $\vec{\thetab}_1=(\thetacb{1},\thetacb{1}\modplus\thetarb{1})$ are independent and uniform  on $[-\pi,\pi)$, from which follows that the components of $\vec{\thetab}_1\modplus\phase{\rvec{s}_1+\rvec{z}_1}$ are also independent and uniform on $[-\pi,\pi)$.

\vspace{.2cm}
\emph{The second term on the \ac{RHS} of \eqref{eq:cap_ub} for $M>2$:} \hfill
\vspace{.2cm}
\begin{align}
    I\paren{\rvec{y}_1;\vec{\thetab}_0\bgiven\rvec{x}_1}&\le     I\paren{\rvec{y}_1,\thetarb{1};\vec{\thetab}_0\bgiven\rvec{x}_1}\nonumber\\
    & = I\paren{\thetarb{1};\vec{\thetab}_0\bgiven\rvec{x}_1}+I\paren{\rvec{y}_1;\vec{\thetab}_0\bgiven\rvec{x}_1,\thetarb{1}}\nonumber\\
    & \note{a}{=} I\paren{\thetarb{1};\thetacb{0},\thetarb{0}\bgiven\rvec{x}_1}+I\paren{\rvec{y}_1;\thetacb{0},\thetarb{0}\bgiven\rvec{x}_1,\thetarb{1}}\nonumber\\
    & \note{b}{=} I\paren{\thetarb{1};\thetarb{0}\bgiven\rvec{x}_1}+I\paren{\rvec{y}_1;\thetacb{0}\bgiven\rvec{x}_1,\thetarb{1}},\label{eq:I_y_theta_M3}
\end{align}
where (a) holds since $(\thetacb{0},\thetarb{0})$ are sufficient statistics for $\vec{\thetab}_0$, and (b) follows because, given $\thetarb{1}$, the pair $(\rvec{y}_1, \thetarb{0})$ is independent, hence $I(\rvec{y}_1; \thetarb{0} \given \rvec{x}_1, \thetarb{1}) = 0$.

The first term on the \ac{RHS} of \eqref{eq:I_y_theta_M3} is
\begin{align}
    I\paren{\thetarb{1};\thetarb{0}\bgiven\rvec{x}_1} &\note{a}{=}  I\paren{\thetarb{1};\thetarb{0}}\nonumber\\
    &= h(\thetarb{1})-h(\thetarb{1}\given\thetarb{0})\nonumber\\
    &= \log(2\pi)-h(\Deltarb{1}),\label{eq:I_y_theta_M3_a}
\end{align}
where (a) holds since $(\thetarb{1},\thetarb{0})$ are independent of the inputs $\rvec{x}_1$.% and (b) follows as that the pair $(\thetarb{1},\thetarb{0})$ are independent of $\thetacb{0}$ .

Recall $\mbf{y}_{1,m}$ and $\mbf{z}_{1,m}$ from \eqref{eq:app:ykm} and \eqref{eq:app:z1}, respectively. Then, define 
\begin{align}
    \tilde{\mbf{y}}_{1,m} &=e^{-jm\thetarb{1}}  \mbf{y}_{1,m}\nonumber\\
    & = e^{j\thetacb{1}+ \phase{\mbf{x}}_{1,m}}(\mbf{s}_{1,m}+\mbf{z}_{1,m}),\label{eq:app:y1_zegond}
\end{align}
and its vector form as $\tvec{\mbf{y}}_1 = (\tilde{\mbf{y}}_{1,0},\dots, \tilde{\mbf{y}}_{1,M-1})$.  Then, the second term on the \ac{RHS} of \eqref{eq:I_y_theta_M3} can be written as 
\begin{align}
I\paren{\rvec{y}_1;\thetacb{0}\bgiven\rvec{x}_1,\thetarb{1}} &= I\fparen{\left\{e^{-jm\thetarb{1}} \mbf{y}_{1,m}\right\}_{m=0}^{M-1};\thetacb{0}\bgiven\rvec{x}_1,\thetarb{1}}\nonumber\\
    & \note{a}{=} I\paren{\tvec{\mbf{y}}_1;\thetacb{0}\bgiven\rvec{x}_1},\label{eq:I_y_theta_M3_b}
\end{align}
where in (a) we used that $\tvec{\mbf{y}}_1$ is independent of $\thetarb{1}$. 

Recall $\norm{\rvec{x}_1} = \norm{\rvec{s}_1}$ and define $\rvec{u}_1 = \rvec{x}_1/\norm{\rvec{x}_1} = \exp{\paren{j\phase{\rvec{x}}_1}}\circ\rvec{s}_1/\norm{\rvec{s}_1}$. Let the matrix $\mbf{U} = (\rvec{u}_1^T,\rvec{u}_2^T,\dots,\rvec{u}_{M}^T)^T$ with $\rvec{u}_2,\dots,\rvec{u}_{M}$ are vectors orthogonal to $\rvec{u}_1^{\dagger}$ and mutually orthonormal i.e., $\mbf{U}^{\dagger}\mbf{U} = I_{M}$. Then the \ac{RHS} of \eqref{eq:I_y_theta_M3_b} can be written as
\begin{align}
I\paren{\tvec{\mbf{y}}_1;\thetacb{0}\bgiven\rvec{x}_1} &=I\fparen{\mbf{U}^{\dagger}\tvec{\mbf{y}}_1;\thetacb{0}\bgiven \mbf{U}^{\dagger}\rvec{x}_1 }\nonumber\\
   &\note{a}{=} I\fparen{\rvec{u}_1^{\dagger}\tvec{\mbf{y}}_1;\thetacb{0}\bgiven \rvec{u}_1^{\dagger}\rvec{x}_1 }\nonumber\\
   &\note{b}{=}I\paren{e^{j\thetacb{1}}(\norm{\rvec{s}_1}+\mbf{v}_1);\thetacb{0}\bgiven\norm{\rvec{s}_1} }\nonumber\\
   & \note{c}{=} I\paren{\thetacb{1}\modplus\phase{\norm{\rvec{s}_1}+\mbf{v}_1};\thetacb{0}\bgiven\norm{\rvec{s}_1},\abs{\norm{\rvec{s}_1}+\mbf{v}_1}}\nonumber\\
   &\note{d}{=} \log(2\pi) -h\paren{\Deltacb{1}\modplus\phase{\norm{\rvec{s}_1}+\mbf{v}_1}\bgiven\norm{\rvec{s}_1},\abs{\norm{\rvec{s}_1}+\mbf{v}_1}}.\label{eq:I_y_theta_M3_c}
\end{align}
Here, (a) holds since $\rvec{u}_1^\dagger\tvec{\mbf{y}}_1$ is sufficient statistics for $\thetacb{0}$ as $\rvec{u}_i^{\dagger} \tvec{\mbf{y}}_1 \sim\CN(0,1)$ and independent of $\thetacb{0}$ for all $i\in\{2,\dots,M\}$; in (b) $\mbf{v}_1 = \rvec{u}_1^{\dagger}\rvec{z}_1$ resulting in $\mbf{v}_1 \sim \CN(0,1)$; in (c) the pair $\left(\thetacb{0},\big\vert\norm{\rvec{s}_1}+\mbf{v}_1\big\vert\right)$ are independent; finally, (d) holds since $\thetacb{1}\sim \Uniform[-\pi,\pi)$. Substituting \eqref{eq:I_y_theta_M3_a}, \eqref{eq:I_y_theta_M3_b} and then \eqref{eq:I_y_theta_M3_c} into \eqref{eq:I_y_theta_M3} we have
\begin{equation}
       I\paren{\rvec{y}_1;\vec{\thetab}_0\bgiven\rvec{x}_1}\le 2\log(2\pi)-h(\Deltarb{1}) -h\paren{\Deltacb{1}\modplus\phase{\norm{\rvec{s}_1}+\mbf{v}_1}\bgiven\norm{\rvec{s}_1},\abs{\norm{\rvec{s}_1}+\mbf{v}_1}}.\label{eq:I_y_theta_M3_final}
\end{equation}

Now, replacing $\rvec{s}_1$ with $\abs{\rvec{x}_1}$ and combining \eqref{eq:I_y_theta_M3_final} and \eqref{eq:I_y_theta_M2_final} gives %\eqref{eq:lemm_I_y_theta} and proof is completed.
\begin{equation}
    I(\rvec{y}_1;\vec{\thetab}_0\given\rvec{x}_1) \le 2\log(2\pi)+F(M,\abs{\rvec{x}_1},\Deltacb{1},\Deltarb{1}),\label{eq:I_y_theta_final}
\end{equation}
where $F(\cdot)$ is defined in \eqref{eq:Fs}.
%============================================================
\subsection{Proof of Lemma \ref{lemma:general_lb}}\label{sec:proof_lemma_general_lb}
To initiate the derivation of the lower bound, we apply the chain rule, leveraging the non-negativity of mutual information. This allows us to write
 \begin{align}
     I(\rvec{x}_1^n;\rvec{y}_1^n) &= \sum_{k=1}^{n} I\paren{\rvec{x}_k;\rvec{y}_1^n\given \rvec{x}_1^{k-1}}\nonumber\\
     &\ge \sum_{k=2}^{n} I\paren{\rvec{x}_k;\rvec{y}_1^k\given \rvec{x}_1^{k-1}}.\label{eq:lb}
 \end{align}
Fix $k\ge 2$ and set 
 \begin{equation}
     \epsilon_k =I\paren{\rvec{x}_k;\vec{\thetab}_{k-1}\given \rvec{x}_{k-1},\rvec{y}_{k-1},\rvec{y}_k}.
 \end{equation}
Following similar footsteps as in \cite{durisi_MIMO_Cap:2013}, we can write 
\begin{align}
    I\paren{\rvec{x}_k;\rvec{y}_1^k\given \rvec{x}_1^{k-1}} &\note{a}{=}I\paren{\rvec{x}_k;\rvec{y}_1^k,\rvec{x}_1^{k-1}}\nonumber\\
    &\note{b}{\ge} I\paren{\rvec{x}_k;\rvec{y}_k,\rvec{y}_{k-1},\rvec{x}_{k-1}}\nonumber\\
    &=I\paren{\rvec{x}_k;\rvec{y}_k,\rvec{y}_{k-1},\rvec{x}_{k-1},\vec{\thetab}_{k-1}}-\epsilon_k\nonumber\\
    &\note{c}{=} I\paren{\rvec{x}_k;\rvec{y}_k,\vec{\thetab}_{k-1}}-\epsilon_k\nonumber\\
    &\note{d}{=}I\paren{\rvec{x}_k;\rvec{y}_k\given\vec{\thetab}_{k-1}}-\epsilon_k\nonumber\\
    &\note{e}{=}I\paren{\rvec{x}_2;\rvec{y}_2\given\vec{\thetab}_{1}}-\epsilon_2,\label{eq:lb_a}
\end{align}
where (a) holds since $\{\rvec{x}_k\}$ are independent; (b) is a consequence of the chain rule and nonnegativity of mutual information; (c) follows because when the pair $(\vec{\thetab}_{k-1},\rvec{y}_k)$ is given, $\rvec{x}_k$ and the pair $(\rvec{y}_{k-1},\rvec{x}_{k-1})$ are conditionally independent; (d) follows because $\rvec{x}_k$ and $\vec{\thetab}_{k-1}$ are independent; finally (e) holds due to stationarity. Substituting \eqref{eq:lb_a} into \eqref{eq:lb} and the result into \eqref{eq:capacity} gives \eqref{eq:cap_lb} for all distributions on $\rvec{x}_2$ such that $\E[\norm{\rvec{x}_2}^{\added{2}}]\le \rho$.

% \bibliography{components/references}
\subsection{Proof of Lemma~\ref{lemma:alpha_prime_power}}\label{app:lemma:alpha_prime}
We can start with the definition of $J(\cdot)$ from \eqref{eq:def_J} and write
\begin{align}
    J(\alpha'_m(\gamma), \gamma) &= \alpha'_m(\gamma) + \frac{e^{-\gamma} \gamma^{\alpha'_m(\gamma)}}{\Gamma(\alpha'_m(\gamma),\gamma)} \nonumber\\
    &\note{a}{\le}     \alpha'_m(\gamma) + \frac{e^{-\gamma} \gamma^{\alpha'_m(\gamma)}}{\Gamma(1,\gamma)}\nonumber\\
    &\note{b}{=} \alpha'_m(\gamma) + \gamma^{\alpha'_m(\gamma)},\label{eq:upper_bound_on_Es}
\end{align}
where (a) follows because $\Gamma(\alpha'(\gamma),\gamma) \ge \Gamma(1,\gamma)$ from Lemma~\ref{lemma:igamma_lb} and (b) follows as $\Gamma(1,\gamma) = e^{-\gamma}$. Note that in (a) we could utilize Lemma~\ref{lemma:igamma_lb} because $\gamma \le x_\mr{max} < x_0$ and $0\le\alpha'(\gamma)\le 1$. 

Now, substituting \eqref{eq:alpha_prime_def} into \eqref{eq:upper_bound_on_Es} we can write
\begin{align}
  \alpha'_m(\gamma) + \gamma^{\alpha'_m(\gamma)} &= \alpha_m^* - c_m(\gamma) + \frac{\gamma^{\alpha_m^*}}{\gamma^{c_m(y)}}\nonumber\\
  & \note{a}{=} \alpha_m^* - c_m(\gamma) +c_m(\gamma)\nonumber\\
  & = \alpha_m^*,
\end{align}
where (a) follows from \eqref{eq:lemma_gamma_pow_c} in Lemma~\ref{lemma:gamma_pow_c}. Thus,
\begin{align}
     J(\alpha'_m(\gamma), \gamma) \le \alpha^*_m,
\end{align}
which proves \eqref{eq:lemma:j_upper_bound}.

Next we prove \eqref{eq:alpha_prime_limit}, which requires to show $\lim_{\gamma \to 0^{+}} c_m(\gamma) = 0$. Based on the definition of the Lambert W function we have
\begin{align}
    W_{\mr{L}}(x) = \frac{x}{e^{W_{\mr{L}}(x)}}. \label{eq:lambert_def}
\end{align}
Thus, we can write
\begin{align}
       \lim_{\gamma \to 0^{+}} c_m(\gamma) &= \lim_{\gamma \to 0^{+}} \frac{\gamma^{\alpha_m^*}\log_{e}\gamma}{e^{W_{\mr{L}}(\gamma^{\alpha_m^*}\log_{e}\gamma)}\log_{e}\gamma} \nonumber\\
       & = \lim_{\gamma \to 0^{+}} \frac{\gamma^{\alpha_m^*}}{e^{W_{\mr{L}}(\gamma^{\alpha_m^*}\log_{e}\gamma)}}\nonumber\\
       & = \frac{0}{1}\nonumber\\
       &= 0,
\end{align}
which proves \eqref{eq:alpha_prime_limit}.

\subsection{Proof of Lemma~\ref{lemma:g_asym}}\label{app:lemma_g_asym}
In this proof, we need to characterize $g(m,\vec{s})$ for a deterministic vector 
\begin{equation}
  \vec{s}>\sqrt{\mu\gamma}.\label{eq:app:sm_lowerbound}
\end{equation}

\label{r1:c2_c}
\added{Starting with the first term on the \ac{RHS} of \eqref{eq:Phi_func} and using \eqref{eq:app:sm_lowerbound}, we have $\phi(m,\vec{s}) \le 6/(\mu \gamma)$. Thus, for $\mu\gamma \ge 3e/(2\pi)$ we can write 
\begin{align}
    \min\fparen{\log(2\pi), \frac{1}{2} \log\fparen{\pi e \phi(m,\vec{s})}} = \frac{1}{2} \log(\pi e)  + \frac{1}{2}\log\fparen{ \phi(m,\vec{s})}. \label{eq:app:gm_a}
\end{align}}

For all $m\in\{0,\dots,M-1\}$, consider \ac{iid} $\mbf{z}_m \sim \CN(0,1)$ and let 
\begin{align}
    \bs{\varphi}_m &= \phase{{s}_m+\mbf{z}_m},\label{eq:app:varphi} \\
    \mbf{r}_m &= \abs{{s}_m+\mbf{z}_m}\nonumber\\
              &= s_m \fparen{1+O\fparen{1/{\sqrt{\mu\gamma}}}}.\label{eq:app:rm}
\end{align}
Then, according to Lemma~\ref{lemma:von_mises}, given $s_m$ and $\mbf{r}_m = r_m$, we have $\bs{\varphi}_m \sim \mathcal{VM}(0,2{s}_m r_m)$ with its \ac{pdf} denoted by $f_{\bs{\varphi}_m}^\mathrm{VM}(\varphi_m; 0, 2{s}_m r_m)$.
Using Lemma~\ref{lemma:von_mises_ent}, the \ac{pdf} of a von Mises distribution can be approximated by a wrapped normal distribution as 
\begin{align}
    f_{\bs{\varphi}_m}^\mathrm{VM}(\varphi_m;0,2{s}_m r_m) &\!\note{a}{=} \!f_{\bs{\varphi}_m}^\mathrm{WN}\!\fparen{\!{\varphi}_m;0,\frac{1}{2{s}_m r_m}\!}\!+\!O\!\bigg(\!\frac{1}{\sqrt{{s}_m r_m}}\!\bigg)\!.
    \label{eq:appc_wn}
\end{align}

\added {The second term on the} \ac{RHS} of \eqref{eq:g_func} can be \added{written} as
\begin{align}
    h\fparen{\phase{{s}_{m}+\mbf{z}_{m}}\bgiven {s}_{m},\abs{{s}_{m}+\mbf{z}_{m}}} &= h\fparen{\bs{\varphi}_m\bgiven {s}_{m},\mbf{r}_m}\nonumber\\
    &\note{a}{=} \frac{1}{2}\log(\pi e)-\frac{1}{2}\log\fparen{s_m r_m} +O\fparen{\frac{1}{\sqrt{s_m r_m}}}\nonumber\\
& \note{b}{=}\frac{1}{2}\log(\pi e)-\frac{1}{2}\log\fparen{s_m^2}+O\fparen{\frac{1}{\sqrt{\mu\gamma}}} ,\label{eq:app:gm_b}
\end{align}
\added{where in (a) we utilized Lemma~\ref{lemma:von_mises_entropy}, and (b) follows from \eqref{eq:app:rm}.}

Finally, substituting \eqref{eq:app:gm_b} \added{ and \eqref{eq:app:gm_a}} into \eqref{eq:g_func} gives \eqref{eq:gm_deterministic}, and the proof is completed. 
\label{r1:c2_c_end}

\bibliography{components/references}

\end{document}